\documentclass[12pt,preprint]{emulateapj}

\usepackage{amsmath,natbib,graphicx}
\usepackage{epsf}
\usepackage{epstopdf}
\input{epsf}
\usepackage{epsfig}
\usepackage{color}
\usepackage{lscape}
\DeclareGraphicsExtensions{.jpg,.pdf,.png,.eps,.ps}
\graphicspath{{FIGURES/}}

\newcommand{\Tcmb}{\mbox{$T_{\mbox{\tiny CMB}}$}}

\newcommand{\ltsima}{$\; \buildrel < \over \sim \;$}
\newcommand{\ltsim}{\lower.5ex\hbox{\ltsima}}
\newcommand{\um}{$\mu \mathrm{m}$}
\newcommand{\beq}{\begin{equation}}
\newcommand{\eeq}{\end{equation}}

\newcommand{\sqdeg}{\ensuremath{\mathrm{deg}^2}}

\newcommand{\alphathresh}{1.66}

\slugcomment{Accepted to ApJ}

\begin{document}

\title{Extragalactic millimeter-wave sources in South Pole Telescope survey data: source counts, catalog, and statistics for an 87~square-degree field}

\shorttitle{Millimeter-wave sources in the SPT survey}


\author{
J.~D.~Vieira,\altaffilmark{1,2,3} T.~M.~Crawford,\altaffilmark{1,4} E.~R.~Switzer,\altaffilmark{1} 
P.~A.~R.~Ade,\altaffilmark{5} K.~A.~Aird,\altaffilmark{6}
M.~L.~N.~Ashby,\altaffilmark{7}
B.~A.~Benson,\altaffilmark{1} L.~E.~Bleem,\altaffilmark{1,2} 
M.~Brodwin,\altaffilmark{7,8} J.~E.~Carlstrom,\altaffilmark{1,2,3,4} 
C.~L.~Chang,\altaffilmark{1,3} H.-M.~Cho,\altaffilmark{9}  
A.~T.~Crites,\altaffilmark{1,4} T.~de~Haan,\altaffilmark{10} M.~A.~Dobbs,\altaffilmark{10} 
W.~Everett,\altaffilmark{6} E.~M.~George,\altaffilmark{9} 
M.~Gladders,\altaffilmark{1,4} N.~R.~Hall,\altaffilmark{13} N.~W.~Halverson,\altaffilmark{11}
F.~W.~High,\altaffilmark{12}  G.~P.~Holder,\altaffilmark{10}
W.~L.~Holzapfel,\altaffilmark{9} J.~D.~Hrubes,\altaffilmark{6}
M.~Joy,\altaffilmark{14} R.~Keisler,\altaffilmark{1,2,3} L.~Knox,\altaffilmark{13} 
A.~T.~Lee,\altaffilmark{9,15} E.~M.~Leitch,\altaffilmark{1,4}
M.~Lueker,\altaffilmark{9} D.~P.~Marrone,\altaffilmark{1,16}
V.~McIntyre,\altaffilmark{17} J.~J.~McMahon,\altaffilmark{1,3,18} J.~Mehl,\altaffilmark{1}
S.~S.~Meyer,\altaffilmark{1,2,3,4} J.~J.~Mohr,\altaffilmark{19,20,21}
T.~E.~Montroy,\altaffilmark{22,23}  
S.~Padin,\altaffilmark{1,4} T.~Plagge,\altaffilmark{1,9}
C.~Pryke,\altaffilmark{1,3,4} C.~L.~Reichardt,\altaffilmark{9}
J.~E.~Ruhl,\altaffilmark{22,23} K.~K.~Schaffer,\altaffilmark{1,3}
L.~Shaw,\altaffilmark{10,24} E.~Shirokoff,\altaffilmark{9} 
H.~G.~Spieler,\altaffilmark{15}
B.~Stalder,\altaffilmark{7} Z.~Staniszewski,\altaffilmark{22,23} A.~A.~Stark,\altaffilmark{7} 
K.~Vanderlinde,\altaffilmark{10} W.~Walsh,\altaffilmark{7} R.~Williamson,\altaffilmark{6}
Y.~Yang,\altaffilmark{25}  
O.~Zahn,\altaffilmark{26} and A.~Zenteno\altaffilmark{25}
}

\altaffiltext{1}{Kavli Institute for Cosmological Physics,
University of Chicago,
5640 South Ellis Avenue, Chicago, IL 60637}
\altaffiltext{2}{Department of Physics,
University of Chicago,
5640 South Ellis Avenue, Chicago, IL 60637}
\altaffiltext{3}{Enrico Fermi Institute,
University of Chicago,
5640 South Ellis Avenue, Chicago, IL 60637}
\altaffiltext{4}{Department of Astronomy and Astrophysics,
University of Chicago,
5640 South Ellis Avenue, Chicago, IL 60637}
\altaffiltext{5}{Department of Physics and Astronomy,
Cardiff University,
CF24 3YB, UK}
\altaffiltext{6}{University of Chicago,
5640 South Ellis Avenue, Chicago, IL 60637}
\altaffiltext{7}{Harvard-Smithsonian Center for Astrophysics,
60 Garden Street, Cambridge, MA 02138}
\altaffiltext{8}{W. M. Keck Postdoctoral Fellow at the Harvard-Smithsonian Center for
Astrophysics}
\altaffiltext{9}{Department of Physics,
University of California,
Berkeley, CA 94720}
\altaffiltext{10}{Department of Physics,
McGill University,
3600 Rue University, Montreal, Quebec H3A 2T8, Canada}
\altaffiltext{11}{Department of Astrophysical and Planetary Sciences and Department of Physics,
University of Colorado,
Boulder, CO 80309}
\altaffiltext{12}{Department of Physics, 
Harvard University, 17 Oxford Street, Cambridge, MA 02138}
\altaffiltext{13}{Department of Physics, University of California, 
One Shields Avenue, Davis, CA 95616}
\altaffiltext{14}{Department of Space Science, VP62,
NASA Marshall Space Flight Center,
Huntsville, AL 35812}
\altaffiltext{15}{Physics Division,
Lawrence Berkeley National Laboratory,
Berkeley, CA 94720}
\altaffiltext{16}{Jansky Fellow, National Radio Astronomy Observatory} 
\altaffiltext{17}{Australia Telescope National Facility, CSIRO, Epping NSW 1710, Australia}
\altaffiltext{18}{Department of Physics,  University of Michigan, 
450 Church Street, Ann Arbor, MI 48109}
\altaffiltext{19}{Department of Physics,
Ludwig-Maximilians-Universit\"{a}t,
Scheinerstr.\ 1, 81679 M\"{u}nchen, Germany}
\altaffiltext{20}{Excellence Cluster Universe,
Boltzmannstr.\ 2, 85748 Garching, Germany}
\altaffiltext{21}{Max-Planck-Institut f\"{u}r extraterrestrische Physik,
Giessenbachstr.\ 85748 Garching, Germany}
\altaffiltext{22}{Physics Department,
Case Western Reserve University,
Cleveland, OH 44106}
\altaffiltext{23}{Center for Education and Research in Cosmology and Astrophysics,
Case Western Reserve University,
Cleveland, OH 44106}
\altaffiltext{24}{Department of Physics, Yale University, P.O. Box 208210, 
New Haven, CT 06520-8120}
\altaffiltext{25}{Department of Astronomy and Department of Physics,
University of Illinois,
1002 West Green Street, Urbana, IL 61801}
\altaffiltext{26}{Berkeley Center for Cosmological Physics,
Department of Physics, University of California, and Lawrence Berkeley
National Labs, Berkeley, CA 94720}

\shortauthors{Vieira, et al.}

\email{vieira@caltech.edu}


\begin{abstract}
We report the results of an 87~\sqdeg \ point-source survey centered at 
R.A. $5^\mathrm{h} 30^\mathrm{m}$, decl. $-55^\circ$ taken with the 
South Pole Telescope (SPT) at $1.4$ and $2.0$~mm wavelengths with arc-minute 
resolution and milli-Jansky depth.
Based on the ratio of flux in the two bands, we 
separate the detected sources into two populations, one consistent
with synchrotron emission from active galactic nuclei (AGN) and one 
consistent with thermal emission from dust.  We present source counts for each population 
from $11$ to $640$~mJy at $1.4$~mm and from $4.4$ to $800$~mJy at $2.0$~mm.
The $2.0$~mm counts are dominated by synchrotron-dominated sources
across our reported flux range; the $1.4$~mm counts are dominated by 
synchroton-dominated sources above $\sim 15$~mJy and by dust-dominated
sources below that flux level.
We detect $141$ synchrotron-dominated sources and $47$ dust-dominated sources at 
S/N $>4.5$ in at least one band.
All of the most significantly detected members of the synchrotron-dominated population are associated
with sources in previously published radio catalogs.  Some of the dust-dominated
sources are associated with nearby ($z\ll1$) galaxies whose dust emission 
is also detected by the Infrared Astronomy Satellite (IRAS).  However, most 
of the bright, dust-dominated sources have no counterparts in any existing catalogs.
We argue that these sources represent the rarest and brightest
members of the population 
commonly referred to as submillimeter galaxies (SMGs).
Because these sources are selected at longer wavelengths than in typical 
SMG surveys, they are expected to have a higher mean redshift distribution 
and may provide 
a new window on galaxy formation in the early universe.
\end{abstract}

\bigskip\bigskip

\section{Introduction}\label{sec:intro}

The 10-meter South Pole Telescope \citep[SPT,][]{carlstrom09} 
is a millimeter/submillimeter (mm/submm) telescope located at the geographic South 
Pole and designed for low-noise observations of diffuse, low-contrast
sources such as anisotropy in the cosmic microwave background (CMB).
The first camera installed on the SPT is a 960-element bolometric receiver 
designed to perform a mass-limited survey of galaxy clusters via their
Sunyaev-Zel'dovich (SZ) signature \citep{sunyaev72} over a large  
area of the southern sky.  This survey is currently underway, and the SPT
team recently published the first-ever discovery of galaxy clusters through their SZ signature
\citep[][hereafter S09]{stani09}. 

The sensitivity and angular resolution of the SPT make it an excellent instrument for 
detecting extragalactic sources of emission.  In this work, we report on source detections 
in a small part of the SPT survey, namely a single 87~\sqdeg \  field centered at 
right ascension (R.A.) $5^\mathrm{h} 30^\mathrm{m}$, declination (decl.) $-55^\circ$ (J2000).  
This field was surveyed by the SPT in the 2008 season to roughly mJy depth at $1.4$~mm and $2.0$~mm (220 and 150~GHz).  
The data presented here represent a major step forward in mm source detection at 
these flux levels, both in area surveyed and in the ability to distinguish between 
source populations using internal estimates of source spectral properties.

Simultaneous information in two bands for each detected source should allow
us to separate our detections into 
distinct source populations.  Based on previous surveys at mm wavelengths
and on surveys in neighboring centimeter (cm) and submm bands, we expect
the sources we detect to fall into two broad categories:  
1) sources with flat or decreasing brightness with decreasing wavelength, 
consistent with synchrotron emission from active galactic nuclei (AGN, 
typically $S\propto\lambda^{\sim1}$); and 2) sources with increasing brightness 
with decreasing wavelength, consistent with thermal emission (typically $S\propto\lambda^{\sim-3}$) from dust-enshrouded star-forming galaxies.

The synchrotron-dominated source population is well-established from radio surveys 
(see \citet{dezotti10} for a recent review).   Despite these sources' decreasing brightness from radio 
to mm wavelengths, simple extrapolations of radio and cm counts of these sources to 
the SPT bands predict that we should detect a significant number of these sources.
This prediction is bolstered by the results of $3$~mm follow-up of $1.5$~cm-detected sources 
presented in \citet{sadler08}, which showed that these sources still emitting strongly at mm 
wavelengths, and by detections of synchrotron-dominated AGN emission made in mm/submm
surveys much smaller than the SPT survey but at similar depths \citep{voss06}.
Measurements of the mm fluxes of a large sample of these sources have the power to
inform both 
astrophysical models of their emission and predictions for the extent of their 
contamination to the CMB power spectrum \citep[e.g.,][]{toffolatti05,reichardt09} and the 
SZ signal from galaxy clusters \citep[e.g.,][]{lin07,sehgal10}.

The dust-enshrouded star-forming galaxy population has been the subject of considerable interest for over thirty years
(see \citet{rieke79} for a pre-IRAS review), but the Infrared Astronomy Satellite (IRAS) was the first 
instrument to systematically discover such objects \citep{sanders96}.  Optical 
and UV observations show these 
sources to be heavily dust-obscured, and to typically have disturbed 
morphologies and high star formation rates, indicative of recent or ongoing 
mergers \citep{lagache05}.  The emission from IR to mm 
wavelengths in these sources arises from short-wavelength photons emitted by stars,
which are absorbed by dust grains and re-radiated at longer wavelengths \citep{draine03}. 

Measurements of the cosmic infrared background (CIB) show that over half the 
energy emitted since the surface of last scattering has 
been absorbed and re-radiated by dust \citep{dwek98}.  IRAS, however,  detected 
mostly low-redshift ($z<1$) objects, and these relatively rare and nearby sources 
contributed only a small fraction of the CIB \citep{lefloch05, caputi07}, 
indicating that the bulk of CIB sources are at high redshift.
The first systematic survey of high redshift sources which contribute
significantly to the CIB --- the population now known as submillimeter galaxies (SMGs) --- was 
carried out a decade ago at 850~\um~by the SCUBA 
instrument on the 15-m JCMT telescope \citep{smail97, hughes98,barger98,holland99}. 
Owing to the spectrum of SMGs---a modified $\sim30$ K blackbody that rises 
steeply with decreasing wavelength, counteracting the expected flux diminution 
with redshift---they can be detected independently of redshift from 
roughly $500$~\um~to $2$~mm \citep{blain02}. This implies that the source 
luminosity is roughly proportional to the brightness from $1<z<10$.

Hundreds of SMGs have now been detected by ground-based telescopes in surveys of blank fields,
but only over a total area on the order of a square degree 
\citep{scott02, borys03, greve04, laurent05, coppin06, bertoldi07, perera08, scott08, austermann10, weiss09}.
Recently, results were published from the Balloon-borne Large-Aperture 
Submillimeter Telescope (BLAST) which surveyed nearly ten square degrees at 250, 350, and 550~\um \ and measured important properties such as dust temperatures and clustering amplitude for SMGs
\citep{devlin09,patanchon09,dye09, viero09}. 
The discovery and study of SMGs has 
revolutionized our understanding of galaxy formation by providing a view of 
galaxy formation which is both unbiased with redshift and inaccessible to optical surveys.
Observations of these objects (see \citet{blain02} for a review) indicate that: 
1) they have dynamical masses of $\sim$$10^{11}$M$_\odot$ and total far-infrared 
luminosities of $\sim$$10^{13}$~L$_\odot$ \citep{swinbank04,greve05,chapman05, kovacs06, pope06}; 
2) they are forming stars prodigiously at $\sim1000$ M$_\odot$/year 
\citep{chapman05, tacconi06}; 
3) their abundance appears to peak at $z\sim2.5$ \citep{pope05,chapman05,aretxaga07,chapin09}; 
4) from observations \citep{tacconi08} and simulation \citep{barnes91,narayanan10}, 
the prodigious star formation rate seen in SMGs is believed to be intrinsically linked to mergers; and 
5) SMGs are an early phase in the formation of the most massive galaxies and are among the 
largest gravitationally collapsed objects in this early epoch of galaxy 
formation \citep{blain04,swinbank08}.

In the context of this broad field of IR/submm/mm-selected, 
dust-enshrouded, star-forming galaxies, 
this paper presents the detection in the SPT data of a 
population of dust-dominated sources 
with surprising and intriguing properties.
These sources are significantly brighter and rarer than the submm-selected population in the literature.  Furthermore, the majority of these sources 
do not have counterparts in IRAS, indicating that they are not members of the 
standard local ultra-luminous infrared galaxy (ULIRG) population.  
This apparently new family of sources represents the most significant new result of this work.

This paper is divided into 
several sections.  Sec.~\ref{sec:obs-reduc} describes the SPT observations, data reduction, 
matched filter for point-source signal,  
and source-finding algorithm.  (The observations and data reduction through the mapmaking 
step are described in greater detail in S09 and \citealt{carlstrom09}.)  Sec.~\ref{sec:mapscat} 
discusses the properties of the filtered maps, presents the source catalog, describes 
our procedures for checking astrometry and estimating completeness and purity,
and discusses basic source properties, including raw spectral classification.  
Sec.~\ref{sec:deboost} 
describes our procedure for estimating each source's intrinsic flux and spectral index, 
which we use to separate 
our sources into two spectrally distinct (synchrotron-dominated and dust-dominated) populations.  
The statistical method used for flux estimation is described in detail in a companion paper, \citet{crawford10}.
Sec.~\ref{sec:counts} presents source counts for each band.  Sec.~\ref{sec:assoc} discusses 
associations with external catalogs.  Finally, Sec.~\ref{sec:populations} presents counts for 
each of the populations and discusses the implications, including the 
potential for a newly discovered population of sources.

In a companion paper \citep{hall10}, we present 
the spatial power spectra of the sources below our detection threshold.


\section{Observations, Data Reduction, and Source Finding}
\label{sec:obs-reduc}

This work is based on observations of a single 
$\sim 100$-\sqdeg \ field.  The timestream 
data for each observation, constituting a single pass over the field, were processed 
and combined to make a map of the field for each observing band.  The maps from 
several hundred individual observations of the field were combined and converted 
to CMB fluctuation temperature units using a calibration from the CMB anisotropy 
as measured by the Wilkinson Microwave Anisotropy Probe (WMAP, \citet{hinshaw09}).  Each single-band map is filtered 
to optimize point-source detection.  A variant of the CLEAN algorithm \citep{hogbom74} 
was then used to search for sources in an 87~\sqdeg \  sub-region of the filtered 
maps.  Finally, amplitudes of the detected peaks in the filtered maps were 
converted from CMB fluctuation temperature units to flux (in units of Jy).

\subsection{Observations}

During the 2008 observing season, the 960-element SPT camera included 
detectors sensitive to radiation within bands centered at approximately 
$1.4$~mm, $2.0$~mm, and $3.2$~mm ($220$~GHz, $150$~GHz, and $95$~GHz).  
The first field mapped to the targeted survey depth was centered at 
R.A.~$5^\mathrm{h} 30^\mathrm{m}$, decl.~$-55^\circ$ (J2000).  
Results in this paper are based on 607 hours of observing time, 
using only the $1.4$~mm and $2.0$~mm data from the 87~\sqdeg \  
portion of the field that was mapped with near-uniform coverage.  (Every 
$1$~arcmin patch in the included area is required to have uniform 
coverage to 10\%.) 
The scanning strategy consisted of constant-elevation 
scans across the 10$^\circ$ wide field. After each single scan back and forth across the field, 
the telescope executes a 0.125$^\circ$ step in elevation.
A complete set of scans covering the entire field takes 
approximately two hours, and we refer to each complete set as an observation.  
Between individual observations of the field, we perform a series of short 
calibration measurements, including measurements of a chopped thermal 
source, $2$ degree elevation nods, and scans across the galactic 
HII regions RCW38 and MAT5a.  This series of regular calibration measurements 
allows us to identify detectors with good performance, assess relative detector 
gains, monitor atmospheric opacity and beam parameters, and constrain 
pointing variations.  S09 and \citet{carlstrom09} describe details of the 
observations of this field and the telescope.  

\subsection{Data Reduction}\label{sec:data}

The reduction of SPT data for this work up to and including the mapmaking step
is very similar to the reduction pipeline used to produce the maps
used in S09, where details of the analysis can be found.  
Broadly, the pipeline consists of filtering the time-ordered data from 
each individual detector, reconstructing the pointing for each detector, 
and combining data from all detectors in a given observing band into 
a map by simple inverse-variance-weighted binning and averaging. 

The filtering used in this analysis differ from that in S09. In this work
the time-ordered detector data were filtered with a $0.18$~Hz Fourier-domain high-pass filter.
With our scan speeds, the high pass filter removes spatial scales 
 $\gtrsim 45\arcmin$. We project out a common mode which consists of three spatial modes (mean, and tilts along two axes) constructed from the 
mean of all working detectors in a single band, weighted by the $x$ and $y$ position
in the focal plane. 
As atmospheric signal is highly correlated
between detectors, removing this common mode
should eliminate the majority of the atmospheric fluctuation power in the detector
timestreams.  
The common-mode subtraction
acts as a spatial high-pass filter with a characteristic scale that 
roughly corresponds to the one degree angular field of view of the array.   This filter option
was demonstrated to remove more atmosphere from the timestream than the method described in S09, but its choice was not critical.
As the common mode is constructed independently for each band, the 
response to large spatial modes on the sky can be slightly different between bands, 
but these modes are heavily de-weighted in the filter we later apply to the maps 
to enhance point-source signal-to-noise.
Finally, in contrast to S09, we did not mask bright point sources when filtering 
the time-ordered data, because we wanted to ensure that the filter transfer 
function would be the same for all sources in the maps.

\subsection{Flux Calibration and Beam Measurements}
\label{sec:calibration}

The relative gains of the detectors and their gain variations over time were estimated using 
measurements of their response to a chopped thermal source.   (These measurements took 
place before each observation of the survey field, or every 2 hours.)  This was the same 
relative calibration method used in S09. 
The absolute calibration is based on the comparison of SPT 
measurements of degree-scale CMB fluctuations at $2.0$~mm directly to the WMAP 5-year 
maps.  This was done using short, dedicated observations of four large fields, totaling 1250~\sqdeg.   Details of the 
cross-calibration with WMAP are given in \citet{lueker10}.
We estimate the uncertainty of this calibration to be $3.6\%$.
We applied this calibration to our $1.4$~mm band by comparing $2.0$~mm 
and $1.4$~mm estimates of CMB anisotropy in our deep survey regions.  This internal 
cross-calibration for $1.4$~mm is 
consistent with a direct absolute calibration from RCW38, but with higher precision.  
We estimate the $1.4$~mm calibration uncertainty to be $7.2\%$.  Because the $1.4$~mm 
calibration is derived from the $2.0$~mm calibration to WMAP, the calibration uncertainties 
in the two bands are correlated with an estimated correlation coefficient of roughly 0.5.

Main-lobe beams were measured using the brightest sources
in the field and are adequately fit by 2d Gaussians with full width at half-maximum 
(FWHM) equal to $1.05^\prime$ and $1.15^\prime$ at $1.4$~mm and $2.0$~mm. 
Large-angle sidelobes were measured using planet observations, but 
the angular scales on which these sidelobes are important are 
heavily downweighted in the filter.  We estimate that beam-shape uncertainties contribute 
roughly $2\%$ and $5\%$ to the absolute flux estimates in our $2.0$~mm and $1.4$~mm 
bands.  This uncertainty was added in quadrature to the calibration uncertainty in 
our flux estimates.  
A visual representation of the SPT $2.0$~mm beam is shown in \citet{padin08}, 
and symmetrized $\ell$-space profiles for the beams at both wavelengths are 
shown in \citet{lueker10}.

A subtlety in estimating the spectral index (Eq.~\ref{eqn:alpha}, below) is that the effective band centers 
(which fold into the index determination) depend upon spectral index.  Using the 
measured passbands for $1.4$~mm and $2.0$~mm, we find that if one were to assume 
an index $\alpha=-1$ in the determination of the band centers, a source with $\alpha=3$ 
would be measured with a $2\%$ bias in the spectral index.  In addition, the 
beam shape (and so flux) will change with the spectral index.  These subtleties can both be 
neglected to the accuracy of the results presented here.

\subsection{Source Extraction}
\label{sec:optfilt}

\subsubsection{Matched Filter}
\label{sec:matchedfilt}

We enhanced the point-source signal-to-noise ratio in the SPT maps by applying a matched 
spatial filter \citep[see e.g.,][]{tegmark98} to each single-band map. The matched filter combines 
knowledge of the instrument beam and any other filtering that has been performed on the data 
with an estimate of noise covariance to optimize the signal-to-noise of a source in the filtered 
map.  This matched filter $\psi$ is applied in the Fourier domain and is given by:
\begin{equation}
\label{eqn:optfilt}
\psi \equiv \frac{\tau^T N^{-1}}{\sqrt{\tau^T N^{-1} \ \tau}}
\end{equation}
where $N$ is the noise covariance matrix
(including astrophysical contaminants such as primary CMB anisotropy), 
and $\tau$ is the assumed source shape in the 
map, which in the case of point sources is a function of beam and filtering only.
Real-space visual representations of  $\tau$ for $1.4$~mm and $2.0$~mm --- 
with filtering very similar to that used in the maps in this work --- are
shown in \citet{plagge10}.

As in S09, the instrumental and atmospheric contributions to the noise
covariance in each band were estimated by computing the average power spectrum of 
hundreds of signal-free maps, constructed by multiplying half of the 320 
individual-observation maps 
of the field by -1, half by +1, and then summing. (Hereafter, this will be 
referred to as the ``difference" map, in which all astrophysical signal has been 
removed from the map, but the atmospheric and detector noise remains.)
The main astrophysical contribution to the noise covariance is 
expected to be primary CMB anisotropy, so an estimate of the CMB power 
spectrum was added to 
the noise covariance.  Adding further astrophysical contributions such as 
the SZ background and point sources below our detection threshold
has a negligible effect on our results.

The source shape used in the matched filter is the convolution of our measured
beam and the map-domain equivalent of any timestream filtering 
we have performed.  
We measured the effect of timestream filtering on the expected shape of point 
sources in our maps by performing signal-only simulations of our data processing, 
using a single model point source projected into our detector timestreams as
the only timestream component.

Observations used here were performed using constant declination scans.  The maps were pixelized using a flat-sky 
projection of the sphere in which the mapping of right ascension to map rows is a 
function of position in the map.\footnote{We chose this pixelization because it 
minimizes beam distortions, which are significant in flat-sky pixelizations
in which pixel rows are at constant declination.}  
As a result, the effects of timestream filtering on 
source shape are also map-position-dependent.  
To account for this, we break the single-band coadded signal maps into nine tiles 
and perform our signal-only simulations nine times --- 
once with the model source located at the center of each tile.  
We broke the map into nine tiles (as opposed to four or sixteen) as it solved 
the problem with the greatest economy.  We estimated 
the noise covariance separately for each tile, as the projection of non-white 
timestream noise into the map is also a function of position.  The noise in the 
outer tiles is within $5\%$ of the noise in the center tile (see Sec.~\ref{sec:mapscat}). We constructed
nine matched filters from these inputs and performed source finding on each map tile
individually with the matched filter constructed from that tile's inputs.  

\subsubsection{Candidate Identification}
\label{sec:extract}

Source candidates were identified in the filtered maps using a variant of the CLEAN 
algorithm from radio astronomy \citep{hogbom74}.  Briefly, the CLEAN procedure
involves iteratively identifying the highest peak in the filtered map and subtracting off a model 
for the source shape centered on that peak until 
no peaks are left above the detection threshold.  To account for several 
non-idealities, including finite-sized map pixels, slightly imperfect source 
shape models, and possibly extended sources, the source model subtraction 
is performed with a multiplicative factor less than unity, usually called the loop 
gain (after the analogous parameter in electronic feedback circuits).  We use a 
loop gain of $0.1$ in this work.

In interferometric radio observations, the source shape template in the CLEAN process
is the interferometer's ``dirty beam"; in our case, it is the source shape in the 
filtered maps:
\begin{equation}
\label{eqn:dirtybeam}
\tau^\prime = \psi \ \tau.
\end{equation}
As discussed in the previous section, the matched filter
$\psi$ was independently calculated for nine different region of each band's map in order to 
account for the map-position-dependent shape of the noise and filtering.  In
constructing the source shape template $\psi \tau$, we used the appropriate 
version of $\psi$ depending on the position of the peak being CLEANed.
For each source peak we independently calculated a pre-matched-filter source shape 
$\tau$ to account for this positional dependence.
In the map pixelization used, this calculation consists of a simple rotation of a fiducial source shape,
so this step was not unduly computationally intensive.

We ran our version of CLEAN on each band's filtered map individually until there 
were no peaks above $3 \sigma$ left in the map.
All map pixels identified above the $3 \sigma$ threshold 
were then sorted by significance and gathered into discrete sources using an association 
radius between $30$~arcsec and $2$~arcmin, depending on the brightness 
of the source.  In other words, the brightest pixel found by CLEAN was declared to be 
the first source, then the list of pixels was examined in descending order of brightness, 
asking if each pixel should be declared a new source or associated with a source already 
identified.  Source fluxes were assigned by converting the value in the filtered 
map of the brightest pixel associated with a source from
CMB fluctuation temperature units to flux (in units of Jy) using the 
following relation:
\beq
\label{eqn:temp2flux}
S [\mathrm{Jy}] = T_\mathrm{peak} \times \Delta \Omega_\mathrm{f} \times 10^{26} \times \frac{2k_{\rm B}}{c^2} \left ( \frac{k_{\rm B} \Tcmb}{h} \right )^2 \frac{x^4 e^x}{(e^x-1)^2},
\eeq
where $x=h\nu/(k_{\rm B} \Tcmb)$ and $\Delta \Omega_\mathrm{f}$ is defined by:
\beq
\label{eqn:area_eff}
\Delta \Omega_\mathrm{f} = \left [ \int du dv \ \psi(u,v) \ \tau(u,v) \right ]^{-1},
\eeq
which can be thought of as the effective solid angle under the 
filtered source template, in that a point source of flux $S$ will
have peak brightness $S/\Delta \Omega_\mathrm{f}$ in the filtered map.
This flux estimate will be biased for extended sources, which are discussed in Sec.~\ref{sec:extend}.
Source positions were obtained by calculating the center of brightness of all pixels (each pixel being $0.25\arcmin\times0.25\arcmin$) associated with a given source.  


\begin{figure*}[h]
 \begin{center}
\epsfig{file=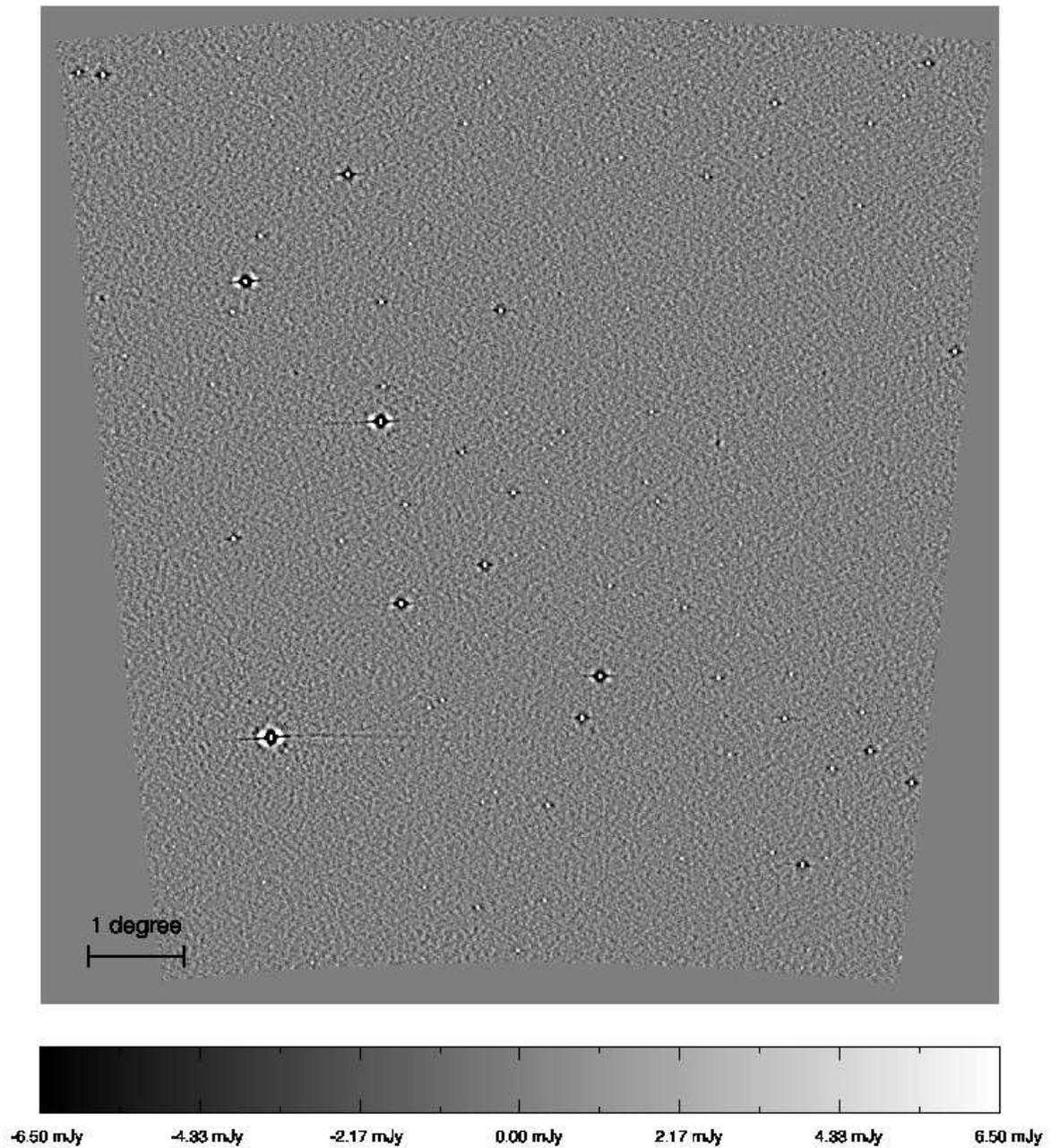, width=16cm} 
\end{center}
\caption{\small The filtered $2.0$~mm map in a flat sky projection.  
The total sky area is $87~\mathrm{deg}^2$ and the field center is 
right ascension R.A.~$5^\mathrm{h} 30^\mathrm{m}$, decl.~$-55^\circ$ (J2000).  
The map is oriented such that north (increasing decl.) is up and east (increasing
R.A.) is left.
Each pixel is $0.25\arcmin\times0.25\arcmin$. 
The RMS in the map is $1.3$~mJy, and the gray scale is $\pm 5\sigma$; 
the brightest source is $>500 \sigma$ ($> 650~\mathrm{mJy}$), 
and the scale saturates for most of the sources visible here.  
Because of time domain filtering, the source signal produces an arc from the 
impulse response of the filter as the detectors scan left and right across the field.  
The azimuthally symmetric ringing around bright sources is due to spatial 
high-pass filtering in the point-source matched filter.
\label{fig:map_2p0mm}}
\end{figure*}


\begin{figure*}[h]
 \begin{center}
\epsfig{file=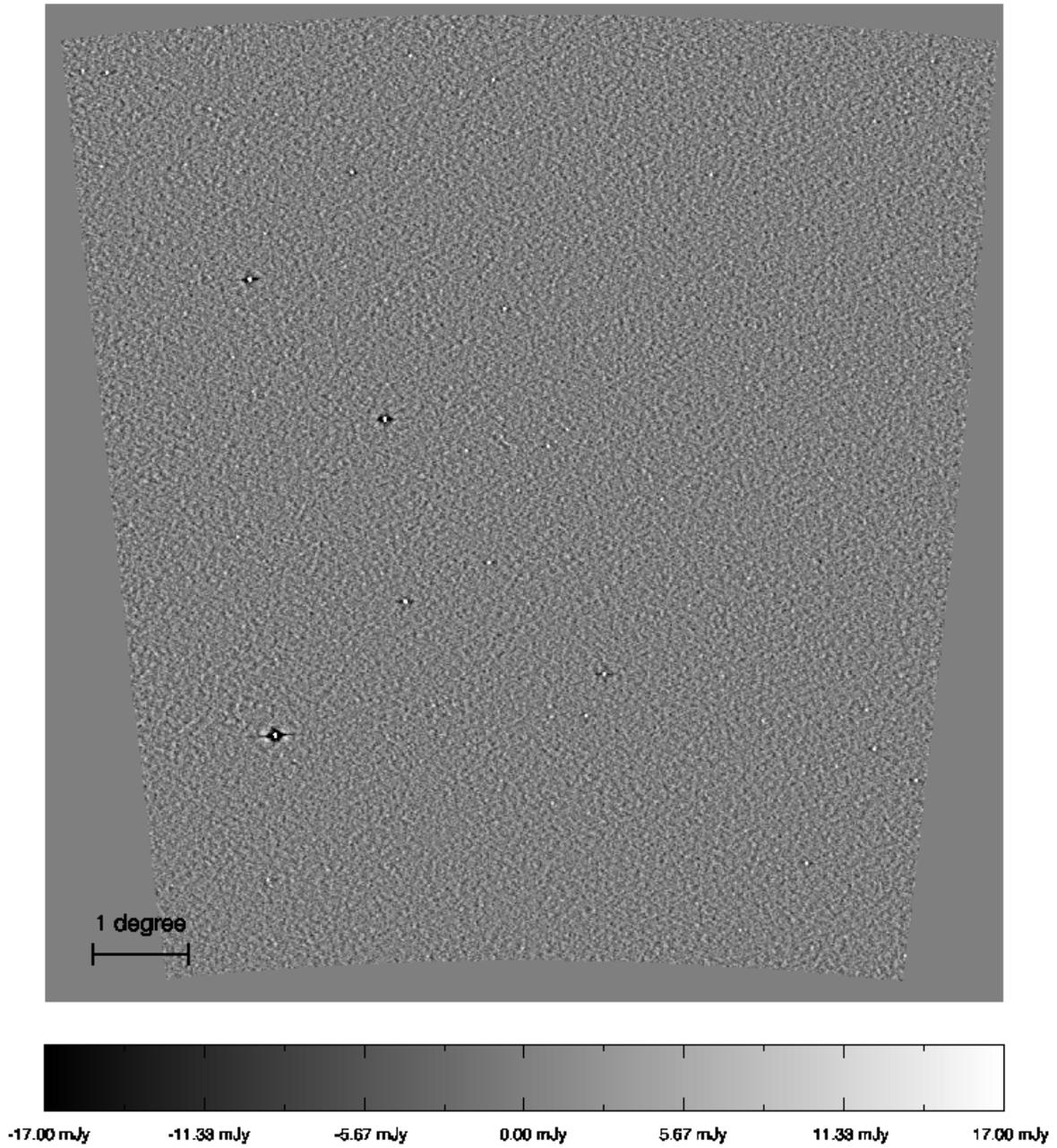, width=16cm} 
\end{center}
\caption{\small The filtered $1.4$~mm map.  
The RMS in the map is $3.4$~mJy, and the 
gray scale is $\pm 5\sigma$; the brightest source is 
$>150\sigma$ ($> 550~\mathrm{mJy}$).  
See the corresponding Fig.~\ref{fig:map_2p0mm} for comments common to both maps.
\label{fig:map_1p4mm}}
\end{figure*}


\begin{figure}[h]
 \begin{center}
\epsfig{file=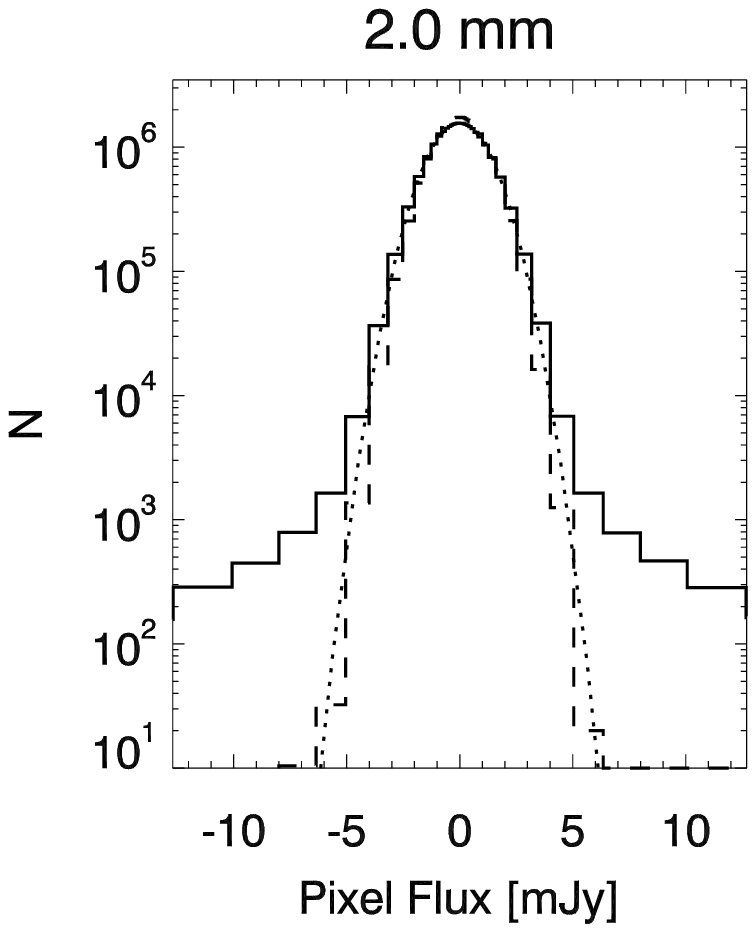, width=7cm} 
\epsfig{file=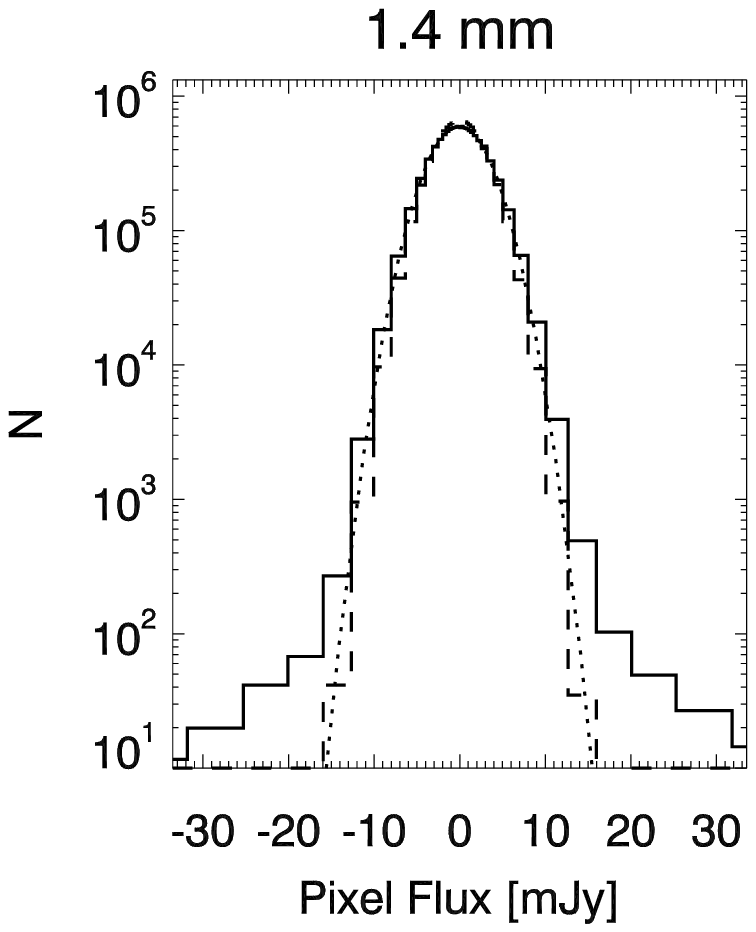, width=7cm} 
\end{center}
\caption{\small The distribution of fluxes in map pixels. 
For each band, the lines are as follows: 
\emph{solid}: The coadded signal map; 
\emph{dashed}: The coadded difference map (see Sec.~\ref{sec:optfilt}); 
\emph{dotted}: Fit to the signal map pixel histogram noise peak. 
For each band, the fit is done to the full signal map and gives $\sigma = 1.3$~mJy for 2.0~mm and $\sigma = 3.4$~mJy for 1.4~mm. 
The noise across the map is Gaussian. 
The negative tails are mainly due to ringing from the various effective high-pass filters on the sources in the map.
\label{fig:pix_hist}}
\end{figure}

\section{Maps and Catalog}
\label{sec:mapscat}

The filtered $2.0$~mm and $1.4$~mm maps used for source candidate identification 
are shown in Fig.~\ref{fig:map_2p0mm}~and Fig.~\ref{fig:map_1p4mm}.  
The total area shown in each map is 87~square~degrees.
As previously noted, the noise varies at the level of $\pm 5\%$ across the maps, mainly as a 
function of declination.  
This trend with declination is due to the fact that 
the coverage is nearly uniform in right ascension, resulting in coverage per
unit solid angle that varies as $\cos(\mathrm{decl.})$
(i.e. the noise is systematically $5\%$ lower at decl.=$-60$ than at decl.=$-55$).
The typical RMS of the map is $1.3$~mJy at $2.0$~mm and 
$3.4$~mJy at $1.4$~mm.  The noise distribution closely approximates a Gaussian, 
as is evident from the central part of the pixel distributions shown in 
Fig.~\ref{fig:pix_hist}. The fact that the maps are so uniform and the noise is so well-understood
makes the analysis much easier and gives us confidence in the robustness of our results.

Detections in both bands are listed in the final catalog as a single source if they are offset $<30$~arcsec between bands. 
For sources detected in both bands, we adopt the position of the more significant detection. 
We are far enough above the confusion limit that this simple and intuitive method is adequate. 
For sources detected in only one band we use the flux in the cleaned map for the second band at the 
position of the detection.   
Table~\ref{tab:cat_both} lists the properties of every source candidate detected above $4.5\sigma$ in either the $2.0$~mm or $1.4$~mm band, while the 
full table listing all 3496 sources above $3\sigma$ in either map
will be available in machine-readable form 
in the electronic version of this article; for now, the full list is available from the public
SPT website.\footnote{http://pole.uchicago.edu/public/data/vieira09/}

\subsection{Catalog field descriptions}

Each column in Table~\ref{tab:cat_both} corresponds to a field in the 
87~\sqdeg \  SPT two-band source catalog.  Descriptions of the catalog 
fields / table columns are as follows:
\begin{enumerate}
\item Source ID:  the IAU designation for the SPT-detected source.
\item RA:  right ascension (J2000) in degrees.  
\item DEC:  declination (J2000) in degrees.
\item S/N ($2.0$~mm):  detection significance (signal-to-noise ratio) in the $2.0$~mm band.
\item $S^\mathrm{raw}$ ($2.0$~mm): raw flux (uncorrected for flux boosting) in the $2.0$~mm band.
\item $S^\mathrm{dist}$ ($2.0$~mm): de-boosted flux values encompassing $16\%$, $50\%$, and $84\%$ ($68\%$ probability enclosed, or $1\sigma$ for the equivalent normal distribution) of the cumulative posterior probability distribution for $2.0$~mm flux, as estimated using the procedure described in Sec.~\ref{sec:deboost}.
\item S/N ($1.4$~mm): same as (4), but for $1.4$~mm.
\item $S^\mathrm{raw}$ ($1.4$~mm): same as (5), but for $1.4$~mm.
\item $S^\mathrm{dist}$ ($1.4$~mm): same as (6), but for $1.4$~mm.
\item $\alpha^\mathrm{raw}$: estimate (from the raw flux in each band) of the $2.0$~mm$-$$1.4$~mm spectral index $\alpha$, where $\alpha$ is the slope of the (assumed) power-law behavior 
of source flux as a function of wavelength:
\beq
S \propto \lambda^{-\alpha}.
\label{eqn:alpha}
\eeq
\item $\alpha^\mathrm{dist}$:  $16\%$, $50\%$, and $84\%$ estimates of the spectral index, based on the probability distributions for spectral index estimated using the procedure described in Sec.~\ref{sec:deboost}.
\item $P(\alpha > \alphathresh)$: fraction of the spectral index posterior probability distribution above the threshold value of $\alphathresh$.  A higher value of $P$ means the source is more likely
to be dust-dominated.
\item Type: source classification (synchrotron- or dust-dominated), 
based on whether $P(\alpha > \alphathresh)$ is greater than or less than $0.5$. 
\item Nearest SUMSS source: angular distance (in arseconds) from the 
nearest source in the 36~cm (843~MHz) Sydney University Molongolo Sky Survey 
(SUMSS) \citep{mauch03}. There are 2731 SUMSS sources in the SPT survey area. 
For a 1~arcmin association radius there is a 2.7\% chance of random association for each SPT source.
\item Nearest RASS source: angular distance (in arseconds) from the nearest source in the 
ROSAT All-Sky Survey (RASS) Bright Source Catalog \citep{voges99} or 
Faint Source Catalog \citep{voges00}. 
There are 1441 RASS sources in the SPT survey area. For a 1~arcmin association 
radius there is a 1.4\% chance of random association for each SPT source. 
\item Nearest IRAS source: angular distance (in arseconds) from the nearest source in
the IRAS Faint-Source Catalog \citep[IRAS-FSC,][]{moshir92}.  There are 493 IRAS sources in the 
SPT survey area. For a 1~arcmin association radius there is a 0.8\% chance 
of random association for each SPT source.
\end{enumerate}
%


\begin{figure}[h]
 \begin{center}
\epsfig{file=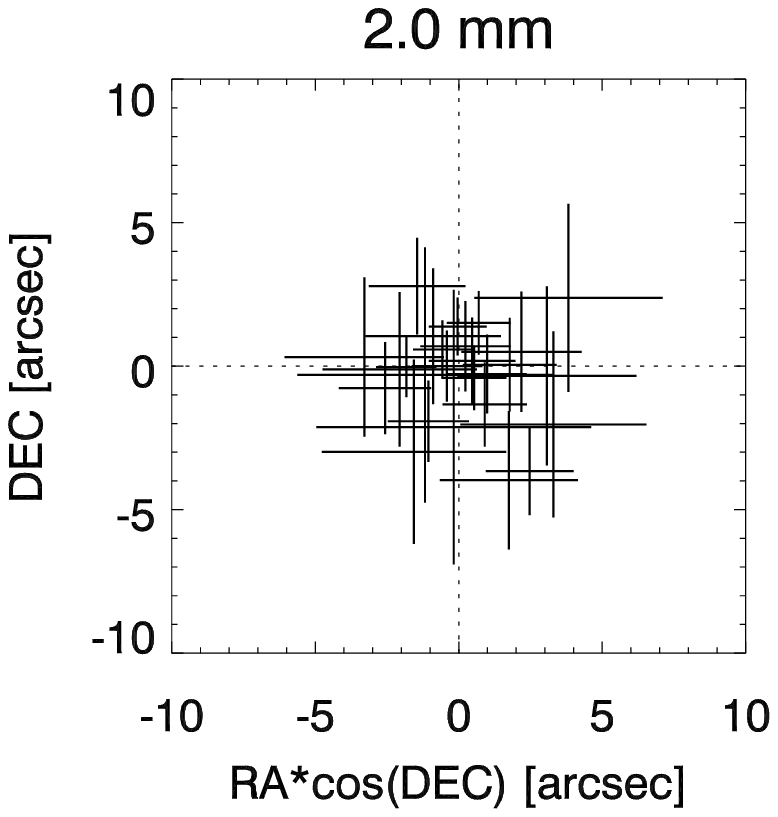, width=7cm} 
\epsfig{file=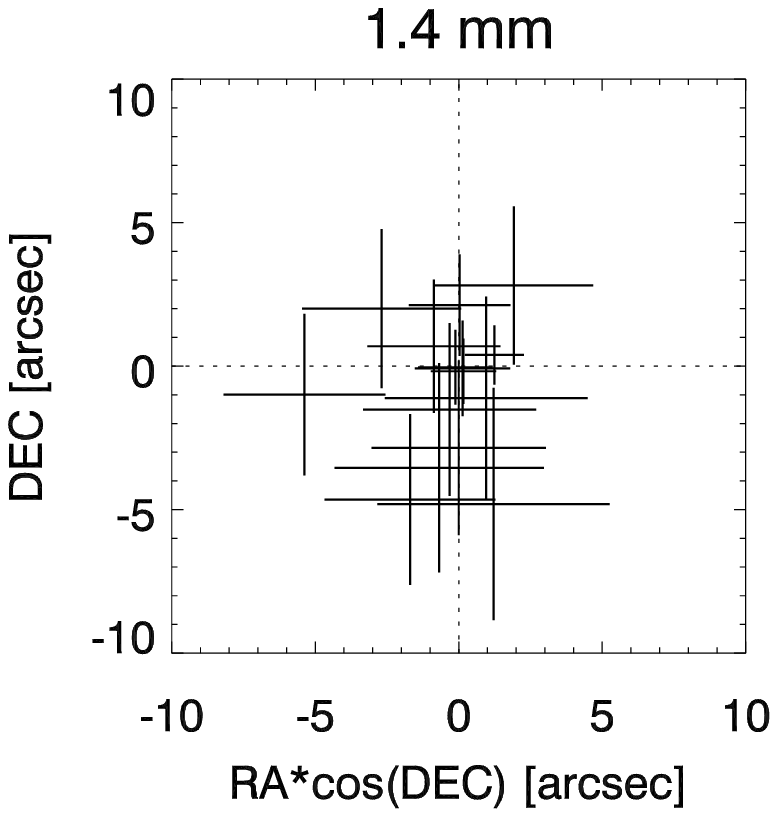, width=7cm} 
\end{center}
\caption{\small 
A comparison of relative pointing offsets between the SPT
bands and AT20G catalog sources. Only sources with S/N~$>10$ which have
a robust counterpart within $20\arcsec$ have been plotted. The errors for the
SPT positions were estimated following Ivison et al. (2007). The
errors plotted here are a quadrature sum of the SPT error and the
quoted error from the AT20G catalog. The RMS for both SPT bands is $<3\arcsec$.
\label{fig:pointing}}
\end{figure}

\subsection{Astrometry}
\label{sec:astrometry}

SPT pointing is reconstructed through a combination of an online pointing model (tied to 
regular observations with optical star cameras), corrections based on observations of
galactic HII regions (performed many times each observing day), and information from 
thermal and linear displacement sensors on the telescope.  The pointing reconstruction 
process is described in more detail in S09 and \citet{carlstrom09}.  In this work we calibrate the absolute
positions in the maps by comparing our best-fit positions 
for bright sources (S/N~$>10$) in our catalog with external determinations of those 
positions from the $1.5$~cm Australia Telescope 20 GHz Survey
(AT20G) catalog, in which the absolute astrometry is tied to VLBI calibrators and
is accurate at the $1$-arcsec level \citep{murphy10}. 
We used 15 point sources for the $1.4$~mm absolute astrometry correction, and 26 point sources for $2.0$~mm.
Fig.~\ref{fig:pointing} shows the distribution of offsets between SPT-determined
positions and AT20G positions.

\subsection{Completeness and Purity}
\label{sec:complpur}


\begin{figure*}[h]
 \begin{center}
\epsfig{file=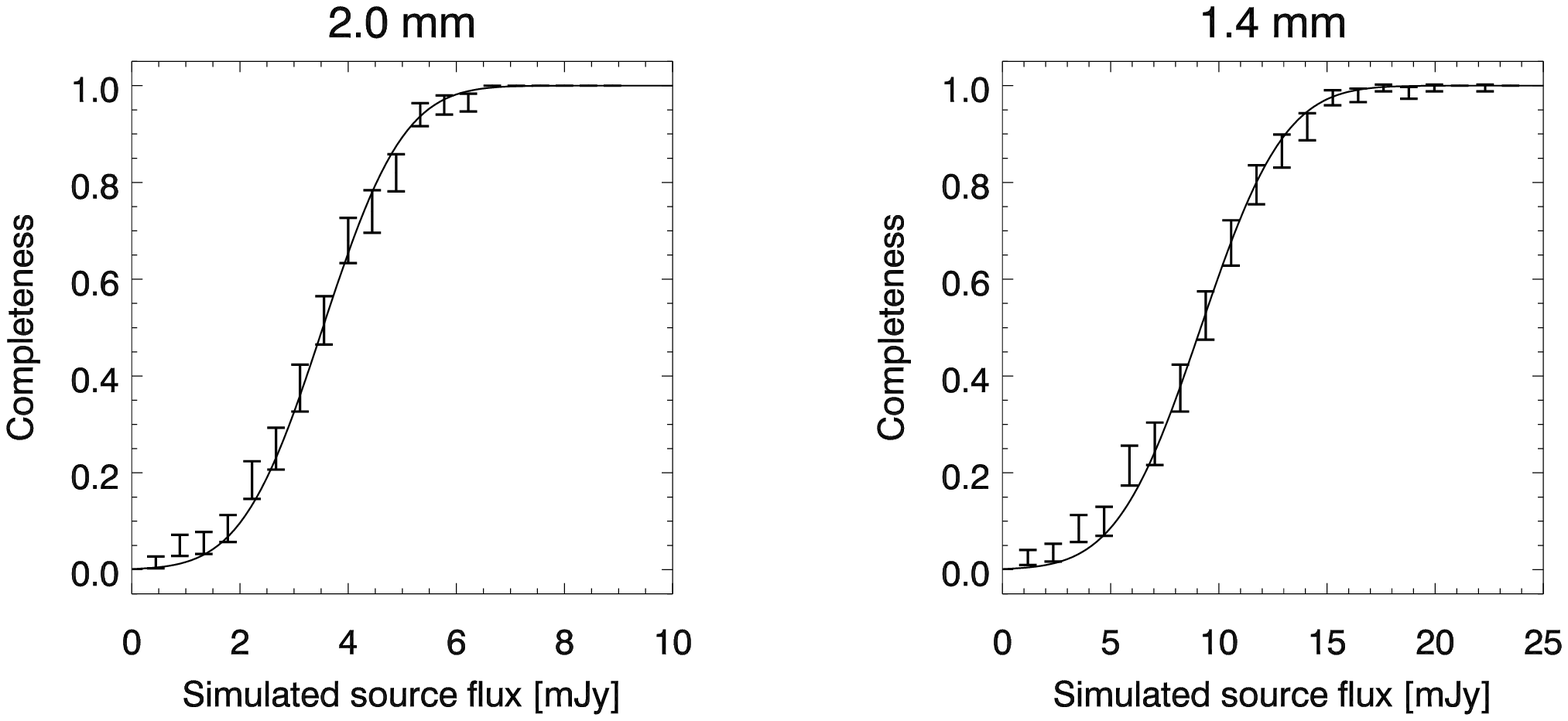, width=18cm} 
\end{center}
\caption{\small The left panel shows the results of the completeness simulation at $2.0$~mm; the right panel shows the results of the completeness simulation at $1.4$~mm.  In each plot, the symbols with error bars show the fraction of sources recovered at $>3\sigma$ with error bars estimated from binomial statistics.  The dashed line shows the best-fit model of the form shown in Eq.~\ref{eqn:compl}.
\label{fig:completeness}}
\end{figure*}


\begin{figure*}[h]
 \begin{center}
\epsfig{file=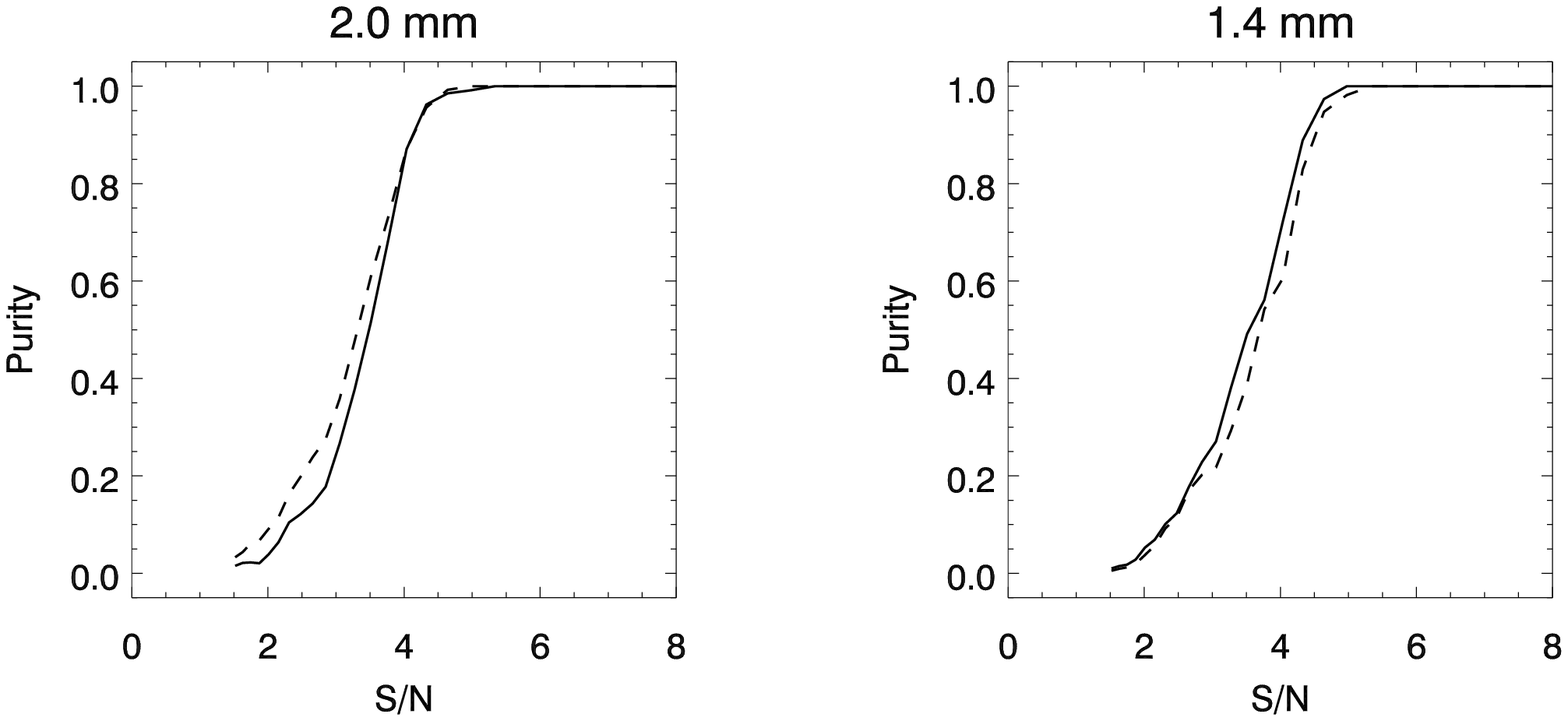, width=18cm} 
\end{center}
\caption{\small Purity in the $2.0$~mm-selected sample (left) and the $1.4$~mm-selected 
sample (right).  In each plot, the \emph{solid line} indicates the purity (see Eq.~\ref{eqn:purity}) 
calculated using the inverted map to estimate the number of false detections, while the \emph{dashed line} indicates the purity calculated using simulated maps to estimate the number of false detections.
The situation is more complicated at $2.0$~mm than at $1.4$~mm due to the presence of SZ (see Sec.~\ref{sec:purity} 
for details).
\label{fig:purity}}
\end{figure*}

\subsubsection{Completeness}
\label{sec:completeness}
We follow \citet{scott08} and estimate our completeness by placing simulated 
sources in the actual signal maps and performing the source extraction as with the real data.  
For the simulated 
source profile, we use the measured beam convolved with the map-domain 
estimate of our timestream filtering and the matched filter.  As with the matched 
filter and the CLEAN process, we use a different simulated source profile in 
each of the nine map tiles (see Sec.~\ref{sec:optfilt} for details).  The simulated 
source is considered detected if it would have made it into our catalog --- i.e., if it is 
detected by the source extraction algorithm at $\ge 3\sigma $.  As expected for maps
whose variance is nearly uniform and is dominated by random, Gaussian-distributed 
noise, our cumulative completeness curves (fraction of simulated sources detected 
above a given flux) are fit well by error functions, as shown in 
Fig.~\ref{fig:completeness}.
The exact functional form used here is
\beq
f_\mathrm{compl}(S) = \frac{1}{\sqrt{2 \pi \sigma^2}} \int_{S}^\infty e^{-\left(S^\prime - S_0 \right)^2/2\sigma^2}dS^\prime.
\label{eqn:compl}
\eeq
On the basis of this test and the error-function fits, we expect the full $\ge 3\sigma$ catalog to be $50\%$ complete at $3.5$ and $9.1$~mJy in the in the $2.0$ and $1.4$~mm bands and to be $95\%$ complete at $5.5$~mJy and $14.1$~mJy in the $2.0$ and $1.4$~mm bands.

\subsubsection{Purity}
\label{sec:purity}
There is some ambiguity in the definition of ``purity" or ``false detection" when one is dealing 
with a source population with steep differential number counts, especially if the detected fluxes are anywhere near the
confusion limit.  In such a situation, there will be at least one source at a non-negligible fraction of the 
detection threshold in every beam.  In this work, we have chosen to define a false detection as a 
fluctuation above the detection threshold in the absence of any mean point source contribution to the 
maps.
We treat the problem of low-flux sources scattering above the detection threshold in the context 
of flux boosting in Sec.~\ref{sec:deboost}.

We estimate our purity using two different methods, both of which are fairly common in the 
SMG literature  \citep[e.g.,][]{perera08}.  First, we invert our maps and run the matched filter and 
source-finding algorithm on the negative maps.  This method is complicated by the 
fact that, at $2.0$~mm, we expect to have real negative signal near the beam scale due to
the thermal SZ signal from galaxy clusters.  To deal with this, we mask the inverted $2.0$~mm 
map around SZ cluster candidates detected at $\ge 4.5\sigma$.  These candidates are identified 
using a filter optimized for extended sources with a particular spatial profile (in this case 
a spherical $\beta$ model, see S09 for details), so we should not be masking point-like noise fluctuations with this 
procedure.  Our second estimate of purity comes from running the matched filter and 
source-finding algorithm on simulated maps. 
These simulated maps contain atmospheric and instrumental noise 
(taken from our difference map -- see Sec.~\ref{sec:optfilt}), 
a realization of the CMB, and a white, Gaussian noise term meant to approximate the contribution from the 
background of sources below the detection threshold.  
The results from both tests are shown in Fig.~\ref{fig:purity}.  In all cases, the quantity plotted is 
\beq
f_\mathrm{pure} = 1 - \frac{N_\mathrm{false}}{N_\mathrm{total}},
\label{eqn:purity}
\eeq
where $N_\mathrm{false}$ is the number of false detections (as estimated, alternately, by one of the 
two methods described above) above a given S/N, and $N_\mathrm{total}$ is the total number of detections above a given S/N
in the real map.
Both methods agree that at S/N~$> 4.5$ 
our sample is  $\gtrsim 90\%$ pure.  \citet{perera08} argue that both of these methods will 
overestimate the true false detection rate, and this hypothesis is supported by the fraction 
of our synchrotron-dominated sources that have clear counterparts in other 
catalogs and/or our ATCA follow-up observations (see Sec.~\ref{sec:assoc} for details).

\subsubsection{Contamination at $2.0$~mm from SZ}
\label{sec:szcontam}
In addition to complicating the purity analysis in the previous section, SZ decrements
from galaxy clusters have the potential to contaminate our source measurements
at $2.0$~mm (though not at $1.4$~mm, which is very close to the thermal SZ null).  We
believe that this contamination will be negligible at the source flux levels considered 
here for two reasons.  One reason is because clusters are expected to be at least 
partially resolved by the SPT at $2.0$~mm, meaning that their contribution to maps
filtered to optimize point-source sensitivity will be diminished.  The other reason is
that the number density of clusters with decrements deep enough to significantly 
affect the source fluxes presented here is expected to be quite low.  The SZ contamination
will be somewhat boosted by the fact that the sources we investigate here are 
expected to be spatially correlated with galaxy clusters at some level 
\citep[e.g.,][]{coble07,bai07}, but we expect the net effect to be negligible even 
after accounting for this correlation.

To make these arguments more quantitative, we investigate the level of 
flux decrements in simulated SZ maps filtered in the same way the SPT data is 
filtered in this work.  We take simulated $2.0$~mm maps created using the 
technique described in \citet{shaw09}, filter them with the estimate of the 
SPT $2.0$~mm beam and filtering discussed in Sec.~\ref{sec:matchedfilt} to 
simulate SPT observation and data processing, and further filter them with the 
matched filter from Sec.~\ref{sec:matchedfilt}.  Finally, we convert the 
filtered map from temperature to flux (as in Sec.~\ref{sec:extract}) and record the 
decrement in Jy at each simulated cluster location.  We find roughly five clusters
per square degree with at least a $1.3$~mJy decrement in the filtered map --- 
equivalent to a $1 \sigma$ noise fluctuation in the $2.0$~mm SPT map.  We 
find roughly one cluster per ten square degrees with at least a $5.8$~mJy decrement 
in the filtered map --- equivalent to a $4.5 \sigma$ noise fluctuation in the 
$2.0$~mm SPT map. 

The SPT beam is roughly $1$~arcmin wide at $2.0$~mm, 
meaning that there are over 1000 independent resolution elements in one square 
degree of the SPT $2.0$~mm map.  Thus, if galaxy clusters were randomly distributed with 
respect to sources in the SPT maps, we would expect fewer than $0.5\%$ of
sources to suffer a $1 \sigma$ or worse systematic flux reduction due to 
cluster SZ signal, and we would expect fewer than $0.01 \%$ of $4.5 \sigma$ sources to be 
completely canceled by an SZ decrement.  \citet{coble07} estimate that radio 
sources are a factor of  $9 \pm 4$ times more likely to be found along a line of sight 
within $0.5$~arcmin of a cluster than in the field.  This boosts the chance of a 
systematic $1 \sigma$ flux error to $\lesssim 5 \%$ and the chance of a $4.5 \sigma$ 
error to $\lesssim 0.1 \%$.  Dust-dominated sources are expected to be less correlated
with galaxy clusters than radio sources \citet{bai07}, so the effect will be even smaller 
for these sources.  The effect of this level of systematic flux error on our final counts is 
completely subdominant to the statistical uncertainty.


\begin{figure*}[h]
\begin{center}
\epsfig{file=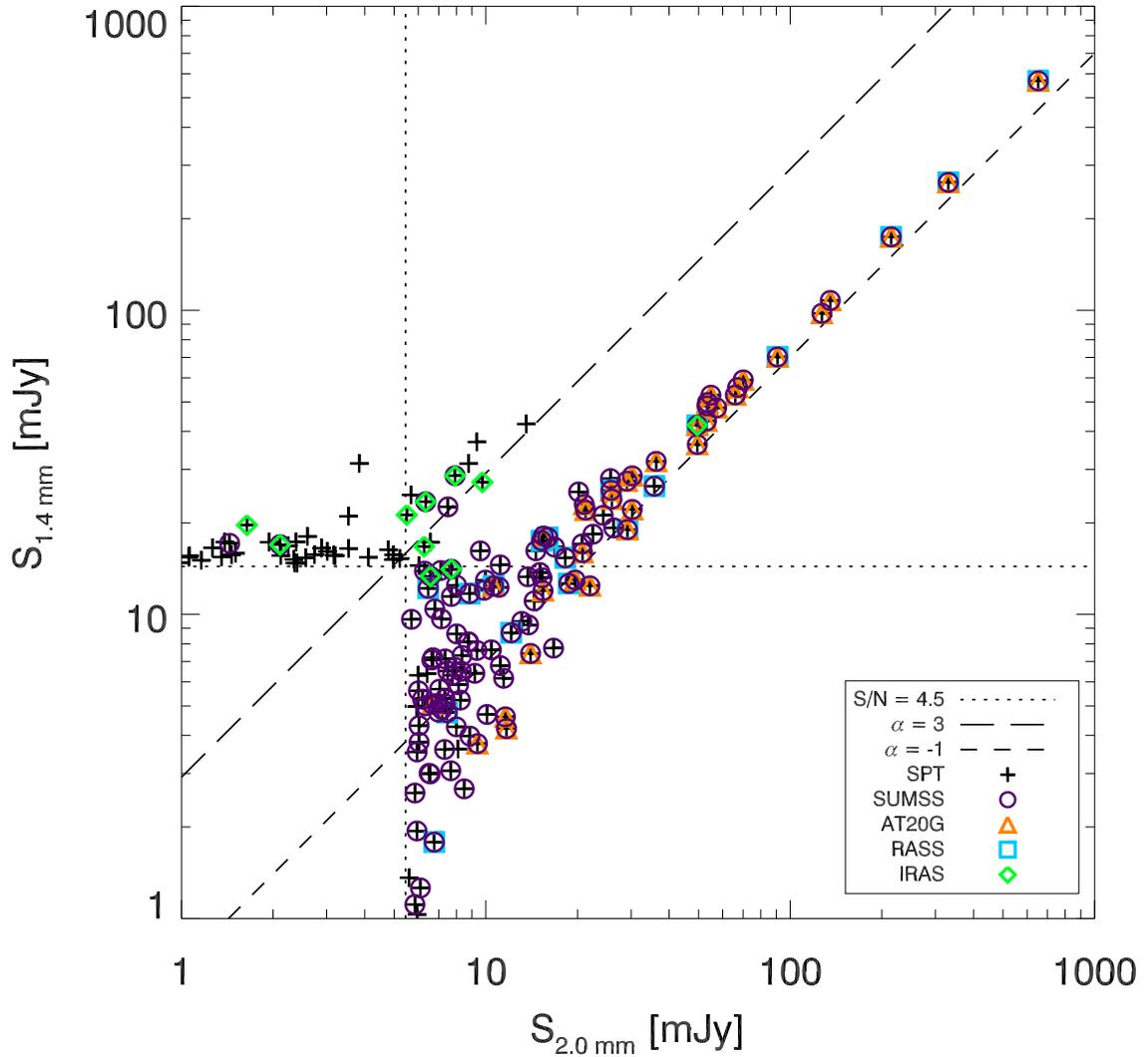, width=17cm}
\end{center}
\caption{\small Raw $1.4$~mm flux versus raw $2.0$~mm flux for sources detected above $4.5 \sigma$ (\emph{dotted lines}) in either band. \emph{long-dashed line}: A spectral index $\alpha=3$ typical for sources dominated by dust emission. \emph{short-dashed line}: A spectral index $\alpha=-1$ typical for sources dominated by synchrotron emission.  By finding associations within 1~arcmin for SUMSS (\emph{purple circles}), AT20G (\emph{orange triangles}), RASS (\emph{light blue squares}), and IRAS (\emph{green diamonds}), we see that most synchrotron-dominated sources are previously known.  The bright dust-dominated population without counterparts in IRAS is discussed in Sec.~\ref{sec:smg}.\label{fig:s_v_s}} 
\end{figure*}

\subsection{Raw Spectral Classification and Source Association}
\label{sec:svs}

Based on previous surveys 
of sources at other wavelengths, we expect most SPT sources to belong to 
one of two populations: one dominated by synchrotron emission, the members 
of which should have an emission spectrum that is flat or falling with decreasing 
wavelength, and one dominated by thermal emission of reprocessed starlight 
by dust, the members of which should have an emission spectrum that 
increases with decreasing wavelength.  Our results confirm this picture.  
Of course, any individual source may have components of each in its emission, and the local slope of the 
spectral energy distribution (SED) will be further modulated by the redshift of the source. 
Though our actual source characterization is based on the integrated posterior 
probability density function (PDF) of the spectral index, estimated using the 
method described in Sec.~\ref{sec:alpha}, a plot of raw $1.4$~mm flux 
vs. raw $2.0$~mm flux, as in Fig.~\ref{fig:s_v_s}, gives the basic picture.  
Of the sources detected above $4.5\sigma$ in both bands, the synchrotron-dominated 
sources occupy a locus of points close to the line $\alpha = -1$, where the 
spectral index $\alpha$ is defined in Eqn.~\ref{eqn:alpha}.
The dust-dominated sources detected in both bands occupy a clearly separated
locus of points close to the $\alpha=3$ line.  Also worth noting in this plot is 
that effectively all of the high-S/N synchrotron-dominated sources have counterparts
in external catalogs, while many of the high-S/N dust-dominated sources do not.
This point is explored in greater detail in Sec.~\ref{sec:assoc} and Sec.~\ref{sec:smg}.

\subsection{Extended Sources and Other Notes}
\label{sec:extend}

As is evident from Eq.~\ref{eqn:temp2flux} and Eq.~\ref{eqn:area_eff}, our 
flux estimates rest on the assumption that all sources 
have the same shape in our filtered maps.  Since the assumed
source shape is just that of our beam and filtering, 
this assumption will only be valid for point-like objects.  
This method will not provide accurate flux estimates for resolved sources. 
For example, our method will underestimate the flux of a source with a 
$\mathrm{FWHM}=0.25$~arcmin Gaussian profile by $3\%$ at $2.0$~mm
and $4\%$ at $1.4$~mm; a $0.5$-arcmin source will be underestimated 
by $10\%$ and $11\%$; a $1$-arcmin source by $31\%$ and $36\%$.

Given the $\sim1\arcmin$ beam of the SPT, we expect that few emissive 
sources will appear extended in our maps.
With $1$-arcmin resolution, a normal galaxy will appear point-like 
at redshifts $z \gtrsim 0.05$ (distances greater than $\sim 200$~Mpc), 
so only very nearby objects or objects with very extended structure
(such as AGN with $100$-kpc-scale jets) would appear extended in our maps.
Furthermore, the matched filter applied to the maps is optimized 
for unresolved sources and will degrade the signal-to-noise on any extended 
source.  

We search for extended sources
by fitting a cut-out of the (unfiltered) map around each detected
source to a model of our measured beam convolved with a Gaussian of 
variable width.  We then identify sources for which the best-fit FWHM is at 
least $0.25$~arcmin and is inconsistent with zero at the $3 \sigma$ level.
We also visually inspect the filtered map at each $\ge 4.5 \sigma$ 
source location for possible extended sources and any other anomalies.

Of the 188 sources detected with S/N$\ge4.5$ in either band, 
11 have a best-fit width of at least $0.25$~arcmin and are inconsistent with
zero width at $\ge 3\sigma$.    
These sources are noted in 
Table~\ref{tab:cat_both} with an ``{\emph a}" next to the source name.
Of these 11 sources, nine fall into our synchrotron-dominated class and have 
counterparts in the SUMSS or Parkes-MIT-NRAO
\citep[PMN,][]{wright94} catalogs.
Three of these nine are also listed in the SUMSS catalog as having detectable 
extent beyond the $\sim 30$~arcsec SUMSS beam.  
The remaining two sources that we identify as extended are nearby 
IR-luminous galaxies that are also detected with IRAS.
Our visual inspection of all sources above $4.5 \sigma$ in either band 
revealed the following cases of note (some of which
are also identified by the quantitative test for extended structure):

\emph{SPT-S~J051614$-$5429.6}:  This detection may be spurious, caused
by sidelobes from the deep SZ decrement at $2.0$~mm from the galaxy cluster 
SPT-CL~J0516$-$5430 (also RXCJ0516.6$-$5430 and Abell S0520).  There is no counterpart
at $1.4$~mm or in external catalogs, the source is classified as extended by the 
method described above, and visual inspection shows it to have an irregular 
shape.  The other bright source very near a galaxy cluster with a deep SZ 
decrement, SPT-S~J050907$-$5339.2 (near SPT-CL~J0509$-$5342) is almost 
certainly not spurious, since it is detected more strongly at $1.4$~mm (which is 
near the SZ null) than at $2.0$~mm.  Evidence of this source is also seen in 
data taken with the Atacama Cosmology Telescope \citep{hincks09}.

\emph{SPT-S~J051217$-$5724.0}: This source is classified as extended, 
and visual inspection reveals a clear offset between the $2.0$~mm and 
$1.4$~mm emission.  The emission in both bands is almost certainly 
associated with the low-redshift ($z=0.0047$) galaxy NGC~1853; 
we appear to be resolving different components of emission 
within the galaxy.  We see a similar configuration in 
SPT-S~J050656$-$5943.2, which is associated with the 
$z=0.0041$ galaxy NGC~1824, and SPT-S~J055116$-$5334.4,
which is associated with the $z=0.015$ galaxy ESO~160-~G~002.
We also see offsets between $1.4$~mm and $2.0$~mm emission in 
SPT-S~J051116$-$5341.9 which has no obvious counterparts
in existing catalogs.

\emph{SPT-S~J052850$-$5300.3}:  We classify this source as dust-dominated, 
and it has no counterpart in the IRAS-FSC.  There is a SUMSS 
source $45$~arcsec away, and visual inspection reveals a low-significance
$2.0$~mm counterpart exactly coinciding with the SUMSS location.  The 
$1.4$~mm emission, however, is clearly offset from both the $2.0$~mm
emission and the SUMSS location, indicating that this may be a chance 
superposition of a known radio source and a previously unknown 
dust-dominated source.

\section{Correcting For flux boosting and Estimating Spectral Behavior}
\label{sec:deboost}

The differential counts of mm-wave selected point sources as a function of source flux
are expected to be very steep, so the measured flux of a point source
in the SPT survey will almost certainly suffer flux boosting.  
In this work, we define flux boosting as the increased
probability that a source we measure
to have flux $S$ is really a dimmer source plus a positive noise
fluctuation relative to the probability that it is a brighter
source plus a negative noise fluctuation.  Because of this
asymmetric probability distribution, raw measurements of source flux 
will be biased high.\footnote{This phenomenon is closely
related to what is referred to in the literature as ``Eddington bias"
\citep[e.g.,][]{teerikorpi04}; however, the consensus use of the term in 
the literature is to describe the bias introduced to estimation of source counts
vs.~brightness, not on the estimated brightness of individual sources.  This
usage is consistent with the original work of \citet{eddington13}.  As such, 
we choose to use ``flux boosting" for the effect on individual source flux 
estimation.}  The standard method in the SMG literature for 
dealing with this problem \citep[e.g.,][]{coppin05} is to construct a posterior
probability distribution for the intrinsic flux of each detection.  The situation
with SPT data is more complicated for two reasons:  1) As discussed 
in \citet{crawford10}, the current implementation of this 
method in the SMG literature is not appropriate for estimating properties of 
individual sources, which is a key aim of this work; 2) We have data in 
more than one observing band, and the prior information that is applied 
to create the posterior flux likelihood will be highly correlated in the two 
bands.

In \citet{crawford10}, we develop a method of correcting for flux boosting (based on 
the Bayesian posterior method used in \citet{coppin05} and others)
which preserves information on individual source properties, and we extend 
that method to estimate the intrinsic multi-band flux of a source based on
the measured flux in each band and the prior knowledge of the source 
populations in the various bands.  In the two-band SPT case, the final 
product for each source is a two-dimensional posterior likelihood, where 
the two variables are either the flux in each band or the flux in one band
and the spectral index between bands.  The two likelihood distributions 
are trivially related by:
\begin{eqnarray}
\label{eqn:psmax_smax}
&&P(S_{\mathrm{max},1},S_{\mathrm{max},2}|S_{p,m,1},S_{p,m,2}) = \\
\nonumber && 
P(S_{\mathrm{max},1},\alpha|S_{p,m,1},S_{p,m,2}) \ \frac{d \alpha}{d S_{\mathrm{max},2}},
\end{eqnarray}
where $S_{p,m,i}$ is the measured flux in a resolution element or pixel 
in band $i$, $S_{\mathrm{max},i}$ is the true flux of the brightest source in 
that resolution element and band, and $d\alpha/dS_{\mathrm{max},2}$ 
is derived from Eqn.~\ref{eqn:alpha}.  If we cast our prior information on 
source behavior in terms of source counts in one band and spectral 
behavior between bands, and we make the assumption that spectral 
index does not depend on flux, then we can write:
\begin{eqnarray}
\label{eqn:psmax_alpha}
&&P(S_{\mathrm{max},1},\alpha|S_{p,m,1},S_{p,m,2}) 
\propto  \\
\nonumber && 
P(S_{p,m,1},S_{p,m,2}|S_{\mathrm{max},1},\alpha)
P(S_{\mathrm{max},1}) P(\alpha)
\end{eqnarray}
(see \citet{crawford10} for details).  Of course, we expect that the spectral 
index distribution of our sources will in fact depend on flux.  But the prior that
we choose to place on $\alpha$ (see Sec.~\ref{sec:priors}) 
is broad enough that it easily encompasses the full expected spectral 
index distribution at all fluxes.

The posterior probability distributions in Eqn.~\ref{eqn:psmax_smax} and 
Eqn.~\ref{eqn:psmax_alpha} are used to calculate most of 
the quantities reported in subsequent sections, including the $16\%$, 
$50\%$, and $84\%$ ($68\%$ probability enclosed about the median) percentiles for de-boosted flux listed 
in Table~\ref{tab:cat_both}; the probability distributions for 
$2.0$~mm--$1.4$~mm spectral index, from which the $16\%$, 
$50\%$, and $84\%$ percentiles for that quantity in Table~\ref{tab:cat_both}
are derived; and the source counts shown in Fig.~\ref{fig:bybandplot}, Fig.~\ref{fig:syncdezotti} and Fig.~\ref{fig:counts_dusty}.

\subsection{Choice of Priors}
\label{sec:priors}

To construct the prior $P(S_{\mathrm{max},1})$ in Eqn.~\ref{eqn:psmax_alpha}, we need
assumptions for the source counts as a function of flux ($dN/dS$) in each of our bands.  We
assume the counts in each band will be the sum of 
dust-dominated and synchrotron-dominated counts, and we use estimates 
of counts for each of these populations from the literature, extrapolated
to our wavelength bands when necessary.  For the dust-dominated population, we use 
the \citet{negrello07} model for counts at $850$~$\mu$m, extrapolated to our wavelengths using a spectral index of $3.0$ for the SMGs and $2.0$ for the IRAS-type galaxies (assuming zero scatter in the index in both cases).  The choice of these spectral indices was taken from an Arp220 SED template and the outcome is not very sensitive to the input.\footnote{While we compare to the integral counts predictions of \citet{lagache04} in Sec.~\ref{sec:smg}, we found that a kink in those differential counts produced a bias toward drawing $\sim 10$~mJy sources from the posterior.  \citet{negrello07} was then used by virtue of its smoothness.}
\citet{dezotti05} make direct predictions for the synchrotron-dominated population counts at $2.0$~mm, which we use without modification.  We extrapolate these predictions to $1.4$~mm using a Gaussian distribution of spectral indices, centered on $-0.5$ with RMS of $0.5$.  We have found that the choice of source-count prior makes only a small difference in the resulting posterior probability distributions (in the S/N range presented in this paper), consistent with the result in \citet{scott08}.  Sec. 2.6 of the companion paper \citet{crawford10} describes the interplay of experimental and prior information.

For the spectral index prior, we have chosen a flat prior between $\alpha=-3$ and $\alpha=5$.  Given what is known about the two populations that are expected to contribute to sources at our wavelengths \citep[e.g.,][]{knox04,mason09}, this estimate conservatively brackets the expected spectral behavior of SPT sources.  (The difference in derived counts by choosing a flat prior from $\alpha=-2$ to $\alpha=4$ is negligible.)  Sec. 3.1 of the companion paper \citet{crawford10} describes a subtlety of choosing the counts prior in parallel with a prior on the spectral index, because the two are interdependent.  In this work, we use a $1.4$~mm counts model prior for the derived $1.4$~mm counts and a $2.0$~mm counts model prior for the derived $2.0$~mm counts.  This means that in the derived $1.4$~mm counts, the information that comes from the $2.0$~mm band is translated to $1.4$~mm flux using the flat prior from $\alpha=-3$ to $\alpha=5$ (and similarly for the cross-band information for the $2.0$~mm counts).  


\begin{figure}[h]
\begin{center}
\epsfig{file=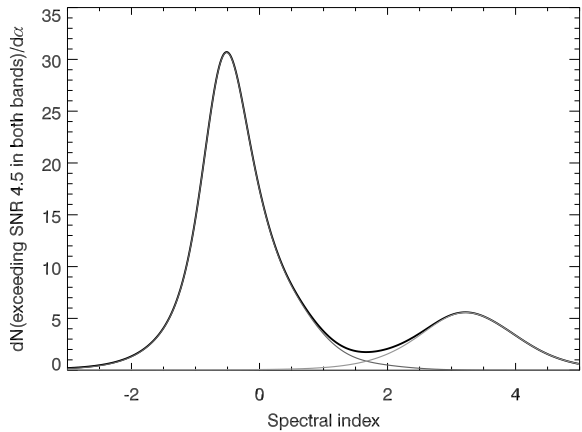, width=9cm}
\end{center}
\caption{\label{fig:alpha_hist} \small 
The distribution of the posterior spectral indices measured between $1.4$~mm and $2.0$~mm for sources with signal-to-noise $>4.5$ in both bands (\emph{thick black line}).  Because we take a flat prior on the spectral index between $-3$ and $5$ (and zero outside), the distribution outside the plotted range here goes to zero.  We sum the $P(\alpha)$ for each source (described in Sec.~\ref{sec:alpha} and normalize it such that the integral over $\alpha$ is equal to $1$ for each source) to give an effective $dN/d\alpha$ for this selection.  We then classify the source by the probability that its posterior spectral index distribution exceeds a classification cut, taken here to be $\alpha=\alphathresh$.  (This threshold is at the minimum of $dN/d\alpha$ between the two populations.)  Sources with $>50\%$ probability of posterior $\alpha>\alphathresh$ are classified as dust dominated and those with $<50\%$ probability of posterior $\alpha > \alphathresh$ are synchrotron-dominated.  There are 11 dust sources (\emph{light gray}) and 41 synchrotron sources (\emph{dark gray}) that contribute to this distribution.  The population split shown here is robust to changes in the signal-to-noise cut.  At lower signal-to-noise cuts, the population features broaden slightly and many sources have poorly-localized $P(\alpha)$ distributions which contribute a floor in $dN/d\alpha$.
}
\end{figure}

\subsection{Spectral Index Estimation and Source Classification Method}
\label{sec:alpha}

By marginalizing the two-dimensional posterior in Eqn.~\ref{eqn:psmax_alpha} over 
the flux in the detection band $P(S_{\mathrm{max},1})$, we obtain a posterior likelihood
for the spectral index of each detected source.  
The $16\%$, $50\%$, and $84\%$ values of $\alpha$ given in Table~\ref{tab:cat_both} 
are taken from the cumulative version of this likelihood distribution for each source.
These individual distributions can be summed to produce the measured $\alpha$ distribution of 
all sources detected in our two bands (which will be the convolution of the intrinsic 
distribution with a complicated function of the noise from instrumental, atmospheric, 
and source background contributions in both bands).
Fig.~\ref{fig:alpha_hist} shows that the posterior spectral index distribution for sources at S/N$>4.5$ in both bands has a clear population split.  We use this split to identify the sources as either dust or synchrotron-dominated through the posterior.  

In Table~\ref{tab:cat_both}, a source with $P(\alpha > \alphathresh) > 0.5$ (having $>50\%$ of its posterior index distribution in excess of $\alphathresh$) is classified as dust-dominated and a source with $P(\alpha > \alphathresh) < 0.5$ (having $<50\%$ of its posterior index distribution in excess of $\alphathresh$) is classified as synchrotron-dominated.  The source counts by population use a probabilistic method based on $P(\alpha)$ that is described in Sec.~\ref{sec:counts}, but there, too, we take $\alpha=\alphathresh$ to be the threshold.  This value of the threshold is the spectral index at the minimum of $dN/d\alpha$ between the two populations (see Fig.~\ref{fig:alpha_hist}).  The source counts presented here are insensitive (within their uncertainty) to this choice of threshold over the range $\alpha=(\alphathresh \pm 0.5)$; this particular value is chosen for definiteness.  


\begin{figure*}[h]
 \begin{center}
\epsfig{file=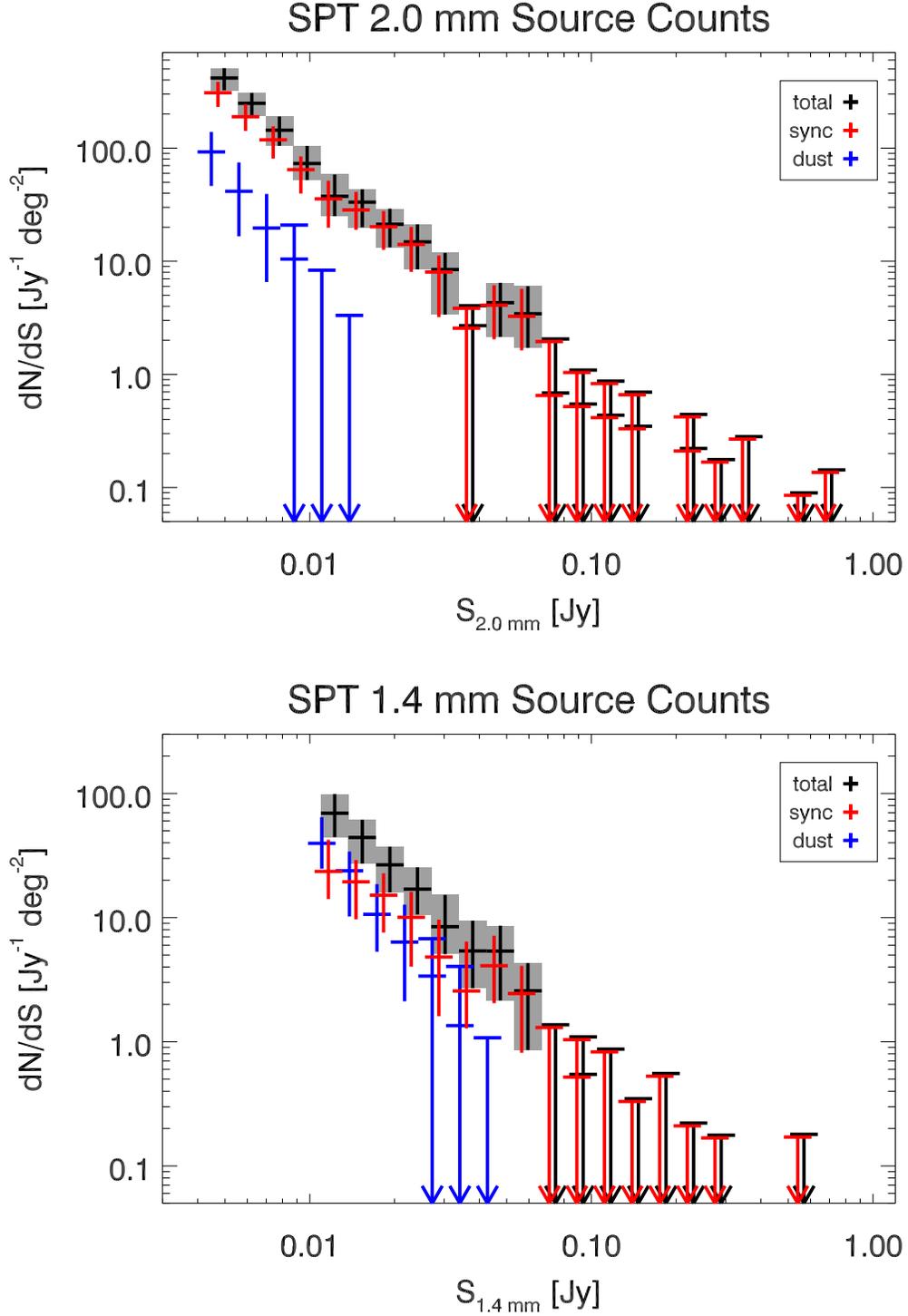, width=15cm} 
\end{center}
\caption{\small Differential source counts by population for the $2.0$~mm (upper plot) and $1.4$~mm (lower plot) bands.  Gray boxes and \emph{black} crosshairs indicate the total counts in that band.  \emph{Red} crosshairs indicate the synchrotron-dominated population counts and \emph{blue} crosshairs indicate the dust-dominated counts.   Crosshairs with full error enclose $68\%$ of the probability about the median and are estimated in the bootstrap over flux described in Sec.~\ref{sec:counts}.  Here we have offset the two populations slightly in flux so that they do not lie on top of one another and the total counts are at zero offset.  
A source is identified as synchrotron-dominated (dust-dominated) if $\alpha < \alphathresh$ ($\alpha \ge \alphathresh$) in the bootstrap resampling from the joint posterior flux distributions, see Sec.~\ref{sec:counts}.  
This splits the populations so that their differential counts sum to the total counts.  A correction for survey completeness from simulations described in Sec.~\ref{sec:complpur} is also applied, and impacts primarily the $1.4$~mm counts in the lowest two flux bins.
\label{fig:bybandplot}}
\end{figure*}

\section{Single-band Source Counts}
\label{sec:counts}


Using the flux measurements in each band, we estimate the probability distribution of intrinsic fluxes for each source by constructing the two-band posterior likelihood $P(S_{\mathrm{max},1},S_{\mathrm{max},2}|S_{p,m,1},S_{p,m,2})$ as described in Sec.~\ref{sec:deboost}.  To derive the counts as a function of flux from these distributions of intrinsic fluxes, we apply a bootstrap method similar to the one described in \citet{austermann10}.  Here, for each source, we randomly draw $5 \times 10^4$ intrinsic fluxes from the two-band posterior, forming effectively $5\times 10^4$ mock catalogs of intrinsic fluxes in both bands.  For each catalog, we draw a subset of these sources with replacement, where the number of sources drawn is a Poisson deviate of the catalog size.  This resampling accounts for sample variance but not cosmological variance (which would require an additional variance term to describe how counts are expected to vary from sky patch to sky patch because of large scale structure).  

For each of these $5 \times 10^4$ resampled catalogs, we estimate $dN/dS$ and $N(>S)$.  In each flux bin, we then find the $16\%$, $50\%$, and $84\%$ percentile points (that is, $68\%$ of the enclosed probability around the median) of the distributions of $dN/dS$ and $N(>S)$ in that bin.  This yields the equivalent of $1\sigma$ normally-distributed errors in each flux bin.  The $N(>S)$ and $dN/dS$ bootstraps account for sample variance and posterior distribution variance (which includes noise, calibration error, deboosting, and cross-band information).  Because the posterior flux distributions per source span several flux bins in the counts, even the errors on differential counts will be correlated.  The counts are corrected for the completeness using the simulations described in Sec.~\ref{sec:completeness}.  Fig.~\ref{fig:bybandplot} shows differential counts in both bands as well as the differential counts of populations that we identify as either dust or synchrotron-dominated.  
The counts are given in numerical form in Tables~\ref{tab:diffcountstable150} through~\ref{tab:countstable220}.

The discrimination into synchrotron and dust-dominated counts is included in the bootstrap method.  Here, each catalog resampling will have drawn intrinsic fluxes at $1.4$~mm and $2.0$~mm, and so each source in that catalog will have a spectral index.  The dust-dominated (synchrotron-dominated) counts are derived from those sources that have $\alpha > \alphathresh$ ($\alpha < \alphathresh$) in the resampling.  A source with $P(\alpha > \alphathresh) > 0.3$ will therefore fall into the dust counts in $30\%$ of the resamplings and into the synchrotron counts $70\%$ of the resamplings.

We also estimate purity in each flux bin of the counts statistics using the resampled catalogs.  The purity is not evaluated for the flux at the bin center (which represents an intrinsic flux), but is instead related to the signal-to-noise of the raw flux of the sources that contribute to that bin.  In each resampling, if a source lies in a flux bin, we find the associated purity from its raw signal-to-noise.  The purity in the bin is then the weighted average of the purity of each source detection that contributes to the bin.  In the counts presented here, the purity in the lowest flux bin is $0.80$ at $2.0$~mm and $0.76$ at $1.4$~mm. 

The Bayesian method accounts for a particular sense of the purity which is slightly different than purity presented in Sec.~\ref{sec:purity}.  In Sec.~\ref{sec:purity}, we take purity to be the fraction of noise fluctuations that are counted as sources.  In principle, one would then suppress this fraction of the counts because it represents spurious detection.  The outlook of the Bayesian method is that each pixel always has some intrinsic source flux -- it may just be a tiny flux on top of a large positive noise fluctuation, and so is not strictly a false detection.  When we determine the counts, we draw such sub-threshold fluxes from the posterior flux distribution.  The effect of this is that in some fraction of the bootstrap samples, a source will scatter downward in flux (representing a tiny intrinsic flux plus a large positive noise fluctuation), out of the flux range presented.  No additional correction for the purity (in the sense of Sec.~\ref{sec:purity}) needs to be applied.  Completeness is not included in this framework and so is corrected for explicitly.  

\begin{deluxetable*}{llllc}
\tabletypesize{\tiny}
\tablecaption{The $2.0$~mm differential counts.
\label{tab:diffcountstable150}}
\tablehead{
\colhead{Flux range} & \colhead{$dN/dS$ total} & \colhead{$dN/dS$ sync} & \colhead{$dN/dS$ dust} & \colhead{completeness} \\
\colhead{Jy} & \colhead{${\rm Jy}^{-1} {\rm deg}^{-2}$} & \colhead{${\rm Jy}^{-1} {\rm deg}^{-2}$} & \colhead{${\rm Jy}^{-1} {\rm deg}^{-2}$} & 
}
\startdata
$4.4\times10^{-3}-5.6\times10^{-3}$ & $(4.17_{-0.9}^{+0.9})\times 10^{2}$ & $(3.24_{-0.8}^{+0.8})\times 10^{2}$ & $(9.26_{-4.6}^{+4.6})\times 10^{1}$ & $0.88$\\
$5.6\times10^{-3}-7.0\times10^{-3}$ & $(2.49_{-0.6}^{+0.6})\times 10^{2}$ & $(1.99_{-0.5}^{+0.6})\times 10^{2}$ & $(4.15_{-2.5}^{+3.3})\times 10^{1}$ & $0.98$\\
$7.0\times10^{-3}-8.7\times10^{-3}$ & $(1.44_{-0.4}^{+0.5})\times 10^{2}$ & $(1.24_{-0.4}^{+0.4})\times 10^{2}$ & $(1.96_{-1.3}^{+2.0})\times 10^{1}$ & $1.00$\\
$8.7\times10^{-3}-1.1\times10^{-2}$ & $(7.32_{-2.1}^{+3.1})\times 10^{1}$ & $(6.79_{-2.6}^{+2.1})\times 10^{1}$ & $(1.05_{-1.0}^{+1.0})\times 10^{1}$ & $1.00$\\
$1.1\times10^{-2}-1.4\times10^{-2}$ & $(3.75_{-1.3}^{+2.1})\times 10^{1}$ & $(3.75_{-1.7}^{+1.7})\times 10^{1}$ & $ 0 _{- 0 }^{+8.3}$ & $1.00$\\
$1.4\times10^{-2}-1.7\times10^{-2}$ & $(3.33_{-1.3}^{+1.0})\times 10^{1}$ & $(2.99_{-1.0}^{+1.3})\times 10^{1}$ & $ 0 _{- 0 }^{+3.3}$ & $1.00$\\
$1.7\times10^{-2}-2.2\times10^{-2}$ & $(2.12_{-0.8}^{+0.8})\times 10^{1}$ & $(2.12_{-0.8}^{+0.8})\times 10^{1}$ &   & $1.00$\\
$2.2\times10^{-2}-2.7\times10^{-2}$ & $(1.48_{-0.6}^{+0.6})\times 10^{1}$ & $(1.48_{-0.6}^{+0.6})\times 10^{1}$ &   & $1.00$\\
$2.7\times10^{-2}-3.4\times10^{-2}$ & $8.45_{-5.1}^{+3.4}$ & $8.45_{-5.1}^{+3.4}$ &   & $1.00$\\
$3.4\times10^{-2}-4.2\times10^{-2}$ & $2.70_{-2.7}^{+1.3}$ & $2.70_{-2.7}^{+1.3}$ &   & $1.00$\\
$4.2\times10^{-2}-5.3\times10^{-2}$ & $4.30_{-2.2}^{+2.2}$ & $4.30_{-2.2}^{+2.2}$ &   & $1.00$\\
$5.3\times10^{-2}-6.7\times10^{-2}$ & $3.43_{-1.7}^{+2.6}$ & $3.43_{-1.7}^{+2.6}$ &   & $1.00$\\
$6.7\times10^{-2}-8.3\times10^{-2}$ & $(6.85_{-6.9}^{+13.7})\times 10^{-1}$ & $(6.85_{-6.9}^{+13.7})\times 10^{-1}$ &   & $1.00$\\
$8.3\times10^{-2}-1.0\times10^{-1}$ & $(5.47_{-5.5}^{+5.5})\times 10^{-1}$ & $(5.47_{-5.5}^{+5.5})\times 10^{-1}$ &   & $1.00$\\
$1.0\times10^{-1}-1.3\times10^{-1}$ & $(4.36_{-4.4}^{+4.4})\times 10^{-1}$ & $(4.36_{-4.4}^{+4.4})\times 10^{-1}$ &   & $1.00$\\
$1.3\times10^{-1}-1.6\times10^{-1}$ & $(3.48_{-3.5}^{+3.5})\times 10^{-1}$ & $(3.48_{-3.5}^{+3.5})\times 10^{-1}$ &   & $1.00$\\
$2.1\times10^{-1}-2.6\times10^{-1}$ & $(2.22_{-2.2}^{+2.2})\times 10^{-1}$ & $(2.22_{-2.2}^{+2.2})\times 10^{-1}$ &   & $1.00$\\
$2.6\times10^{-1}-3.2\times10^{-1}$ & $ 0 _{- 0 }^{+0.2}$ & $ 0 _{- 0 }^{+0.2}$ &   & $1.00$\\
$3.2\times10^{-1}-4.1\times10^{-1}$ & $ 0 _{- 0 }^{+0.3}$ & $ 0 _{- 0 }^{+0.3}$ &   & $1.00$\\
$5.1\times10^{-1}-6.4\times10^{-1}$ & $ 0 _{- 0 }^{+9.0\times10^{-2}}$ & $ 0 _{- 0 }^{+9.0\times10^{-2}}$ &   & $1.00$\\
$6.4\times10^{-1}-8.0\times10^{-1}$ & $ 0 _{- 0 }^{+0.1}$ & $ 0 _{- 0 }^{+0.1}$ &   & $1.00$\\
\end{deluxetable*}
 
\begin{deluxetable*}{llllc}
\tabletypesize{\tiny}
\tablecaption{The $1.4$~mm differential counts.
\label{tab:diffcountstable220}}
\tablehead{
\colhead{Flux range} & \colhead{$dN/dS$ total} & \colhead{$dN/dS$ sync} & \colhead{$dN/dS$ dust} & \colhead{completeness} \\
\colhead{Jy} & \colhead{${\rm Jy}^{-1} {\rm deg}^{-2}$} & \colhead{${\rm Jy}^{-1} {\rm deg}^{-2}$} & \colhead{${\rm Jy}^{-1} {\rm deg}^{-2}$} & 
}
\startdata
$1.1\times10^{-2}-1.4\times10^{-2}$ & $(6.93_{-2.5}^{+3.0})\times 10^{1}$ & $(2.48_{-1.0}^{+2.0})\times 10^{1}$ & $(3.96_{-1.5}^{+2.5})\times 10^{1}$ & $0.84$\\
$1.4\times10^{-2}-1.7\times10^{-2}$ & $(4.42_{-1.7}^{+1.7})\times 10^{1}$ & $(2.04_{-1.0}^{+1.0})\times 10^{1}$ & $(2.38_{-1.4}^{+1.0})\times 10^{1}$ & $0.97$\\
$1.7\times10^{-2}-2.2\times10^{-2}$ & $(2.66_{-1.1}^{+1.1})\times 10^{1}$ & $(1.59_{-0.8}^{+0.8})\times 10^{1}$ & $(1.06_{-0.5}^{+0.8})\times 10^{1}$ & $1.00$\\
$2.2\times10^{-2}-2.7\times10^{-2}$ & $(1.69_{-0.6}^{+0.8})\times 10^{1}$ & $(1.06_{-0.6}^{+0.6})\times 10^{1}$ & $6.35_{-4.2}^{+6.4}$ & $1.00$\\
$2.7\times10^{-2}-3.4\times10^{-2}$ & $8.45_{-3.4}^{+6.8}$ & $5.07_{-3.4}^{+5.1}$ & $3.38_{-3.4}^{+3.4}$ & $1.00$\\
$3.4\times10^{-2}-4.2\times10^{-2}$ & $5.39_{-2.7}^{+4.0}$ & $2.70_{-1.3}^{+4.0}$ & $1.35_{-1.3}^{+2.7}$ & $1.00$\\
$4.2\times10^{-2}-5.3\times10^{-2}$ & $5.38_{-3.2}^{+3.2}$ & $4.30_{-2.2}^{+3.2}$ & $ 0 _{- 0 }^{+1.1}$ & $1.00$\\
$5.3\times10^{-2}-6.7\times10^{-2}$ & $2.58_{-1.7}^{+1.7}$ & $2.58_{-1.7}^{+1.7}$ &   & $1.00$\\
$6.7\times10^{-2}-8.3\times10^{-2}$ & $ 0 _{- 0 }^{+1.4}$ & $ 0 _{- 0 }^{+1.4}$ &   & $1.00$\\
$8.3\times10^{-2}-1.0\times10^{-1}$ & $(5.47_{-5.5}^{+5.5})\times 10^{-1}$ & $(5.47_{-5.5}^{+5.5})\times 10^{-1}$ &   & $1.00$\\
$1.0\times10^{-1}-1.3\times10^{-1}$ & $ 0 _{- 0 }^{+0.9}$ & $ 0 _{- 0 }^{+0.9}$ &   & $1.00$\\
$1.3\times10^{-1}-1.6\times10^{-1}$ & $ 0 _{- 0 }^{+0.3}$ & $ 0 _{- 0 }^{+0.3}$ &   & $1.00$\\
$1.6\times10^{-1}-2.1\times10^{-1}$ & $ 0 _{- 0 }^{+0.6}$ & $ 0 _{- 0 }^{+0.6}$ &   & $1.00$\\
$2.1\times10^{-1}-2.6\times10^{-1}$ & $ 0 _{- 0 }^{+0.2}$ & $ 0 _{- 0 }^{+0.2}$ &   & $1.00$\\
$2.6\times10^{-1}-3.2\times10^{-1}$ & $ 0 _{- 0 }^{+0.2}$ & $ 0 _{- 0 }^{+0.2}$ &   & $1.00$\\
$5.1\times10^{-1}-6.4\times10^{-1}$ & $ 0 _{- 0 }^{+0.2}$ & $ 0 _{- 0 }^{+0.2}$ &   & $1.00$\\
\end{deluxetable*}
 
\begin{deluxetable*}{llllc}
\tabletypesize{\tiny}
\tablecaption{The $2.0$~mm cumulative counts.  The reported purity is the weighted average of the purity of source detections that contribute to a bin, see Sec.~\ref{sec:counts}.
\label{tab:countstable150}}
\tablehead{
\colhead{Flux range (Jy)} & \colhead{$N(>S)$ total} & \colhead{$N(>S)$ sync} & \colhead{$N(>S)$ dust} & \colhead{purity} \\
\colhead{Jy} & \colhead{${\rm deg}^{-2}$} & \colhead{${\rm deg}^{-2}$} & \colhead{${\rm deg}^{-2}$} & 
}
\startdata
$4.4\times10^{-3}-5.6\times10^{-3}$ & $1.90_{-0.2}^{+0.2}$ & $1.65_{-0.2}^{+0.2}$ & $(2.47_{-0.6}^{+0.7})\times 10^{-1}$ & $0.80$\\
$5.6\times10^{-3}-7.0\times10^{-3}$ & $1.43_{-0.1}^{+0.1}$ & $1.29_{-0.1}^{+0.1}$ & $(1.39_{-0.5}^{+0.5})\times 10^{-1}$ & $0.92$\\
$7.0\times10^{-3}-8.7\times10^{-3}$ & $1.08_{-0.1}^{+0.1}$ & $1.00_{-0.1}^{+0.1}$ & $(6.92_{-2.3}^{+4.6})\times 10^{-2}$ & $0.99$\\
$8.7\times10^{-3}-1.1\times10^{-2}$ & $(8.30_{-1.0}^{+1.0})\times 10^{-1}$ & $(7.84_{-0.9}^{+1.0})\times 10^{-1}$ & $(3.46_{-2.3}^{+2.3})\times 10^{-2}$ & $1.00$\\
$1.1\times10^{-2}-1.4\times10^{-2}$ & $(6.57_{-0.9}^{+0.9})\times 10^{-1}$ & $(6.46_{-0.9}^{+0.8})\times 10^{-1}$ & $(1.15_{-1.2}^{+1.2})\times 10^{-2}$ & $1.00$\\
$1.4\times10^{-2}-1.7\times10^{-2}$ & $(5.42_{-0.8}^{+0.8})\times 10^{-1}$ & $(5.42_{-0.8}^{+0.8})\times 10^{-1}$ & $ 0 _{- 0 }^{+1.2\times10^{-2}}$ & $1.00$\\
$1.7\times10^{-2}-2.2\times10^{-2}$ & $(4.27_{-0.7}^{+0.8})\times 10^{-1}$ & $(4.27_{-0.7}^{+0.8})\times 10^{-1}$ &   & $1.00$\\
$2.2\times10^{-2}-2.7\times10^{-2}$ & $(3.34_{-0.6}^{+0.7})\times 10^{-1}$ & $(3.34_{-0.6}^{+0.7})\times 10^{-1}$ &   & $1.00$\\
$2.7\times10^{-2}-3.4\times10^{-2}$ & $(2.54_{-0.5}^{+0.6})\times 10^{-1}$ & $(2.54_{-0.5}^{+0.6})\times 10^{-1}$ &   & $1.00$\\
$3.4\times10^{-2}-4.2\times10^{-2}$ & $(2.08_{-0.5}^{+0.5})\times 10^{-1}$ & $(2.08_{-0.5}^{+0.5})\times 10^{-1}$ &   & $1.00$\\
$4.2\times10^{-2}-5.3\times10^{-2}$ & $(1.85_{-0.5}^{+0.5})\times 10^{-1}$ & $(1.85_{-0.5}^{+0.5})\times 10^{-1}$ &   & $1.00$\\
$5.3\times10^{-2}-6.7\times10^{-2}$ & $(1.38_{-0.3}^{+0.5})\times 10^{-1}$ & $(1.38_{-0.3}^{+0.5})\times 10^{-1}$ &   & $1.00$\\
$6.7\times10^{-2}-8.3\times10^{-2}$ & $(9.23_{-3.5}^{+3.5})\times 10^{-2}$ & $(9.23_{-3.5}^{+3.5})\times 10^{-2}$ &   & $1.00$\\
$8.3\times10^{-2}-1.0\times10^{-1}$ & $(6.92_{-2.3}^{+2.3})\times 10^{-2}$ & $(6.92_{-2.3}^{+2.3})\times 10^{-2}$ &   & $1.00$\\
$1.0\times10^{-1}-1.3\times10^{-1}$ & $(5.77_{-2.3}^{+2.3})\times 10^{-2}$ & $(5.77_{-2.3}^{+2.3})\times 10^{-2}$ &   & $1.00$\\
$1.3\times10^{-1}-1.6\times10^{-1}$ & $(4.61_{-2.3}^{+2.3})\times 10^{-2}$ & $(4.61_{-2.3}^{+2.3})\times 10^{-2}$ &   & $1.00$\\
$1.6\times10^{-1}-2.1\times10^{-1}$ & $(3.46_{-2.3}^{+2.3})\times 10^{-2}$ & $(3.46_{-2.3}^{+2.3})\times 10^{-2}$ &   & $1.00$\\
$2.1\times10^{-1}-2.6\times10^{-1}$ & $(3.46_{-2.3}^{+1.2})\times 10^{-2}$ & $(3.46_{-2.3}^{+2.3})\times 10^{-2}$ &   & $1.00$\\
$2.6\times10^{-1}-3.2\times10^{-1}$ & $(2.31_{-1.2}^{+1.2})\times 10^{-2}$ & $(2.31_{-1.2}^{+1.2})\times 10^{-2}$ &   & $1.00$\\
$3.2\times10^{-1}-4.1\times10^{-1}$ & $(1.15_{-1.2}^{+2.3})\times 10^{-2}$ & $(1.15_{-1.2}^{+2.3})\times 10^{-2}$ &   & $1.00$\\
$4.1\times10^{-1}-5.1\times10^{-1}$ & $(1.15_{-1.2}^{+1.2})\times 10^{-2}$ & $(1.15_{-1.2}^{+1.2})\times 10^{-2}$ &   & 1.00\\
$5.1\times10^{-1}-6.4\times10^{-1}$ & $(1.15_{-1.2}^{+1.2})\times 10^{-2}$ & $(1.15_{-1.2}^{+1.2})\times 10^{-2}$ &   & $1.00$\\
$6.4\times10^{-1}-8.0\times10^{-1}$ & $ 0 _{- 0 }^{+2.3\times10^{-2}}$ & $ 0 _{- 0 }^{+2.3\times10^{-2}}$ &   & $1.00$\\
\end{deluxetable*}
 
\begin{deluxetable*}{llllc}
\tabletypesize{\tiny}
\tablecaption{The $1.4$~mm cumulative counts.
\label{tab:countstable220}}
\tablehead{
\colhead{Flux range (Jy)} & \colhead{$N(>S)$ total} & \colhead{$N(>S)$ sync} & \colhead{$N(>S)$ dust} & \colhead{purity} \\
\colhead{Jy} & \colhead{${\rm deg}^{-2}$} & \colhead{${\rm deg}^{-2}$} & \colhead{${\rm deg}^{-2}$} & 
}
\startdata
$1.1\times10^{-2}-1.4\times10^{-2}$ & $(8.39_{-1.1}^{+1.2})\times 10^{-1}$ & $(4.99_{-0.8}^{+0.9})\times 10^{-1}$ & $(3.37_{-0.8}^{+0.8})\times 10^{-1}$ & $0.76$\\
$1.4\times10^{-2}-1.7\times10^{-2}$ & $(6.38_{-0.9}^{+1.0})\times 10^{-1}$ & $(4.18_{-0.7}^{+0.8})\times 10^{-1}$ & $(2.20_{-0.6}^{+0.6})\times 10^{-1}$ & $0.86$\\
$1.7\times10^{-2}-2.2\times10^{-2}$ & $(4.84_{-0.8}^{+0.8})\times 10^{-1}$ & $(3.46_{-0.6}^{+0.7})\times 10^{-1}$ & $(1.38_{-0.5}^{+0.5})\times 10^{-1}$ & $0.96$\\
$2.2\times10^{-2}-2.7\times10^{-2}$ & $(3.69_{-0.7}^{+0.7})\times 10^{-1}$ & $(2.77_{-0.6}^{+0.7})\times 10^{-1}$ & $(8.07_{-2.3}^{+4.6})\times 10^{-2}$ & $1.00$\\
$2.7\times10^{-2}-3.4\times10^{-2}$ & $(2.65_{-0.6}^{+0.6})\times 10^{-1}$ & $(2.19_{-0.5}^{+0.6})\times 10^{-1}$ & $(4.61_{-2.3}^{+3.5})\times 10^{-2}$ & $1.00$\\
$3.4\times10^{-2}-4.2\times10^{-2}$ & $(1.96_{-0.5}^{+0.6})\times 10^{-1}$ & $(1.85_{-0.5}^{+0.5})\times 10^{-1}$ & $(2.31_{-2.3}^{+1.2})\times 10^{-2}$ & $1.00$\\
$4.2\times10^{-2}-5.3\times10^{-2}$ & $(1.50_{-0.3}^{+0.5})\times 10^{-1}$ & $(1.50_{-0.5}^{+0.5})\times 10^{-1}$ & $ 0 _{- 0 }^{+1.2\times10^{-2}}$ & $1.00$\\
$5.3\times10^{-2}-6.7\times10^{-2}$ & $(9.23_{-3.5}^{+4.6})\times 10^{-2}$ & $(9.23_{-3.5}^{+4.6})\times 10^{-2}$ &   & $1.00$\\
$6.7\times10^{-2}-8.3\times10^{-2}$ & $(6.92_{-3.5}^{+2.3})\times 10^{-2}$ & $(6.92_{-3.5}^{+2.3})\times 10^{-2}$ &   & $1.00$\\
$8.3\times10^{-2}-1.0\times10^{-1}$ & $(5.77_{-2.3}^{+2.3})\times 10^{-2}$ & $(5.77_{-2.3}^{+2.3})\times 10^{-2}$ &   & $1.00$\\
$1.0\times10^{-1}-1.3\times10^{-1}$ & $(4.61_{-2.3}^{+2.3})\times 10^{-2}$ & $(4.61_{-2.3}^{+2.3})\times 10^{-2}$ &   & $1.00$\\
$1.3\times10^{-1}-1.6\times10^{-1}$ & $(3.46_{-2.3}^{+2.3})\times 10^{-2}$ & $(3.46_{-2.3}^{+2.3})\times 10^{-2}$ &   & $1.00$\\
$1.6\times10^{-1}-2.1\times10^{-1}$ & $(2.31_{-1.2}^{+2.3})\times 10^{-2}$ & $(2.31_{-1.2}^{+2.3})\times 10^{-2}$ &   & $1.00$\\
$2.1\times10^{-1}-2.6\times10^{-1}$ & $(2.31_{-1.2}^{+1.2})\times 10^{-2}$ & $(2.31_{-1.2}^{+1.2})\times 10^{-2}$ &   & $1.00$\\
$2.6\times10^{-1}-3.2\times10^{-1}$ & $(1.15_{-1.2}^{+2.3})\times 10^{-2}$ & $(1.15_{-1.2}^{+2.3})\times 10^{-2}$ &   & $1.00$\\
$3.2\times10^{-1}-4.1\times10^{-1}$ & $(1.15_{-1.2}^{+1.2})\times 10^{-2}$ & $(1.15_{-1.2}^{+1.2})\times 10^{-2}$ &   & $1.00$\\
$4.1\times10^{-1}-5.1\times10^{-1}$ & $(1.15_{-1.2}^{+1.2})\times 10^{-2}$ & $(1.15_{-1.2}^{+1.2})\times 10^{-2}$ &   & $1.00$\\
$5.1\times10^{-1}-6.4\times10^{-1}$ & $(1.15_{-1.2}^{+1.2})\times 10^{-2}$ & $(1.15_{-1.2}^{+1.2})\times 10^{-2}$ &   & $1.00$\\
\end{deluxetable*}
 
\section{Associations with External Catalogs and Follow-up Observations with ATCA}
\label{sec:assoc}

Where possible, we identify candidate counterparts to the SPT-detected sources in several external catalogs
and databases.  We have queried the NED\footnote{http://nedwww.ipac.caltech.edu/} and 
SIMBAD\footnote{http://simbad.u-strasbg.fr/simbad}
databases for counterparts within $2.0$~arcmin 
of all 3496 of our $\ge 3 \sigma$ sources. We have also searched catalogs
from five individual observatories for counterparts:  1) the SUMSS catalog; 
2) the IRAS-FSC; 3) the RASS-BSC and RASS-FSC; 4) the PMN catalog; 
and 5) the AT20G catalog.
We search these catalogs, in particular, 
because these observatories are especially relevant
for extragalactic sources in the Southern Hemisphere.
Additionally, as mentioned in Section ~\ref{sec:purity}, we have 
performed follow-up observations at 6~cm on many of our brightest sources with the Australia Telescope Compact Array (ATCA). 

ATCA observations were performed at 6 cm, 12 mm, and 7 mm during August 2008 under program C1563. 
At 6 cm, 55 sources in this field were observed and 52 sources were detected at a typical RMS noise of 0.5~mJy.
One of the sources not detected (SPT-S J053250-5047.1) we classify as a dusty source (see Section ~\ref{sec:smg}). 
The other two undetected sources (SPT-S J053412-5924.3 and SPT-S J055232-5349.4) are extended in the SUMMS 
catalog (45\arcsec  ~resolution) and thus heavily resolved at the typical 
9\arcsec$\times$3\arcsec \ 6 cm resolution.

The majority of our synchrotron-dominated sources have a clear counterpart either in an external catalog, 
in our 6~cm ATCA observations, or both.  Because these sources are expected
to have fluxes that vary significantly over time \citep{kellerman68}, and because
the radio catalogs in which we search for counterparts are not significantly 
deeper than our own measurements, one might expect to not find counterparts 
for sources that were caught at peak brightness in the SPT observations.

Despite this caveat, of the 107 sources above 
$5\sigma$ that we classify as synchrotron-dominated, only three of them do not 
have SUMSS counterparts within $30$~arcsec.  The brightest of these three 
sources (SPT-S J053345-5818.1) is classified as extended using the method 
of Sec.~\ref{sec:extend}, and we find counterparts for this source within $10$~arcsec in 
both the PMN catalog and our ATCA follow-up observations.
The extended nature of this source and the offset 
between the SUMSS and SPT/PMN/ATCA positions for this source are expected 
if this source is an AGN with extended jets.
The synchrotron emission from the source is presumably
dominated by the radio-lobe (jet) contribution in the 36~cm SUMSS observations 
but dominated by emission from the core at shorter 
wavelengths.  This frequency-dependent core-to-lobe flux ratio is commonly seen
in radio-loud AGN \citep[e.g.,][]{kharb08,dezotti10}
and is predicted by certain unified AGN models \citep[e.g.,][]{jackson99}.
Indeed, visual inspection of this source reveals one SUMSS source 
on either side of the SPT location (each within $40$~arcsec) and a RASS-BSC object 
(also presumably dominated by emission from the AGN core) directly 
on top of the SPT location.  

Of the remaining two $\ge 5 \sigma$ synchrotron-dominated sources with no SUMSS 
counterpart within $30$~arcsec, 
one (SPT-S J050334-5244.8)
has a SUMSS counterpart and a counterpart in our ATCA follow-up, both 
within $35$~arcsec, and the other (SPT-S J053726-5434.4)
has a counterpart $11$~arcsec away
in the \citet{veron06} and Hamburg-ESO \citep{wisotzki91} catalogs as well 
as a possibly associated SUMSS source $1.3$~arcmin away. 
We thus believe that every one of the sources that we detect at $\ge 5 \sigma$ 
and classify as synchrotron-dominated is a real source.  This is consistent with our 
estimates of purity in Sec.~\ref{sec:purity}, which predict a false detection 
rate of effectively zero above $5 \sigma$.

The situation with our dust-dominated sources is very different.  
Of the 23 (47) sources 
above $5 \sigma$ ($4.5 \sigma$) that we classify as dust-dominated, only 10 (12) have 
counterparts (in any catalog) within $30$~arcsec, and only 12 (15) have counterparts 
within $1$~arcmin.  
Given the studies summarized in Sec.~\ref{sec:purity} and the counterparts found for 
the synchrotron-dominated sources, there is almost no chance that all (or even a majority of) 
these detections without counterparts are spurious.  
Of the dust-dominated sources above $5 \sigma$ that do have counterparts within
$30$~arcsec, all but three are nearby galaxies detected with IRAS.  Two of the 
remaining three are associated with SUMSS sources, while the other is 
SPT-S~J054716$-$5104.1 which is associated with the debris disk around 
the star $\beta$ Pictoris.  In this field there are no 
other SPT sources within $30$~arcsec of a SIMBAD database star.

\section{Individual-Population Source Counts and Implications}
\label{sec:populations}
 
To briefly summarize the results of the last several sections, three broad classes 
of point sources are detected with high significance in the SPT $1.4$~mm and 
$2.0$~mm maps: 
\begin{enumerate}
\item Sources with $1.4$~mm-to-$2.0$~mm flux ratios consistent with 
synchrotron emission, the vast majority of which appear in radio 
catalogs and/or in our centimeter-wave follow-up observations with ATCA and
which we generically refer to as AGN;  
\item Sources with $1.4$~mm-to-$2.0$~mm flux ratios consistent with 
dust emission which have low-redshift $(z \ll 1)$ counterparts in the IRAS-FSC, 
and which we will generically call IRAS sources; 
\item Previously undetected sources with $1.4$~mm-to-$2.0$~mm flux ratios 
consistent with dust emission.
\end{enumerate}
Here we present the measured source counts as a function of flux 
for each of these populations and discuss the implications of our measurements.


\begin{figure*}[h]
 \begin{center}
 \epsfig{file=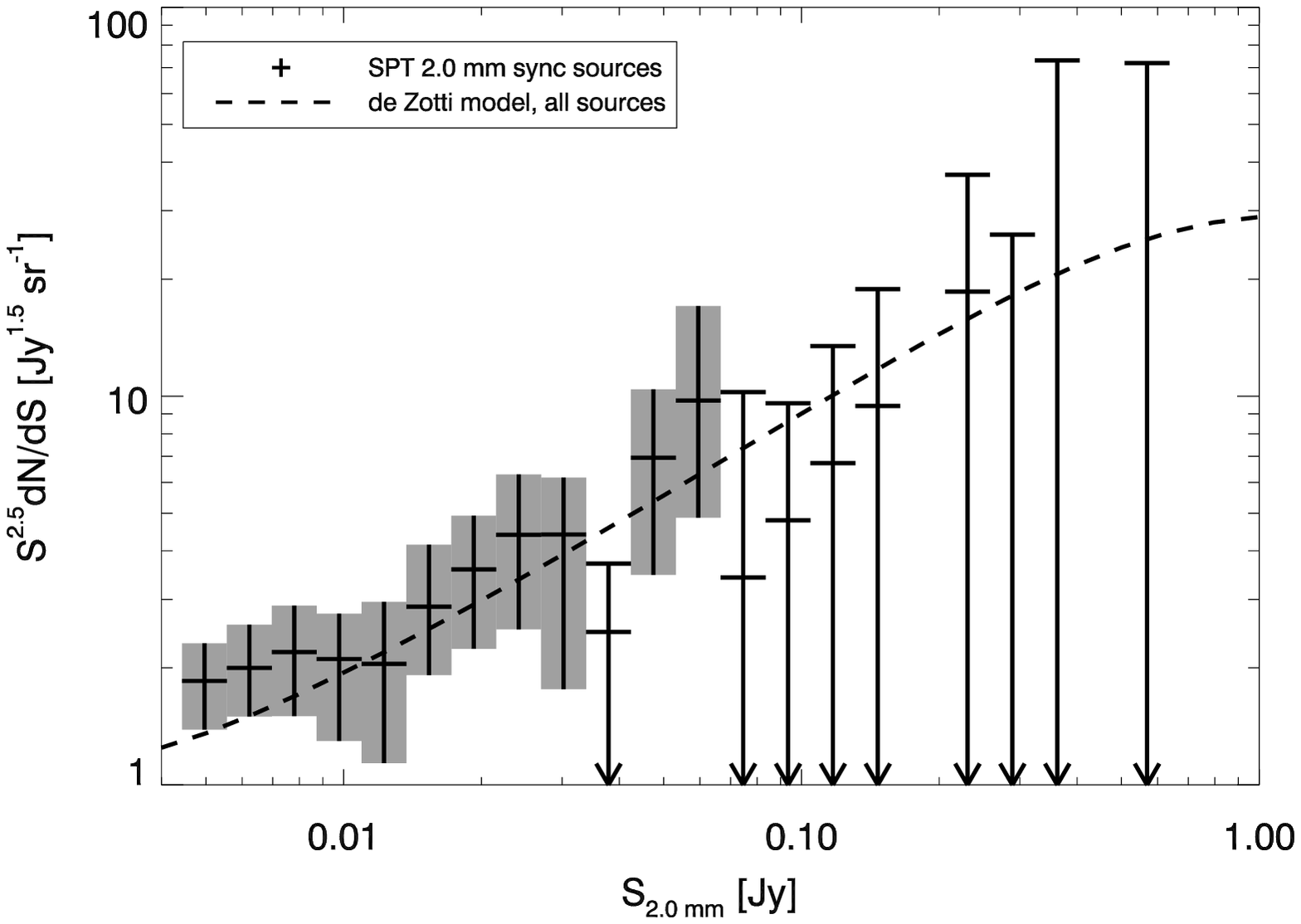, width=16cm} 
\end{center}
\caption{\small Differential counts for the population of sources identified as synchrotron-dominated compared to the \citet{dezotti05} model.  Here the counts are scaled by $S^{2.5}$ relative to by-band differential counts shown in Fig.~\ref{fig:bybandplot} to match the (geometrical) convention in AGN literature.  The error regions enclose $68\%$ of the probability centered about the median counts, and are calculated using the bootstrap over the two-band posterior intrinsic flux (at $2.0$~mm) that is described in Sec.~\ref{sec:counts}. 
In these counts, a source is identified as synchrotron-dominated if $\alpha < \alphathresh$ in the resampling. 
\label{fig:syncdezotti}}
\end{figure*}

\subsection{Synchrotron-Dominated Source Counts}

From associations with radio catalogs and from our $6$~cm followup
observations with ATCA, we conclude that SPT sources with $1.4$~mm-to-$2.0$~mm spectral
indices less than $\alphathresh$ are consistent with being members of the 
classical radio-source population (see \citet{dezotti10} for a recent review). 
Although contributions to this population can come from synchrotron and free-free 
emission in normal and starburst galaxies \citep{condon92,dezotti10}, 
the population is dominated at short radio wavelengths ($30$~cm and below) 
and moderate to high fluxes ($10$~mJy to $1$~Jy) by synchrotron emission from AGN
\citep{dezotti10}.  At even shorter wavelengths ($1$~cm and below), 
the moderate-to-high-flux counts are expected to be dominated by the 
sub-class of AGN known as flat-spectrum radio quasars (FSRQs)
\citep{dezotti05}.

The behavior of this source population at mm wavelengths is interesting for 
several reasons.  Astrophysically, mm measurements of AGN have 
the potential to inform models of AGN emission mechanisms and evolution, 
particularly whether FSRQs undergo spectral 
steepening at shorter wavelengths.  This short-wavelength behavior of 
FSRQs is also of interest to the CMB and SZ communities, as the predictions of
contamination of mm CMB power spectrum measurements and SZ galaxy cluster 
surveys by AGN emission depend heavily on extrapolations of 
long-wavelength source 
properties to mm wavelengths \citep{lin07,reichardt09}.  Finally, 
the compact angular size of FSRQs (along with their short-wavelength 
brightness) make them attractive candidates for phase calibration sources for 
the Atacama Large Millimeter Array (ALMA).

Two obvious questions that can be addressed with multi-band SPT 
observations of classical radio sources are:  1) Are the sources selected 
at $2.0$~mm consistent with a population of FSRQs; and 2) Is the 
spectral behavior of these sources between $2.0$~mm and $1.4$~mm
consistent with their longer-wavelength (flat-spectrum) behavior, or is 
there evidence of a spectral break or turnover?

The first question is made somewhat ambiguous by the lack of a  clear 
definition of the FSRQ population in terms of synchrotron spectral index.
\citet{dezotti05} use a single spectral index of $\alpha = -0.1$ for their
model of the FSRQ population, while surveys targeting ``flat spectrum
radio sources" have generally chosen $\alpha \ge -0.4$ or $-0.5$ as 
the defining threshold \citep{jackson02,healey07}.  Nevertheless, we 
can ask whether the behavior of these sources between $2.0$~mm and 
longer wavelengths is more consistent with a mean spectral index such 
as \citet{dezotti05} use or with a value more typical of steep-spectrum
sources such as $\alpha \sim -0.8$ \citep{dezotti10}.

We investigate this question using two lines of evidence.  
First, we have the comparison of the $2.0$~mm SPT flux 
with fluxes measured at longer wavelengths.  A serious caveat to any 
comparison of non-simultaneous observations of radio sources is 
that the known variability of such sources will add scatter --- and, potentially, 
bias --- to estimates of spectral behavior.  Our ATCA $6$~cm follow-up
observations took place approximately two months after the SPT 
observations, which will limit the effect of variability on our
$6$~cm-to-$2.0$~mm comparisons to timescales shorter than this 
separation.  According to \citet{murphy10}, the AT20G observations 
of sources between $-60^\circ$ and $-50^\circ$ declination took place 
in September and October, 2005, so $1.5$~cm-to-$2.0$~mm comparisons
should be viewed with greater caution.

With this caveat in mind, we find that our highest-significance 
$2.0$~mm-selected, synchrotron-dominated sources are completely
consistent with flat spectral behavior (or $\alpha \sim -0.1$)
between $6$~cm and $2.0$~mm.
If we take the 57 sources for which we have robust ($\ge 5 \sigma$) 
detections at both $6$~cm and $2.0$~mm, and we calculate a 
single spectral index for each source using the best-fit $2.0$~mm 
flux from Table~\ref{tab:cat_both} and the raw $6$~cm flux, we find
these sources have a distribution of spectral indices characterized 
by $\alpha = -0.13 \pm 0.21$.  Because this sample of 57 sources is 
not purely $2.0$~mm-selected (we are omitting sources without a robust
$6$~cm detection) and because the $6$~cm data have not been corrected 
for flux boosting, one might worry that this result is biased toward high 
$6$~cm fluxes and, hence, steeper spectral indices.  If we repeat the 
calculation using only the top 10 brightest $2.0$~mm sources --- which are 
all detected above $50 \sigma$ at $6$~cm --- we find a spectral index
distribution of $\alpha = -0.08 \pm 0.22$.

The second line of evidence that supports the hypothesis that our
$2.0$~mm-selected, synchrotron-dominated sources are FSRQs ---
or, more specifically, have a mean spectral index near $\bar{\alpha}=-0.1$ ---
is the agreement between the $2.0$~mm synchrotron-dominated 
counts and the predictions of \citet{dezotti05}. 
The \citet{dezotti05} 
model includes contributions from many populations of radio sources, including
normal and star-forming galaxies and many types of AGN, but at $1$~cm
and below, the $> 10$~mJy model counts are dominated by FSRQs.  
Fig.~\ref{fig:syncdezotti} compares our synchrotron-dominated $2.0$~mm 
counts to the \citet{dezotti05} $2$~mm model.  The model is roughly 
consistent with our measured counts, indicating that the $2.0$~mm 
synchrotron-dominated counts are consistent with a power-law extrapolation
of the long-wavelength FSRQ counts, and that the spectral
behavior used to extrapolate the long-wavelength FSRQ counts in the 
model --- in this case, simply assuming $\alpha=-0.1$ for all FSRQs --- 
is reasonably accurate down to $2.0$~mm.  

The answer to the second question --- whether the radio sources 
selected at $2.0$~mm show flat spectral behavior all the way to $1.4$~mm --- 
is addressed by our simultaneous $1.4$~mm and $2.0$~mm observations of 
these sources.  These observations indicate that a mean spectral index of
$\bar{\alpha}=-0.1$ is not an accurate description of the spectral behavior 
of these sources between $1.4$~mm and $2.0$~mm.
The distribution of spectral 
indices in Fig.~\ref{fig:alpha_hist} shows that our synchrotron-dominated
sources have $1.4$~mm-to-$2.0$~mm spectral indices peaked around
$\alpha=-0.5$, and a similar result is found for the sum of the posterior
spectral index PDFs for the 57 sources detected 
above $5 \sigma$ at both $6$~cm and $2.0$~mm.
If we take the ten brightest $2.0$~mm-selected sources, 
we find that the sum of the posterior
spectral index PDFs of these sources
is inconsistent with $\alpha = -0.1$ at $97\%$.

Interestingly, if we perform the same calculation using the published 
$1.5$~cm AT20G fluxes for our top ten $2.0$~mm-selected sources, 
the $2.0$~mm-to-$1.5$~cm spectral index distribution is 
$\alpha = -0.31 \pm 0.29$ 
while the $1.5$~cm-to-$6$~cm spectral index distribution is 
$\alpha = 0.25 \pm 0.36$ 
--- i.e., there is some evidence that the flux of these sources peaks
between $6$~cm and $2.0$~mm.  Biases induced by source variability 
are unlikely to have a significant effect on this measurement for two reasons.
First, these sources are so far above the detection threshold at $2.0$~mm
that they would have been significant detections at any epoch.  Second, 
biases due to variability 
would tend to drive estimates of this spectral index in the opposite direction ---
because sources are likely to be selected at their peak brightness, 
measurements taken at any other epoch should produce systematically lower 
fluxes.  This implies that if we had $1.5$~cm measurements taken simultaneously
with the SPT measurements, we would tend to find even higher $1.5$~cm fluxes
than in the AT20G catalog (assuming the variability is simultaneous at the two 
wavelengths).  These arguments imply that the evidence for these sources' flux peaking
between $6$~cm and $2.0$~mm is fairly robust, if not overly statistically significant.

\subsection{Dust-Dominated Source Counts}
\label{sec:smg}

\subsubsection{The Dust-Dominated Assumption, Spectral Indices, and Self-Absorbed Synchrotron}
\label{sec:gps}
\begin{figure*}[h]
 \begin{center}
 \epsfig{file=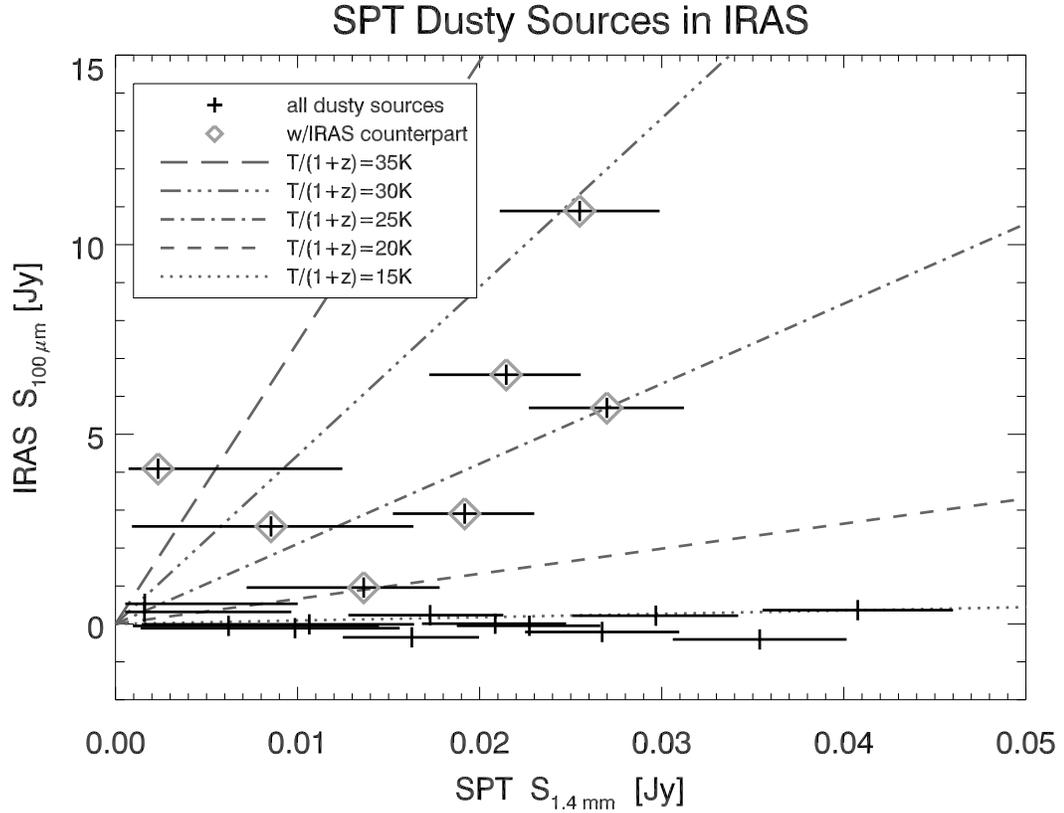, width=16cm} 
\end{center}
\caption{\small IRAS 100~\um \ flux vs.~SPT 1.4~mm flux for all SPT dust-dominated sources at S/N~$>5$.  
IRAS flux is taken from a version of the ISSA \citep{wheelock94}
100~\um \ map which has been filtered to enhance point-source
signal-to-noise.  Horizontal error bars are the $68 \%$ enclosed interval in the 
posterior $1.4$~mm flux distribution (as in Table~\ref{tab:cat_both}).
Vertical error bars are the width of the noise distribution in the filtered
IRAS map.
SPT sources with counterparts within $1$ arcmin in the IRAS FSC are shown
with diamond symbols.  Lines of constant 100~\um--1.4~mm flux ratio are shown for five emission
models, all modified blackbody laws with a dust emissivity index of $\beta=1.5$ 
(consistent with the value of $\beta$ used in \citet{dunne01}, \citet{chapman05}, and \citet{kovacs06})
and with 
dust temperatures of $15$, $20$, $25$, $30$, and $35~\mathrm{K}$ (if the emitter is nearby) or
those temperatures times $1+z$ (if the emitter lies at redshift $z$).
There is a clear distinction between the locus of sources with IRAS-FSC
counterparts --- which have flux ratios consistent with warm, low-redshift dust emission --- 
and the points which lie along the $x$ axis and have no counterparts, which must be either 
at moderate to high redshift or have anomalously cold dust. \label{fig:iras100um}}
\end{figure*}

We have referred throughout 
this work to the population of sources with $1.4$~mm-to-$2.0$~mm spectral 
indices greater than $\alphathresh$ as ``dust-dominated."  
For the sources in this class that have IRAS-FSC counterparts, the assumption
that thermal emission from dust is the dominant mechanism at mm wavelengths
is reasonable given their large IR fluxes.  The $1.4$~mm-to-$2.0$~mm
spectral indices for these sources are consistent with dust emission ---
the sum of the posterior spectral index PDFs for the 5 brightest sources with
IRAS counterparts peaks at $\alpha = 3.2$
--- and comparisons of $1.4$~mm SPT fluxes and $100 \ \mu \mathrm{m}$ IRAS 
fluxes of these sources show that their emission is consistent with thermal 
dust at moderate temperatures ($20$ to $40$~K) from the mm through the far-IR
(see Sec.~\ref{sec:dust_newpop} and Fig.~\ref{fig:iras100um} for details).

While thermal dust emission is also
the most natural candidate for explaining the sources with 
$1.4$~mm-to-$2.0$~mm spectral 
indices greater than $\alphathresh$ that do not have IRAS-FSC
counterparts, we cannot 
\emph{a priori} rule out self-absorbed synchrotron emission from AGN as the 
dominant emission mechanism for these sources.  Self-absorbed synchrotron
is the leading emission model for the population of 
gigahertz-peaked-spectrum (GPS) radio sources (see \citealt{odea98} for a review), 
and it is the only emission mechanism other than thermal emission from dust
that could plausibly produce $1.4$~mm-to-$2.0$~mm spectral indices well above $0$.
As the name suggests, GPS sources typically peak around wavelengths of 
$30$~cm \citep[e.g.,][]{stanghellini98}, but GPS sources have been observed
with peaks at wavelengths as short as $1$~cm \citep{edge98}, and there is no 
fundamental physics that rules out self-absorbed synchrotron emission peaking at 
much shorter wavelengths.  However, several lines of reasoning suggest 
that the $\alpha > \alphathresh$ sources without IRAS-FSC counterparts are not dominated by 
self-absorbed synchrotron.

The first argument against GPS radio sources as the explanation for 
our ``dust-dominated" counts without IRAS-FSC counterparts is that their spectral behavior is too steep 
even for self-absorbed synchrotron, while it is perfectly consistent with 
thermal dust emission.  \citet{stanghellini98} show that, even well longward of 
the peak wavelength, the mean spectral index of GPS sources is $\alpha \sim 0.8$,
and rarely do they find spectral indices as high as $2.0$.  In contrast, the sum of 
the posterior spectral index PDFs for the brightest 
$5$ dust-dominated sources without IRAS counterparts peaks at 
$\alpha = 3.3$ and is inconsistent with $\alpha=2.0$ at $97\%$ confidence.  
This peak value of $\alpha=3.3$ is consistent (within the width of the two 
distributions)
with the peak of $\alpha=3.2$ for the top five sources with IRAS-FSC 
counterparts, sources which we are confident are truly dust-dominated.
The peak value of these $\alpha$ distributions are also
consistent with the best-fit mean spectral index 
of  $\bar{\alpha}=3.7 \pm 0.2$ found by \citet{hall10} for Poisson-distributed 
sources below the SPT detection threshold --- the vast majority of which are 
expected to be dust-dominated.  Interestingly, all of these peak $\alpha$ values 
are actually steeper than typical
values predicted by models for spectral indices of dust-dominated sources 
\citep[e.g.,][]{lagache04}, a point discussed in detail in \citet{hall10}.

Another argument against the 
GPS explanation for these sources is the lack of radio and x-ray counterparts.
\citet{siemiginowska08} found that GPS sources have $2$-$10$~keV fluxes of 
up to $10^{46} \ \mathrm{erg} \ \mathrm{s}^{-1}$, easily detectable in the ROSAT 
All-Sky Survey, and our brightest ``dust-dominated" sources would have to 
be almost an order of magnitude dimmer at $36$~cm than at $1.4$~mm to 
evade detection in SUMSS.  Finally, \citet{kellerman81} argue that the peak
wavelength of a GPS source should be proportional to flux density to the 
$-0.4$ power, meaning that sources that peak at mm wavelengths should be 
$2.5$ orders of magnitude brighter than sources that peak at cm wavelengths, 
so they should be much rarer as well.  Based on this set of arguments, we
conclude that our sources with $1.4$~mm-to-$2.0$~mm spectral 
indices greater than $\alphathresh$ are indeed dominated by thermal dust emission.


\begin{figure*}[h]
\begin{center}
\epsfig{file=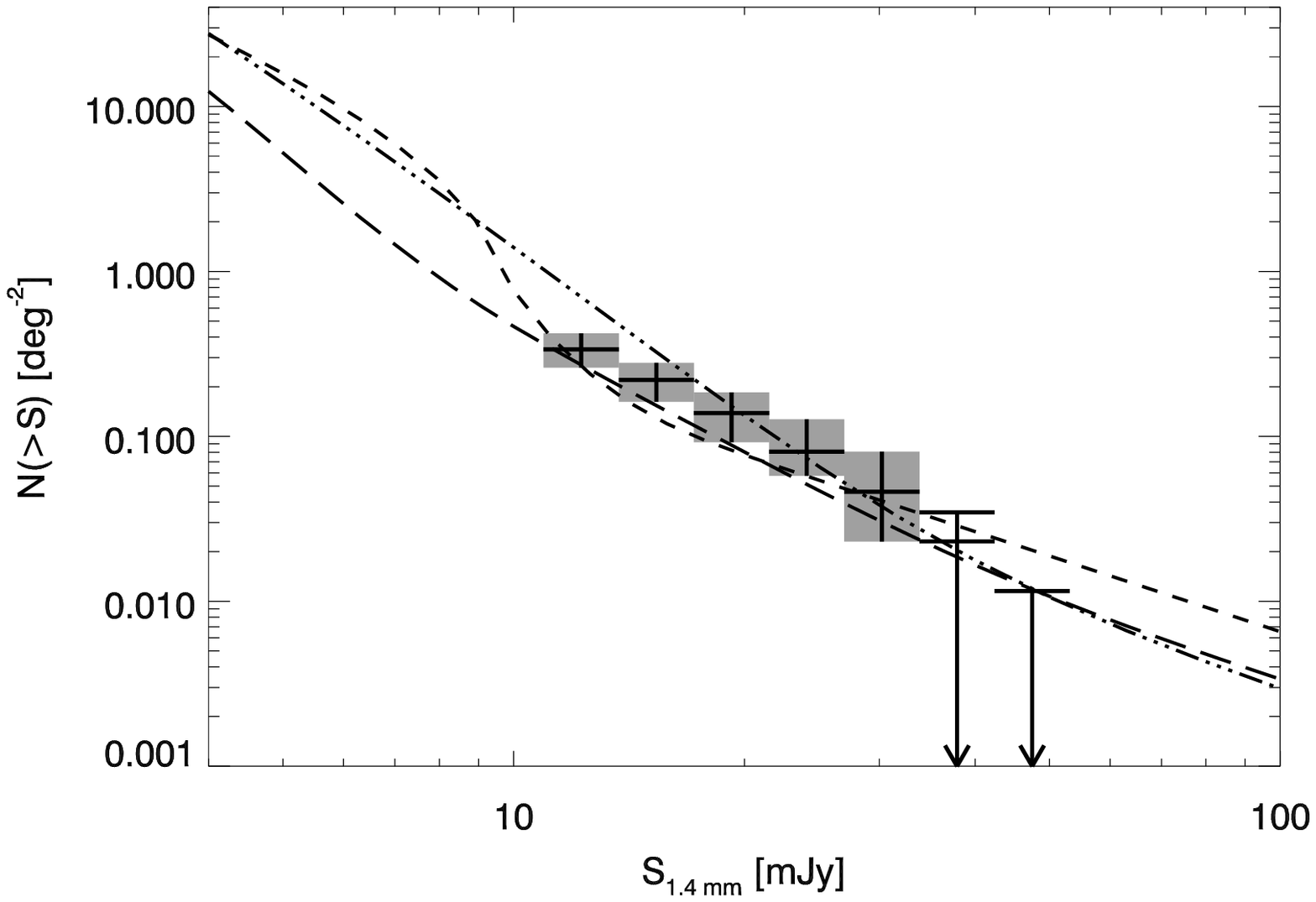, width=12cm} 
\epsfig{file=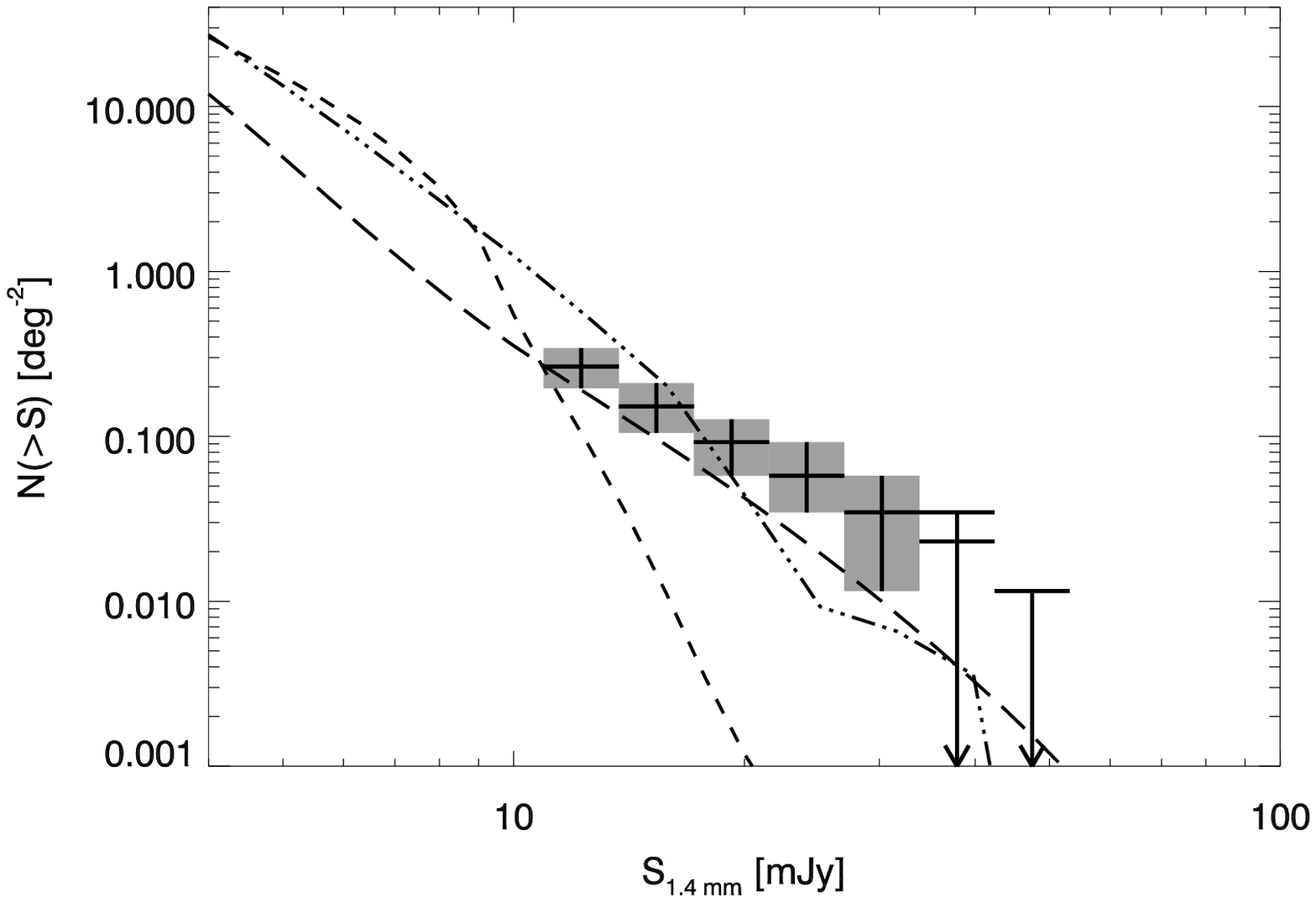, width=12cm} 
\end{center}
\caption{\small Cumulative SPT dust-dominated source counts vs.~$1.4$~mm flux, 
with models overplotted. 
Model curves are as follows:  
\emph{short dashed line}: \citet{lagache04} $1.38$~mm prediction; 
\emph{long dashed line}: \citet{negrello07} $1.4$~mm prediction; 
\emph{dot-dashed line}: \citet{pearson09} $1.38$~mm prediction.  
The error regions enclose $68\%$ of the probability centered about the median counts, and are calculated using the bootstrap over the two-band posterior intrinsic flux (at $1.4$~mm) that is described in Sec.~\ref{sec:counts}. 
In these counts, a source is identified as dust-dominated if $\alpha \ge \alphathresh$ in the resampling.
\emph{Top Panel}: This plot shows counts and models with all dust-dominated sources included. 
\emph{Bottom Panel}: This plot shows counts calculated excluding sources that have IRAS-FSC 
counterparts (within 1 arcmin) and models calculated excluding populations that 
should be detectable in the IRAS $60$~\um \ band above the typical FSC limit 
of $0.2$~Jy. The SPT detects sources in excess of what is predicted by the models shown once the IRAS sources have been removed. 
\label{fig:counts_dusty}}
\end{figure*}

\subsubsection{SPT Dust-Dominated Source Counts and Arguments 
for a New Population of mm Sources}
\label{sec:dust_newpop}
One might presume that the dust-dominated sources presented here
are exact analogues to the submm-selected SMG population measured
by SCUBA and other instruments.  However, 
SMG source counts at 850~\um \ are found to drop precipitously at 
flux levels above 
5-10 mJy \citep[e.g.,][]{coppin06}.  Assuming an average spectral 
index of $\alpha=2.8$ (the estimate for mean SMG spectral index from
\citet{knox04}), this drop in the counts would occur at $\sim 2$~mJy
at 1.4~mm, leading one to expect a 1.4~mm survey to see very few SMGs
above the $\sim 10$~mJy level needed for a robust detection in the SPT 
1.4~mm band.  Indeed, extrapolating two of the models used to fit SCUBA 
counts in \citet{coppin06} to 1.4~mm and 10~mJy indicates that there should 
be $\ll 1$ SMG above $10$~mJy at $1.4$~mm 
in the SPT 87 \sqdeg\ field; a similar extrapolation of the 
Schechter function fit to the AzTEC $1.1$~mm counts in \citet{austermann10}
yields a similar prediction.  Using a spectral index closer to the mean 
of the dust-dominated part of the distribution shown in Fig.~\ref{fig:alpha_hist} would 
drive these predictions even lower.
Quite contrary to these predictions, SPT detects 20 dust-dominated 
sources above $10$~mJy at 1.4~mm, including 9 above $20$~mJy.  What are 
these sources if not SMGs?  

One possibility is that they are nearby 
ULIRGs, the low-redshift analogues of SMGs.  These sources 
are rare enough to not contribute significantly to the submm
counts estimated from very small patches of sky.
As noted in Sec.~\ref{sec:assoc}, a fraction of the bright 
dust-dominated sources detected by the SPT 
have counterparts in the IRAS-FSC, and most of these sources are
indeed low-redshift ULIRGs.
The majority of SPT dust-dominated sources, however, 
do not have IRAS-FSC counterparts, or counterparts in any existing catalog.  
Likewise, in deep ($\sim 24.4$ AB mag) $griz$ optical data taken by the Blanco Cosmology Survey (BCS)\footnote{http://cosmology.uiuc.edu/BCS/} 
(which encompass roughly half of the $87$~\sqdeg\ described here) 
there are no obvious counterparts.

It is unlikely that these sources are just 
below the threshold for inclusion in IRAS-FSC, given that 
the brightest three SPT dust-dominated sources 
do not have IRAS-FSC counterparts. Fig.~\ref{fig:iras100um} makes 
an even stronger argument that these sources are a very different population 
than the SPT sources with IRAS-FSC counterparts.  This figure shows a 
scatter plot of IRAS 100~\um \ flux --- estimated from an IRAS Sky Survey 
Atlas \citep[ISSA,][]{wheelock94} 100~\um \ map filtered to enhance point-source 
S/N --- vs.~SPT 1.4~mm flux.\footnote{We chose to plot the IRAS $100$~\um~channel (as opposed to $60$~\um) because it is the closest band to the SED peak and to the SPT 1.4~mm band.}
The sources with and without IRAS-FSC counterparts occupy 
clearly distinct loci of points.  The sources with IRAS-FSC counterparts are consistent with nearby
sources that have typical dust temperatures of $25-35~\mathrm{K}$ \citep{dunne01}, while the 
sources without IRAS-FSC counterparts have ISSA 100~\um \ flux consistent
with zero and thus must either be at moderate to high redshift or have 
anomalously cold dust ($\lesssim 15~\mathrm{K}$).  A stacking analysis of the 
13 objects classified as dust-dominated with $1.4$~mm S/N greater than 4.5 and
no IRAS-FSC counterpart shows that these sources have mean IRAS $100$~\um \ flux
of $15.9$~mJy, with a $2 \sigma$ upper limit for that mean of $106$~mJy.  These same
sources have mean $1.4$~mm flux of $18$~mJy.  If we assume an Arp220 SED for these
sources, this stacking analysis implies these sources lie at a mean redshift of 
$\langle z \rangle = 3.1$, with a $2 \sigma$ lower limit of $\langle z \rangle = 2.0$.  Assuming a modified 
blackbody spectrum with $\beta=1.5$, we obtain a $2 \sigma$ upper 
limit of the combination of dust temperature and redshift of 
$T_d/(1+z) = 14.3 \mathrm{K}$, implying a $2 \sigma$ lower limit on mean redshift
of $\langle z \rangle = 1.1$ if $T_d = 30 \mathrm{K}$.

The mean dust temperature of high redshift SMGs measured from previous work is $\sim35$~K \citep{chapman05, kovacs06}, 
but there have been detections of a few galaxies with dust temperatures $\sim20$K (e.g. \citet{kovacs06}). 
There is expected to be both a hot ($\sim40$K) dust component from the ISM surrounding actively star 
forming regions with many young stars, as well as a cold ($\sim20$K) dust component surrounding 
the diffuse quiescently evolving population of old red stars \citep{dunne01,vlahakis05,coppin08}. 
IRAS would be largely insensitive to the cold dust component, which might explain why these SPT sources escaped previous detection. 
However, given the brightness of these objects, if these sources were galaxies at $z<1$ 
then we would expect to see them in BCS, 
Digitized Sky Survey,\footnote{http://cadcwww.dao.nrc.ca/cadcbin/getdss} or Two Micron All Sky Survey images,\footnote{http://irsa.ipac.caltech.edu/applications/2MASS/} and we do not.
This test depends on the assumption of a particular SED for the galaxies, 
but the most pessimistic galaxy SED assumption --- namely an Arp220-like SED with a high
mm-to-optical brightness ratio --- still predicts that such objects would be detected in 
the optical and near-IR (NIR).  

Throughout this work we have assumed that the majority of sources detected in the 
mm waveband by the SPT have been of extragalactic origin, as this field is at high 
galactic latitude and has explicitly been chosen for its low contamination by galactic cirrus. 
How robust of a statement can be made 
as to the extragalactic nature on the dusty sources not detected in any external catalog? 
A similar question was posed in \citet{lawrence01} and it was demonstrated that the majority of SCUBA sources could not be 
of galactic origin.
The SPT sources cannot be galactic HII regions as the accompanying free-free emission would have been detected by low 
frequency radio catalogs such as PMN and SUMSS. They could be nearby, cold ($T_d<10K$), 
pre-star-forming clouds in the galaxy similar to sources described in \citet{desert08}. 
However, as this field is at high galactic latitude and the clouds would presumably be located in the galactic disk and not the halo, the sources would therefore be nearby (within $\sim 1$~kpc). 
Since these clouds are expected to be parsec-scale objects \citep{egan98}, they would have an angular extent ($\sim 5 \arcmin$) much larger than the 
SPT beam and would be obviously detected by eye in the unfiltered map. 
We have also tested the dark cloud hypothesis by measuring optical extinction along the lines of sight 
towards the SPT dusty sources. 
A comparison of the optical color distribution of BCS objects toward 
dust-dominated sources to the colors toward random directions in the field shows 
no excess extinction at the positions of these sources.
Although we cannot definitively confirm that \textit{none} of these sources are of galactic origin, 
these tests demonstrate that they are predominantly extragalactic. 


Comparisons of SPT dust-dominated source counts with model predictions 
strengthen the conclusion that our bright, dust-dominated sources without IRAS
counterparts are a new, possibly high-redshift and lensed, population.
Fig.~\ref{fig:counts_dusty} shows the cumulative SPT dust-dominated 
counts vs.~flux and predictions from three models.  
\citet{lagache04}, \citet{negrello07}, and \citet{pearson09} make predictions for counts at or very near our $1.4$~mm band.
All three models have two basic components: moderate-to-high-redshift starburst 
galaxies (which account for basically all the counts seen by SCUBA at 
$850$~\um) and nearby galaxies (including the LIRGs and ULIRGs
seen in IRAS). 
The \citet{negrello07} model also includes a component of the high-redshift starburst population 
which has been strongly lensed by foreground galaxies.
At first glance, all models agree fairly well with the SPT counts at $1.4$~mm.
However, the counts at fluxes above $10$-$20$~mJy are dominated in all 
models by sources that should be detectable in the IRAS-FSC above the $60$~\um \
flux cut of $200$~mJy, while our measured counts are dominated by sources 
without IRAS-FSC counterparts.  We have modified
the publicly available \citet{lagache04} code to exclude such sources from their 
model,\footnote{Using the template SED models supplied by \citet{lagache04}, 
we find that any source detected at 
$>10$~mJy at $1.4$~mm below redshift $z=0.2$ should have $60$~\um \ flux 
above $200$~mJy, so our modification to their model is effectively just a redshift cut 
at $z=0.2$.} and \citet{pearson09} and \citet{negrello07} have supplied us with model counts 
excluding sources with $60$~\um \ flux greater than $200$~mJy.
We then re-calculate our $1.4$~mm dust-dominated counts excluding sources
with IRAS-FSC counterparts and compare these modified counts 
to the modified predictions in the bottom panel of Fig.~\ref{fig:counts_dusty}.

There are significant discrepancies between our measured 
counts without IRAS counterparts and the \citet{lagache04} and \citet{pearson09} model predictions in both the slope of the counts and the number of sources above 20 mJy. 
The \citet{negrello07} model also under-predicts the SPT counts but the shape of the counts are in good agreement. 
A modest adjustment along either axis would bring the model into excellent agreement. As stated above, the main difference between the  \citet{negrello07} model and the  \citet{lagache04} and \citet{pearson09} models is the inclusion of a population of high redshift, strongly lensed dusty star forming galaxies.

Among the possible explanations for this new population, the hypothesis
that these sources are at high redshift is particularly compelling.
The longer-wavelength SPT 
observations will be sensitive to a higher-redshift population than the 
submm surveys (due to the stronger negative K-correction), and there 
is considerable evidence that the very brightest SMGs are at the
highest redshifts \citep{ivison02,pope05,greve08}. This
empirically observed trend of SMG brightness with redshift is plausible 
both because more distant systems 
have a higher probability of being gravitationally lensed \citep{blain96b,blain99b} and because 
evolution in star formation as a function of environment, called ``cosmic downsizing", 
is consistent with a higher star-formation rate in massive systems at high redshift \citep{cowie96, cole00}.  Models exist in the literature that predict the existence of a mm/submm-bright population
similar to ours resulting from both lensing \citep{blain96b, perrotta03, negrello07} and intrinsically luminous
high-$z$ sources \citep{devriendt10}.
The reason that such objects would have been missed by previous 
mm/submm instruments 
is simply that SPT surveys so much more area (87 square degrees 
for this small subset of the SPT survey to $\sim 1$ square degree for the 
total area surveyed by SCUBA) and is hence much more likely to find 
rare, bright systems (due to strong lensing or intrinsic luminosity) that a smaller 
survey might miss.

If this subset of the SPT-identified dusty sources are indeed at high redshift, 
they represent an intriguing new class of mm sources, whether they are strongly 
lensed or intrinsically ultra-luminous.  
Strongly lensed systems allow observers to detect fainter background sources at 
higher redshift than would otherwise be obtainable. 
Because lensing is achromatic, these sources will be brighter at all wavelengths, 
facilitating detailed studies which have otherwise been difficult to achieve.  
Equally exciting would be the identification of high-redshift galaxies which are 
more massive and forming stars more prodigiously than any systems yet identified.
The identification of such galaxies would be a strong test of models of galaxy 
formation and evolution.
Regardless of whether the sources are lensed or intrinsically luminous, a sample 
of high-redshift, dust-enshrouded star-forming galaxies has the potential to be a 
useful tool for the study of very early epochs of star and galaxy formation.

\section{Conclusions}\label{sec:conclusions}

The SPT has detected $188$ sources above $4.5\sigma$ 
(over $3000$ above $3\sigma$) in two-band data over a small ($87~\mathrm{deg}^2$) 
subset of the full survey region.  
This is the first survey of its kind at mm wavelengths and mJy flux levels, both 
in survey area --- the 87~\sqdeg \  presented here is 
over an order of magnitude more area than previous mm surveys at these flux 
levels --- and in the ability to spectrally discriminate between different source 
populations.  
Some of the sources detected in this work appear to be members of a new and intriguing population 
of dust-enshrouded, star-forming galaxies.

We use the ratio of flux in the $1.4$~mm and $2.0$~mm bands to estimate the 
spectral index for every detected source.  Using the posterior PDF for this spectral index, 
we classify each source as either synchrotron-dominated or dust-dominated.
At high flux levels (above $\sim 15$~mJy) in both bands, the majority of sources 
we detect are synchrotron-dominated; the synchrotron-dominated population 
continues to dominate the $2.0$~mm counts down to the detection threshold, but
the dust-dominated sources begin to take over the $1.4$~mm counts near the 
$15$~mJy level.

The synchrotron-dominated sources we detect are consistent with the population 
of radio sources that is well-established from radio and cm surveys.
All of our $>5\sigma$ sources in this class have a clear counterpart in existing radio catalogs.
The number counts as a function of $2.0$~mm flux that we derive for this population 
are consistent with the predictions of \citet{dezotti05}.  
Spectral comparisons
within the SPT data and with ATCA data at $6$~cm (taken in follow-up observations)
and $1.5$~cm \citep{murphy10} show that the synchrotron-dominated sources we 
detect are consistent with FSRQs (as predicted), but that the flat-spectrum behavior
has turned over by mm wavelengths.  This conclusion is important for AGN models
and for predictions of radio source contamination in CMB and SZ measurements.

A fraction of our dust-dominated sources have counterparts in the IRAS-FSC
and are typically associated with low-redshift ($z\ll 1$) ULIRGs.
The majority of our dust-dominated sources, however, have no counterpart in existing catalogs,
and we argue that they represent a new and exciting population of mm sources.  Comparisons
of source counts for this population with model predictions and IRAS limits on
the $100$~\um~flux of these sources demonstrate that 
these sources are inconsistent with either simple extrapolations of the submm-selected
population at lower fluxes or with low-redshift galaxies with normal dust properties.
Further, the mm-wavelength selection of such high-flux sources over a 
large survey area suggests a high-redshift population.
Possibilities for explaining this new family 
of sources include strong lensing of dimmer background sources, ultra-luminous
starburst galaxies at moderate-to-high redshift, and ($z\sim1$) galaxies with extremely 
cold dust.  Comparisons to models favors the hypothesis for these sources being strongly lensed.
Any one of these explanations would have interesting implications 
for models of star and galaxy formation and be of potential cosmological interest.
Properly locating these sources in the broader context of the IR/submm/mm galaxy population is
a major challenge for future work. 
An extensive program to determine the SEDs, morphologies, and redshift distribution of these objects is underway.
Multi-wavelength follow-up imaging facilities such as the Hubble Space Telescope, the Spitzer Space Telescope, the Herschel Space Observatory, and the Atacama Large Millimeter Array, will be crucial to disentangling the various possibilities and uncovering the nature of these objects.

The point source results presented here use only a small fraction of the complete SPT data set. 
SPT is continuing to take data and has already observed 800~\sqdeg \ to similar depths as this work at $1.4$ and $2.0$~mm. 
600~\sqdeg \ of this area also has $3.2$~mm coverage. 
The complete SPT survey is expected to cover over 2000~\sqdeg \ with these three wavelengths and will produce a catalog containing thousands of additional sources.

\acknowledgments

We would like to thank A.\ Blain and M.\ Malkan
for advice and guidance, C.\ Lacey, G.\ Lagache, M.\ Negrello, 
C.\ Pearson, and G.\ De Zotti for providing us with their models, and the anonymous referee for useful suggestions.
 
The SPT team gratefully acknowledges the
contributions to the design and construction of the telescope by S.\ Busetti, E.\ Chauvin,
T.\ Hughes, P.\ Huntley, and E.\ Nichols and his team of iron workers. We also thank the
National Science Foundation (NSF) Office of Polar Programs, the United States Antarctic Program and the Raytheon Polar Services Company for their support of the project.  We are grateful for professional\
 support from the staff of the South Pole station. We thank C.\ Greer, T.\ Lanting, J.\ Leong, W.\ Lu, M.\ Runyan, D.\ Schwan, and M.\ Sharp  for their early contributions to the SPT project and J.\ Joseph and C.\ Vu \
for their contributions to the electronics. We acknowledge S.\ Alam, W.\ Barkhouse, S.\ Bhattacharya, L.\ Buckley-Geer, S.\ Hansen, H.\ Lin, Y-T Lin, A.\ Rest, C.\ Smith and D.\ Tucker for their contribution to BCS data acquisition.

The South Pole Telescope is supported by the 
NSF through grants ANT-0638937 and ANT-0130612.  Partial support is also provided by the  
NSF Physics Frontier Center grant PHY-0114422 to the Kavli Institute of Cosmological Physics 
(KICP) at the University of Chicago, the Kavli Foundation and the Gordon and Betty Moore Foundation. 
The McGill group acknowledges funding from the National Sciences and
Engineering Research Council of Canada, the Quebec Fonds de recherche
sur la nature et les technologies, and the Canadian Institute for
Advanced Research. The following individuals acknowledge additional support:
B.\ Stalder from the Brinson Foundation, B.\ A.\ Benson, K.\ K.\ Schaffer and E.\ R.\ Switzer from a KICP Fellowship,  J.\ McMahon from a Fermi Fellowship,  Z.\ Staniszewski from a GAAN Fellowship, and A.\ T.\ Lee from the Miller Institute for Basic Research in Science,
University of California Berkeley.  N.\ W.\ Halverson acknowledges support from an Alfred P.\ Sloan Research Fellowship.  A.\ A.\ Stark and W.\ Walsh received grant support from the Smithsonian Institution Office of the Under Secretary for Science. 
Support for M.\ Brodwin was provided by the W.\ M.\ Keck Foundation.  

This research has made use of the SIMBAD database,
operated at CDS, Strasbourg, France, the NASA/IPAC Extragalactic Database (NED) which 
is operated by the Jet Propulsion Laboratory, California Institute of Technology, 
under contract with the National Aeronautics and Space Administration, 
and the NASA/ IPAC Infrared Science Archive, which is operated by the 
Jet Propulsion Laboratory, California Institute of Technology, 
under contract with the National Aeronautics and Space Administration. 
This research used the facilities of the Canadian Astronomy Data Centre 
operated by the National Research Council of Canada with the support of the Canadian Space Agency. 
The ATCA is part of the Australia Telescope, which is funded by the Commonwealth
of Australia for operation as a national facility managed by the CSIRO.
Some of the results in this paper have been derived using the HEALPix \citep{gorski05} package. We acknowledge the use of the Legacy Archive for Microwave Background Data Analysis (LAMBDA). Support for LAMBDA is provided by the NASA Office of Space Science.

\bibliography{spt_smg}

\clearpage
\LongTables 
\pagestyle{empty}
\begin{landscape}
\begin{center}
\def\arraystretch{1.500}
\tabletypesize{\scriptsize}
\begin{deluxetable}{lccccccccccccccccccc}
\tablecaption{Point sources above $4.5 \sigma$ at $1.4$~mm or $2.0$~mm in an $87$ square degree SPT field centered at R.A.~$5^\mathrm{h} 30^\mathrm{m}$, decl.~$-55^\circ$ (J2000)
\label{tab:cat_both}}
\tablehead{
\multicolumn{4}{l}{{\bf ID \& coordinates:}} & 
\multicolumn{4}{l}{\bf $2.0$~mm data:} &
\multicolumn{4}{l}{\bf $1.4$~mm data:} &
\multicolumn{5}{l}{\bf Spectral index \& type:} &
\multicolumn{3}{l}{\bf Nearest source in:} \\
\colhead{SPT ID} & 
\colhead{RA} & 
\colhead{DEC}  & & 
\colhead{S/N} & 
\colhead{$S^\mathrm{raw}$} & 
\colhead{$S^\mathrm{dist}$} & & 
\colhead{S/N} & 
\colhead{$S^\mathrm{raw}$} & 
\colhead{$S^\mathrm{dist}$} & & 
\colhead{$\alpha^\mathrm{raw}$} & 
\colhead{$\alpha^\mathrm{dist}$} & 
\colhead{$P(\alpha > 1.66)$} &  
\colhead{Type} & & 
\colhead{SUMSS} & 
\colhead{RASS} & 
\colhead{IRAS} \\
\colhead{} & \colhead{[deg]} & \colhead{[deg]} & & 
\colhead{} & 
\colhead{[mJy]} & \colhead{[mJy]} & & 
\colhead{} & 
\colhead{[mJy]} & \colhead{[mJy]} & & 
\colhead{} & \colhead{} & \colhead{} & \colhead{} & & 
\colhead{[arcsec]} & \colhead{[arcsec]} & \colhead{[arcsec]} \\ 
}

\startdata 

SPT-S J045913$-$5942.4 & $ 74.804$ & $ -59.708$ & & $ 6.20$ & $ 7.51$ & $7.26^{+1.27}_{-1.27}$ & & $ 7.05$ & $ 22.58$ & $20.85^{+3.90}_{-4.02}$ & & $ 3.0$ & $2.9^{+0.7}_{-0.7}$ & $ 0.95$ & dust & & $ 43$ & $ 575$ & $ 815$ \\
SPT-S J050000$-$5752.5 & $ 75.000$ & $ -57.875$ & & $ 44.01$ & $ 53.28$ & $52.99^{+2.50}_{-2.50}$ & & $ 15.15$ & $ 48.52$ & $47.80^{+5.39}_{-5.39}$ & & $ -0.3$ & $-0.3^{+0.3}_{-0.3}$ & $ 0.00$ & sync & & $ 3$ & $ 313$ & $ 486$ \\
SPT-S J050003$-$5229.0 & $ 75.015$ & $ -52.485$ & & $ 5.33$ & $ 6.70$ & $5.91^{+1.38}_{-1.47}$ & & $ 2.15$ & $ 7.20$ & $5.57^{+2.88}_{-2.47}$ & & $ 0.2$ & $0.0^{+1.4}_{-1.7}$ & $ 0.08$ & sync & & $ 18$ & $ 923$ & $1979$ \\
SPT-S J050019$-$5321.3 & $ 75.081$ & $ -53.356$ & & $ 39.42$ & $ 49.52$ & $49.26^{+2.32}_{-2.41}$ & & $ 10.82$ & $ 36.21$ & $35.52^{+4.64}_{-4.64}$ & & $ -0.9$ & $-0.9^{+0.3}_{-0.4}$ & $ 0.00$ & sync & & $ 4$ & $ 627$ & $ 189$ \\
SPT-S J050211$-$5040.7 & $ 75.546$ & $ -50.679$ & & $ 6.76$ & $ 8.86$ & $8.15^{+1.40}_{-1.41}$ & & $ 1.15$ & $ 3.98$ & $4.71^{+2.51}_{-1.48}$ & & $ -2.2$ & $-1.5^{+1.2}_{-1.0}$ & $ 0.00$ & sync & & $ 7$ & $1184$ & $1459$ \\
SPT-S J050321$-$5328.3 & $ 75.839$ & $ -53.472$ & & $ 5.56$ & $ 6.98$ & $6.16^{+1.36}_{-1.45}$ & & $ 1.43$ & $ 4.77$ & $4.35^{+2.64}_{-1.70}$ & & $ -1.0$ & $-0.9^{+1.5}_{-1.4}$ & $ 0.03$ & sync & & $ 8$ & $ 694$ & $1312$ \\
SPT-S J050329$-$5735.6\tablenotemark{a} & $ 75.874$ & $ -57.595$ & & $ 45.35$ & $ 54.90$ & $54.60^{+2.57}_{-2.57}$ & & $ 16.43$ & $ 52.62$ & $51.83^{+5.76}_{-5.66}$ & & $ -0.1$ & $-0.1^{+0.3}_{-0.3}$ & $ 0.00$ & sync & & $ 2$ & $1058$ & $ 583$ \\
SPT-S J050334$-$5244.8\tablenotemark{b} & $ 75.892$ & $ -52.747$ & & $ 5.96$ & $ 7.49$ & $6.80^{+1.35}_{-1.38}$ & & $ 1.95$ & $ 6.51$ & $5.43^{+2.86}_{-2.19}$ & & $ -0.4$ & $-0.5^{+1.3}_{-1.5}$ & $ 0.02$ & sync & & $ 32$ & $ 581$ & $1611$ \\
SPT-S J050401$-$5023.2 & $ 76.006$ & $ -50.387$ & & $ 40.49$ & $ 53.07$ & $52.79^{+2.49}_{-2.49}$ & & $ 12.50$ & $ 43.27$ & $42.54^{+5.17}_{-5.27}$ & & $ -0.6$ & $-0.6^{+0.3}_{-0.3}$ & $ 0.00$ & sync & & $ 1$ & $ 998$ & $1301$ \\
SPT-S J050424$-$5711.9 & $ 76.100$ & $ -57.199$ & & $ 16.05$ & $ 20.17$ & $19.94^{+1.47}_{-1.52}$ & & $ 7.56$ & $ 25.28$ & $24.53^{+4.08}_{-4.08}$ & & $ 0.6$ & $0.6^{+0.4}_{-0.5}$ & $ 0.00$ & sync & & $ 11$ & $ 158$ & $ 493$ \\
SPT-S J050437$-$5818.1 & $ 76.156$ & $ -58.302$ & & $ 4.83$ & $ 5.85$ & $4.23^{+1.67}_{-3.55}$ & & $ 0.81$ & $ 2.59$ & $2.69^{+2.20}_{-1.66}$ & & $ -2.2$ & $-0.8^{+2.1}_{-1.5}$ & $ 0.13$ & sync & & $ 36$ & $ 895$ & $ 211$ \\
SPT-S J050508$-$5346.4 & $ 76.284$ & $ -53.775$ & & $ 4.76$ & $ 5.98$ & $4.56^{+1.63}_{-3.38}$ & & $ 1.49$ & $ 4.98$ & $3.54^{+2.70}_{-1.95}$ & & $ -0.5$ & $-0.2^{+2.0}_{-1.8}$ & $ 0.15$ & sync & & $ 120$ & $ 744$ & $ 605$ \\
SPT-S J050511$-$5346.0 & $ 76.297$ & $ -53.767$ & & $ 2.06$ & $ 2.59$ & $2.72^{+1.08}_{-1.42}$ & & $ 5.39$ & $ 18.05$ & $10.64^{+5.75}_{-9.20}$ & & $ 5.3$ & $3.6^{+1.0}_{-3.5}$ & $ 0.74$ & dust & & $ 82$ & $ 766$ & $ 568$ \\
SPT-S J050523$-$5808.5 & $ 76.346$ & $ -58.143$ & & $ 4.56$ & $ 5.52$ & $1.44^{+3.44}_{-1.19}$ & & $ 0.22$ & $ 0.72$ & $1.57^{+1.86}_{-1.26}$ & & $ -5.6$ & $-0.6^{+2.9}_{-1.7}$ & $ 0.21$ & sync & & $ 302$ & $ 795$ & $ 408$ \\
SPT-S J050526$-$5044.8 & $ 76.358$ & $ -50.747$ & & $ 12.82$ & $ 16.81$ & $16.50^{+1.48}_{-1.48}$ & & $ 4.81$ & $ 16.65$ & $15.65^{+3.78}_{-3.84}$ & & $ -0.0$ & $-0.1^{+0.6}_{-0.8}$ & $ 0.00$ & sync & & $ 3$ & $ 717$ & $ 924$ \\
SPT-S J050528$-$5056.6 & $ 76.369$ & $ -50.944$ & & $ 1.71$ & $ 2.24$ & $1.63^{+1.45}_{-1.17}$ & & $ 4.87$ & $ 16.84$ & $1.94^{+9.28}_{-1.29}$ & & $ 5.5$ & $1.9^{+2.3}_{-3.2}$ & $ 0.52$ & dust & & $ 253$ & $ 464$ & $ 216$ \\
SPT-S J050607$-$5844.8 & $ 76.533$ & $ -58.748$ & & $ 7.25$ & $ 8.77$ & $8.28^{+1.28}_{-1.30}$ & & $ 2.54$ & $ 8.12$ & $6.93^{+3.03}_{-2.65}$ & & $ -0.2$ & $-0.4^{+1.1}_{-1.4}$ & $ 0.01$ & sync & & $ 6$ & $ 373$ & $ 786$ \\
SPT-S J050617$-$5748.7 & $ 76.573$ & $ -57.813$ & & $ 21.73$ & $ 26.30$ & $26.06^{+1.63}_{-1.63}$ & & $ 6.01$ & $ 19.26$ & $18.47^{+3.70}_{-3.65}$ & & $ -0.8$ & $-0.9^{+0.5}_{-0.6}$ & $ 0.00$ & sync & & $ 3$ & $ 212$ & $ 963$ \\
SPT-S J050620$-$5741.0 & $ 76.584$ & $ -57.685$ & & $ 5.61$ & $ 6.79$ & $6.16^{+1.31}_{-1.35}$ & & $ 3.25$ & $ 10.42$ & $9.00^{+3.26}_{-3.42}$ & & $ 1.2$ & $1.1^{+1.1}_{-1.4}$ & $ 0.28$ & sync & & $ 3$ & $ 456$ & $ 544$ \\
SPT-S J050624$-$5024.8 & $ 76.600$ & $ -50.415$ & & $ 5.37$ & $ 7.04$ & $6.16^{+1.44}_{-1.56}$ & & $ 1.64$ & $ 5.68$ & $4.73^{+2.81}_{-1.98}$ & & $ -0.6$ & $-0.6^{+1.5}_{-1.5}$ & $ 0.04$ & sync & & $ 23$ & $ 343$ & $ 954$ \\
SPT-S J050656$-$5943.2\tablenotemark{c} & $ 76.735$ & $ -59.721$ & & $ 5.18$ & $ 6.27$ & $5.94^{+1.25}_{-1.29}$ & & $ 5.22$ & $ 16.73$ & $13.63^{+4.17}_{-6.43}$ & & $ 2.7$ & $2.2^{+1.0}_{-1.7}$ & $ 0.67$ & dust & & $ 385$ & $1497$ & $ 12$ \\
SPT-S J050658$-$5435.0 & $ 76.742$ & $ -54.585$ & & $ 5.15$ & $ 6.46$ & $5.72^{+1.37}_{-1.42}$ & & $ 3.64$ & $ 12.18$ & $10.79^{+3.50}_{-3.70}$ & & $ 1.7$ & $1.8^{+1.1}_{-1.3}$ & $ 0.52$ & dust & & $ 8$ & $ 14$ & $ 654$ \\
SPT-S J050716$-$5954.8 & $ 76.820$ & $ -59.914$ & & $ 4.91$ & $ 5.94$ & $4.62^{+1.53}_{-3.02}$ & & $ 1.10$ & $ 3.53$ & $3.14^{+2.40}_{-1.57}$ & & $ -1.4$ & $-0.7^{+1.9}_{-1.5}$ & $ 0.11$ & sync & & $ 11$ & $1818$ & $ 267$ \\
SPT-S J050725$-$5013.3 & $ 76.855$ & $ -50.223$ & & $ 7.16$ & $ 9.38$ & $8.70^{+1.39}_{-1.41}$ & & $ 1.09$ & $ 3.76$ & $4.79^{+2.43}_{-1.41}$ & & $ -2.5$ & $-1.6^{+1.2}_{-0.9}$ & $ 0.00$ & sync & & $ 2$ & $1050$ & $ 183$ \\
SPT-S J050732$-$5104.2 & $ 76.884$ & $ -51.071$ & & $ 19.68$ & $ 25.79$ & $25.56^{+1.64}_{-1.69}$ & & $ 7.39$ & $ 25.59$ & $24.79^{+4.21}_{-4.16}$ & & $ -0.0$ & $-0.1^{+0.4}_{-0.5}$ & $ 0.00$ & sync & & $ 4$ & $ 13$ & $ 556$ \\
SPT-S J050747$-$5156.1 & $ 76.948$ & $ -51.936$ & & $ 11.75$ & $ 15.40$ & $15.06^{+1.47}_{-1.44}$ & & $ 3.46$ & $ 11.97$ & $10.96^{+3.64}_{-3.32}$ & & $ -0.7$ & $-0.9^{+0.8}_{-1.0}$ & $ 0.00$ & sync & & $ 2$ & $1407$ & $1139$ \\
SPT-S J050758$-$5850.5\tablenotemark{a} & $ 76.995$ & $ -58.842$ & & $ 47.34$ & $ 57.31$ & $57.01^{+2.58}_{-2.58}$ & & $ 14.92$ & $ 47.79$ & $47.01^{+5.37}_{-5.27}$ & & $ -0.5$ & $-0.5^{+0.3}_{-0.3}$ & $ 0.00$ & sync & & $ 3$ & $ 961$ & $1474$ \\
SPT-S J050831$-$5449.4 & $ 77.132$ & $ -54.823$ & & $ 0.84$ & $ 1.06$ & $0.72^{+1.08}_{-0.50}$ & & $ 4.66$ & $ 15.58$ & $1.10^{+2.02}_{-0.67}$ & & $ 7.3$ & $1.9^{+2.3}_{-2.8}$ & $ 0.52$ & dust & & $ 391$ & $ 540$ & $ 770$ \\
SPT-S J050847$-$5754.3 & $ 77.199$ & $ -57.907$ & & $ 0.71$ & $ 0.86$ & $0.64^{+0.96}_{-0.44}$ & & $ 4.56$ & $ 14.60$ & $1.03^{+1.75}_{-0.63}$ & & $ 7.7$ & $1.9^{+2.2}_{-2.8}$ & $ 0.54$ & dust & & $ 669$ & $ 55$ & $1086$ \\
SPT-S J050907$-$5339.2\tablenotemark{d} & $ 77.283$ & $ -53.653$ & & $ 3.27$ & $ 4.11$ & $3.48^{+1.30}_{-1.33}$ & & $ 4.61$ & $ 15.42$ & $4.76^{+9.05}_{-3.24}$ & & $ 3.6$ & $1.3^{+2.4}_{-3.2}$ & $ 0.46$ & sync & & $ 268$ & $ 205$ & $ 314$ \\
SPT-S J051003$-$5651.9 & $ 77.516$ & $ -56.865$ & & $ 10.93$ & $ 13.74$ & $13.41^{+1.39}_{-1.39}$ & & $ 3.97$ & $ 13.30$ & $12.19^{+3.55}_{-3.55}$ & & $ -0.1$ & $-0.2^{+0.8}_{-1.0}$ & $ 0.00$ & sync & & $ 6$ & $ 722$ & $ 688$ \\
SPT-S J051018$-$5719.7 & $ 77.578$ & $ -57.329$ & & $ 23.24$ & $ 29.19$ & $28.93^{+1.75}_{-1.70}$ & & $ 5.68$ & $ 19.01$ & $18.20^{+3.78}_{-3.72}$ & & $ -1.2$ & $-1.3^{+0.5}_{-0.6}$ & $ 0.00$ & sync & & $ 5$ & $ 30$ & $ 990$ \\
SPT-S J051029$-$5624.8 & $ 77.623$ & $ -56.414$ & & $ 6.37$ & $ 8.00$ & $7.30^{+1.33}_{-1.36}$ & & $ 1.28$ & $ 4.27$ & $4.52^{+2.55}_{-1.54}$ & & $ -1.7$ & $-1.3^{+1.3}_{-1.1}$ & $ 0.01$ & sync & & $ 6$ & $ 577$ & $1578$ \\
SPT-S J051030$-$5844.4 & $ 77.628$ & $ -58.740$ & & $ 12.43$ & $ 15.04$ & $14.77^{+1.35}_{-1.38}$ & & $ 4.29$ & $ 13.74$ & $12.79^{+3.47}_{-3.47}$ & & $ -0.2$ & $-0.4^{+0.7}_{-0.9}$ & $ 0.00$ & sync & & $ 15$ & $ 761$ & $ 891$ \\
SPT-S J051040$-$5558.3 & $ 77.670$ & $ -55.972$ & & $ 2.04$ & $ 2.56$ & $1.77^{+1.35}_{-1.21}$ & & $ 4.59$ & $ 15.36$ & $1.86^{+7.34}_{-1.15}$ & & $ 4.9$ & $1.2^{+2.7}_{-2.8}$ & $ 0.45$ & sync & & $ 145$ & $1014$ & $1357$ \\
SPT-S J051116$-$5341.9\tablenotemark{e} & $ 77.819$ & $ -53.700$ & & $ 5.23$ & $ 6.57$ & $5.91^{+1.33}_{-1.36}$ & & $ 5.16$ & $ 17.27$ & $16.27^{+3.69}_{-3.79}$ & & $ 2.6$ & $2.8^{+0.9}_{-0.9}$ & $ 0.89$ & dust & & $ 203$ & $1025$ & $ 287$ \\
SPT-S J051121$-$5920.0 & $ 77.840$ & $ -59.333$ & & $ 6.62$ & $ 8.01$ & $7.52^{+1.29}_{-1.31}$ & & $ 3.89$ & $ 12.47$ & $11.28^{+3.39}_{-3.51}$ & & $ 1.2$ & $1.1^{+0.9}_{-1.1}$ & $ 0.27$ & sync & & $ 5$ & $ 742$ & $ 733$ \\
SPT-S J051135$-$5809.5 & $ 77.899$ & $ -58.158$ & & $ 4.72$ & $ 5.71$ & $0.71^{+3.80}_{-0.51}$ & & $ 0.00$ & $ -4.35$ & $0.89^{+1.50}_{-0.70}$ & & $ NaN$ & $-1.0^{+3.0}_{-1.4}$ & $ 0.18$ & sync & & $ 21$ & $ 608$ & $ 881$ \\
SPT-S J051140$-$5846.7 & $ 77.919$ & $ -58.779$ & & $ 5.17$ & $ 6.26$ & $4.89^{+1.49}_{-2.65}$ & & $ 0.09$ & $ 0.28$ & $2.64^{+1.69}_{-1.16}$ & & $ -8.5$ & $-1.5^{+1.7}_{-1.1}$ & $ 0.06$ & sync & & $ 13$ & $ 439$ & $ 459$ \\
SPT-S J051217$-$5724.0\tablenotemark{a,f} & $ 78.073$ & $ -57.400$ & & $ 4.54$ & $ 5.50$ & $5.28^{+1.19}_{-1.12}$ & & $ 6.63$ & $ 21.24$ & $19.18^{+3.82}_{-3.94}$ & & $ 3.7$ & $3.5^{+0.8}_{-0.8}$ & $ 0.98$ & dust & & $ 118$ & $ 765$ & $ 4$ \\
SPT-S J051254$-$5413.3 & $ 78.226$ & $ -54.223$ & & $ 4.73$ & $ 5.94$ & $3.76^{+2.03}_{-3.36}$ & & $ 0.58$ & $ 1.94$ & $2.36^{+2.15}_{-1.79}$ & & $ -3.0$ & $-0.8^{+2.5}_{-1.5}$ & $ 0.16$ & sync & & $ 18$ & $ 259$ & $ 414$ \\
SPT-S J051259$-$5935.6 & $ 78.247$ & $ -59.594$ & & $ 4.69$ & $ 5.68$ & $5.54^{+1.20}_{-1.02}$ & & $ 7.72$ & $ 24.71$ & $22.73^{+3.89}_{-3.99}$ & & $ 4.0$ & $3.9^{+0.6}_{-0.7}$ & $ 1.00$ & dust & & $ 204$ & $ 540$ & $ 162$ \\
SPT-S J051355$-$5055.7 & $ 78.483$ & $ -50.928$ & & $ 23.09$ & $ 30.27$ & $29.99^{+1.82}_{-1.76}$ & & $ 8.25$ & $ 28.56$ & $27.78^{+4.37}_{-4.31}$ & & $ -0.2$ & $-0.2^{+0.4}_{-0.5}$ & $ 0.00$ & sync & & $ 2$ & $ 578$ & $1047$ \\
SPT-S J051358$-$5826.0 & $ 78.496$ & $ -58.435$ & & $ 5.18$ & $ 6.28$ & $4.71^{+1.55}_{-3.60}$ & & $ 0.00$ & $ -1.31$ & $2.39^{+1.44}_{-1.25}$ & & $ NaN$ & $-1.7^{+1.7}_{-1.0}$ & $ 0.06$ & sync & & $ 22$ & $ 262$ & $ 673$ \\
SPT-S J051400$-$5046.2 & $ 78.503$ & $ -50.771$ & & $ 5.16$ & $ 6.77$ & $5.46^{+1.56}_{-2.23}$ & & $ 0.52$ & $ 1.78$ & $3.15^{+2.13}_{-1.31}$ & & $ -3.6$ & $-1.3^{+1.7}_{-1.2}$ & $ 0.06$ & sync & & $ 24$ & $ 8$ & $ 491$ \\
SPT-S J051404$-$5555.2 & $ 78.521$ & $ -55.920$ & & $ 0.92$ & $ 1.16$ & $0.72^{+1.05}_{-0.48}$ & & $ 4.51$ & $ 15.07$ & $1.06^{+1.81}_{-0.65}$ & & $ 7.0$ & $1.7^{+2.3}_{-2.8}$ & $ 0.51$ & dust & & $ 282$ & $ 546$ & $1119$ \\
SPT-S J051406$-$5124.6 & $ 78.527$ & $ -51.410$ & & $ 6.29$ & $ 8.25$ & $7.54^{+1.39}_{-1.42}$ & & $ 1.51$ & $ 5.23$ & $5.00^{+2.81}_{-1.79}$ & & $ -1.2$ & $-1.1^{+1.3}_{-1.2}$ & $ 0.01$ & sync & & $ 5$ & $1148$ & $ 197$ \\
SPT-S J051418$-$5629.5 & $ 78.577$ & $ -56.492$ & & $ 7.44$ & $ 9.35$ & $8.84^{+1.33}_{-1.35}$ & & $ 2.28$ & $ 7.61$ & $6.65^{+3.14}_{-2.47}$ & & $ -0.6$ & $-0.8^{+1.2}_{-1.3}$ & $ 0.01$ & sync & & $ 1$ & $ 728$ & $ 390$ \\
SPT-S J051424$-$5744.8 & $ 78.604$ & $ -57.747$ & & $ 9.45$ & $ 11.44$ & $11.00^{+1.31}_{-1.31}$ & & $ 1.92$ & $ 6.17$ & $6.21^{+2.70}_{-1.78}$ & & $ -1.7$ & $-1.6^{+1.0}_{-0.9}$ & $ 0.00$ & sync & & $ 4$ & $ 362$ & $1007$ \\
SPT-S J051433$-$5532.3 & $ 78.638$ & $ -55.539$ & & $ 6.61$ & $ 8.30$ & $7.70^{+1.34}_{-1.35}$ & & $ 1.94$ & $ 6.50$ & $5.69^{+2.96}_{-2.16}$ & & $ -0.7$ & $-0.8^{+1.3}_{-1.3}$ & $ 0.01$ & sync & & $ 10$ & $ 449$ & $ 566$ \\
SPT-S J051439$-$5329.0 & $ 78.666$ & $ -53.483$ & & $ 6.39$ & $ 8.14$ & $7.48^{+1.36}_{-1.39}$ & & $ 1.74$ & $ 5.87$ & $5.29^{+2.87}_{-1.98}$ & & $ -0.9$ & $-0.9^{+1.3}_{-1.3}$ & $ 0.01$ & sync & & $ 23$ & $1436$ & $ 941$ \\
SPT-S J051456$-$5114.7 & $ 78.735$ & $ -51.246$ & & $ 4.66$ & $ 6.01$ & $1.11^{+4.07}_{-0.89}$ & & $ 0.00$ & $ -1.50$ & $1.36^{+1.77}_{-1.12}$ & & $ NaN$ & $-0.9^{+3.0}_{-1.6}$ & $ 0.19$ & sync & & $ 28$ & $1797$ & $ 571$ \\
SPT-S J051459$-$5130.0 & $ 78.748$ & $ -51.501$ & & $ 5.68$ & $ 7.32$ & $6.44^{+1.40}_{-1.47}$ & & $ 1.04$ & $ 3.60$ & $4.05^{+2.48}_{-1.45}$ & & $ -1.9$ & $-1.2^{+1.4}_{-1.2}$ & $ 0.02$ & sync & & $ 5$ & $ 990$ & $ 668$ \\
SPT-S J051506$-$5344.2 & $ 78.779$ & $ -53.738$ & & $ 6.05$ & $ 7.70$ & $7.12^{+1.37}_{-1.37}$ & & $ 4.18$ & $ 14.09$ & $12.93^{+3.59}_{-3.75}$ & & $ 1.6$ & $1.6^{+0.9}_{-1.0}$ & $ 0.48$ & sync & & $ 8$ & $ 971$ & $ 12$ \\
SPT-S J051517$-$5500.2 & $ 78.824$ & $ -55.004$ & & $ 4.60$ & $ 5.86$ & $2.02^{+3.30}_{-1.76}$ & & $ 0.26$ & $ 0.87$ & $1.75^{+1.99}_{-1.44}$ & & $ -5.2$ & $-0.7^{+2.9}_{-1.7}$ & $ 0.20$ & sync & & $ 275$ & $1168$ & $ 347$ \\
SPT-S J051535$-$5657.1 & $ 78.898$ & $ -56.953$ & & $ 16.31$ & $ 20.78$ & $20.51^{+1.52}_{-1.52}$ & & $ 4.74$ & $ 15.97$ & $15.06^{+3.74}_{-3.62}$ & & $ -0.7$ & $-0.8^{+0.6}_{-0.8}$ & $ 0.00$ & sync & & $ 15$ & $1865$ & $ 959$ \\
SPT-S J051550$-$5546.5 & $ 78.958$ & $ -55.776$ & & $ 6.56$ & $ 8.36$ & $7.76^{+1.35}_{-1.38}$ & & $ 2.18$ & $ 7.33$ & $6.21^{+3.04}_{-2.44}$ & & $ -0.4$ & $-0.6^{+1.2}_{-1.4}$ & $ 0.01$ & sync & & $ 4$ & $ 773$ & $1328$ \\
SPT-S J051614$-$5429.6\tablenotemark{g} & $ 79.059$ & $ -54.494$ & & $ 4.67$ & $ 5.94$ & $2.96^{+2.60}_{-2.66}$ & & $ 0.31$ & $ 1.03$ & $1.98^{+1.99}_{-1.60}$ & & $ -4.8$ & $-0.8^{+2.8}_{-1.6}$ & $ 0.18$ & sync & & $ 104$ & $ 193$ & $ 783$ \\
SPT-S J051618$-$5417.8 & $ 79.077$ & $ -54.298$ & & $ 4.78$ & $ 6.09$ & $3.91^{+2.02}_{-3.52}$ & & $ 0.37$ & $ 1.26$ & $2.33^{+2.01}_{-1.77}$ & & $ -4.3$ & $-0.9^{+2.4}_{-1.4}$ & $ 0.15$ & sync & & $ 12$ & $ 505$ & $ 943$ \\
SPT-S J051639$-$5920.4 & $ 79.165$ & $ -59.341$ & & $ 1.95$ & $ 2.38$ & $2.55^{+1.04}_{-1.40}$ & & $ 5.42$ & $ 17.32$ & $9.86^{+5.75}_{-8.48}$ & & $ 5.4$ & $3.6^{+1.0}_{-3.4}$ & $ 0.75$ & dust & & $ 189$ & $ 266$ & $ 897$ \\
SPT-S J051659$-$5736.9 & $ 79.249$ & $ -57.616$ & & $ 0.13$ & $ 0.15$ & $0.48^{+0.72}_{-0.30}$ & & $ 4.60$ & $ 14.70$ & $0.90^{+1.38}_{-0.52}$ & & $ 12.4$ & $2.2^{+2.0}_{-2.7}$ & $ 0.58$ & dust & & $ 253$ & $ 257$ & $ 763$ \\
SPT-S J051744$-$5851.4 & $ 79.434$ & $ -58.858$ & & $ 9.13$ & $ 11.15$ & $10.79^{+1.31}_{-1.34}$ & & $ 4.55$ & $ 14.55$ & $13.54^{+3.46}_{-3.55}$ & & $ 0.7$ & $0.6^{+0.7}_{-0.9}$ & $ 0.06$ & sync & & $ 5$ & $ 251$ & $1436$ \\
SPT-S J051811$-$5144.1 & $ 79.547$ & $ -51.735$ & & $ 22.54$ & $ 29.06$ & $28.80^{+1.74}_{-1.74}$ & & $ 7.92$ & $ 27.44$ & $26.67^{+4.30}_{-4.25}$ & & $ -0.2$ & $-0.2^{+0.4}_{-0.5}$ & $ 0.00$ & sync & & $ 3$ & $ 652$ & $ 665$ \\
SPT-S J051825$-$5614.2 & $ 79.608$ & $ -56.237$ & & $ 16.27$ & $ 20.72$ & $20.46^{+1.55}_{-1.52}$ & & $ 5.06$ & $ 17.05$ & $16.14^{+3.77}_{-3.73}$ & & $ -0.5$ & $-0.6^{+0.6}_{-0.7}$ & $ 0.00$ & sync & & $ 13$ & $ 95$ & $ 650$ \\
SPT-S J051841$-$5006.1 & $ 79.673$ & $ -50.103$ & & $ 12.94$ & $ 16.68$ & $16.28^{+1.47}_{-1.44}$ & & $ 2.24$ & $ 7.75$ & $8.15^{+2.88}_{-1.91}$ & & $ -2.1$ & $-1.9^{+0.8}_{-0.7}$ & $ 0.00$ & sync & & $ 14$ & $ 634$ & $ 423$ \\
SPT-S J051909$-$5609.6 & $ 79.788$ & $ -56.161$ & & $ 5.11$ & $ 6.51$ & $5.32^{+1.50}_{-2.08}$ & & $ 0.89$ & $ 3.00$ & $3.35^{+2.34}_{-1.43}$ & & $ -2.1$ & $-1.0^{+1.7}_{-1.3}$ & $ 0.06$ & sync & & $ 4$ & $ 497$ & $ 714$ \\
SPT-S J051939$-$5044.8 & $ 79.914$ & $ -50.748$ & & $ 5.78$ & $ 7.46$ & $6.66^{+1.38}_{-1.44}$ & & $ 1.37$ & $ 4.75$ & $4.54^{+2.71}_{-1.71}$ & & $ -1.2$ & $-1.0^{+1.4}_{-1.3}$ & $ 0.02$ & sync & & $ 9$ & $ 7$ & $ 532$ \\
SPT-S J052039$-$5329.8 & $ 80.164$ & $ -53.498$ & & $ 3.90$ & $ 4.97$ & $4.40^{+1.35}_{-1.33}$ & & $ 4.66$ & $ 15.70$ & $8.38^{+6.77}_{-6.30}$ & & $ 3.1$ & $1.6^{+1.8}_{-3.5}$ & $ 0.50$ & sync & & $ 297$ & $1078$ & $ 865$ \\
SPT-S J052044$-$5508.4 & $ 80.187$ & $ -55.141$ & & $ 16.69$ & $ 21.26$ & $21.00^{+1.53}_{-1.53}$ & & $ 6.54$ & $ 22.02$ & $21.20^{+3.96}_{-3.92}$ & & $ 0.1$ & $0.0^{+0.5}_{-0.6}$ & $ 0.00$ & sync & & $ 3$ & $1012$ & $1653$ \\
SPT-S J052124$-$5412.5 & $ 80.353$ & $ -54.210$ & & $ 10.84$ & $ 13.81$ & $13.45^{+1.40}_{-1.42}$ & & $ 2.74$ & $ 9.23$ & $8.56^{+3.26}_{-2.64}$ & & $ -1.1$ & $-1.2^{+0.9}_{-1.0}$ & $ 0.00$ & sync & & $ 9$ & $1109$ & $ 128$ \\
SPT-S J052137$-$5456.7 & $ 80.407$ & $ -54.946$ & & $ 8.67$ & $ 11.05$ & $10.65^{+1.36}_{-1.38}$ & & $ 3.65$ & $ 12.28$ & $11.02^{+3.48}_{-3.55}$ & & $ 0.3$ & $0.1^{+0.8}_{-1.1}$ & $ 0.01$ & sync & & $ 8$ & $ 514$ & $ 829$ \\
SPT-S J052206$-$5016.3 & $ 80.528$ & $ -50.272$ & & $ 4.59$ & $ 5.92$ & $4.01^{+1.92}_{-3.48}$ & & $ 1.44$ & $ 4.98$ & $3.12^{+2.85}_{-2.21}$ & & $ -0.5$ & $-0.0^{+2.3}_{-1.9}$ & $ 0.20$ & sync & & $ 395$ & $ 784$ & $ 592$ \\
SPT-S J052215$-$5607.4 & $ 80.565$ & $ -56.124$ & & $ 5.24$ & $ 6.68$ & $5.73^{+1.43}_{-1.60}$ & & $ 1.50$ & $ 5.07$ & $4.28^{+2.69}_{-1.80}$ & & $ -0.8$ & $-0.7^{+1.6}_{-1.5}$ & $ 0.06$ & sync & & $ 13$ & $ 676$ & $ 757$ \\
SPT-S J052309$-$5822.9 & $ 80.789$ & $ -58.383$ & & $ 1.96$ & $ 2.40$ & $1.67^{+1.30}_{-1.13}$ & & $ 4.61$ & $ 14.74$ & $1.87^{+6.85}_{-1.16}$ & & $ 4.9$ & $1.4^{+2.6}_{-2.9}$ & $ 0.47$ & sync & & $ 219$ & $ 995$ & $ 376$ \\
SPT-S J052318$-$5308.5 & $ 80.826$ & $ -53.143$ & & $ 9.18$ & $ 11.69$ & $11.12^{+1.38}_{-1.36}$ & & $ 1.25$ & $ 4.21$ & $5.55^{+2.32}_{-1.36}$ & & $ -2.8$ & $-1.9^{+1.0}_{-0.8}$ & $ 0.00$ & sync & & $ 1$ & $ 661$ & $1064$ \\
SPT-S J052400$-$5133.9 & $ 81.003$ & $ -51.566$ & & $ 7.74$ & $ 9.97$ & $9.52^{+1.39}_{-1.39}$ & & $ 3.74$ & $ 12.97$ & $11.66^{+3.62}_{-3.72}$ & & $ 0.7$ & $0.6^{+0.9}_{-1.1}$ & $ 0.08$ & sync & & $ 7$ & $ 503$ & $1247$ \\
SPT-S J052405$-$5602.2 & $ 81.021$ & $ -56.037$ & & $ 11.32$ & $ 14.41$ & $14.07^{+1.43}_{-1.41}$ & & $ 3.29$ & $ 11.07$ & $10.12^{+3.49}_{-3.14}$ & & $ -0.7$ & $-0.9^{+0.9}_{-1.0}$ & $ 0.00$ & sync & & $ 6$ & $ 728$ & $1083$ \\
SPT-S J052440$-$5658.8 & $ 81.168$ & $ -56.981$ & & $ 106.29$ & $ 135.38$ & $134.67^{+5.84}_{-5.58}$ & & $ 32.01$ & $ 107.78$ & $106.49^{+10.15}_{-10.15}$ & & $ -0.6$ & $-0.6^{+0.2}_{-0.3}$ & $ 0.00$ & sync & & $ 2$ & $1078$ & $ 798$ \\
SPT-S J052500$-$5135.2 & $ 81.250$ & $ -51.588$ & & $ 8.09$ & $ 10.42$ & $9.94^{+1.37}_{-1.39}$ & & $ 2.21$ & $ 7.64$ & $6.89^{+3.22}_{-2.42}$ & & $ -0.8$ & $-1.0^{+1.1}_{-1.2}$ & $ 0.00$ & sync & & $ 5$ & $ 79$ & $1217$ \\
SPT-S J052558$-$5725.1 & $ 81.495$ & $ -57.419$ & & $ 54.06$ & $ 66.02$ & $65.67^{+2.97}_{-2.97}$ & & $ 16.50$ & $ 52.71$ & $51.93^{+5.69}_{-5.69}$ & & $ -0.6$ & $-0.6^{+0.3}_{-0.3}$ & $ 0.00$ & sync & & $ 2$ & $ 789$ & $ 347$ \\
SPT-S J052731$-$5546.0 & $ 81.883$ & $ -55.767$ & & $ 1.84$ & $ 2.34$ & $1.47^{+1.39}_{-1.02}$ & & $ 4.51$ & $ 15.18$ & $1.58^{+4.90}_{-0.97}$ & & $ 5.1$ & $1.1^{+2.8}_{-2.7}$ & $ 0.44$ & sync & & $ 128$ & $1413$ & $1607$ \\
SPT-S J052743$-$5426.3 & $ 81.931$ & $ -54.438$ & & $ 16.52$ & $ 21.04$ & $20.77^{+1.55}_{-1.51}$ & & $ 6.83$ & $ 23.01$ & $22.24^{+3.97}_{-4.01}$ & & $ 0.2$ & $0.2^{+0.5}_{-0.6}$ & $ 0.00$ & sync & & $ 5$ & $ 406$ & $1340$ \\
SPT-S J052759$-$5232.2 & $ 81.998$ & $ -52.537$ & & $ 2.49$ & $ 3.17$ & $2.48^{+1.32}_{-1.47}$ & & $ 4.65$ & $ 15.67$ & $2.75^{+9.52}_{-1.76}$ & & $ 4.4$ & $1.3^{+2.6}_{-3.0}$ & $ 0.47$ & sync & & $ 276$ & $1773$ & $1001$ \\
SPT-S J052832$-$5433.8 & $ 82.137$ & $ -54.564$ & & $ 7.53$ & $ 9.59$ & $9.15^{+1.36}_{-1.36}$ & & $ 4.80$ & $ 16.18$ & $15.18^{+3.67}_{-3.79}$ & & $ 1.4$ & $1.4^{+0.7}_{-0.9}$ & $ 0.35$ & sync & & $ 13$ & $ 269$ & $1582$ \\
SPT-S J052834$-$5820.3\tablenotemark{a} & $ 82.145$ & $ -58.339$ & & $ 19.89$ & $ 24.29$ & $24.07^{+1.55}_{-1.59}$ & & $ 6.63$ & $ 21.17$ & $20.42^{+3.73}_{-3.78}$ & & $ -0.4$ & $-0.5^{+0.5}_{-0.5}$ & $ 0.00$ & sync & & $ 23$ & $ 409$ & $ 134$ \\
SPT-S J052841$-$5726.0 & $ 82.174$ & $ -57.433$ & & $ 1.93$ & $ 2.35$ & $1.65^{+1.30}_{-1.11}$ & & $ 4.62$ & $ 14.76$ & $1.84^{+6.72}_{-1.13}$ & & $ 5.0$ & $1.4^{+2.6}_{-2.9}$ & $ 0.47$ & sync & & $ 190$ & $1438$ & $ 442$ \\
SPT-S J052846$-$5919.9 & $ 82.192$ & $ -59.333$ & & $ 12.49$ & $ 15.25$ & $14.99^{+1.38}_{-1.41}$ & & $ 5.47$ & $ 17.47$ & $16.62^{+3.57}_{-3.64}$ & & $ 0.4$ & $0.3^{+0.6}_{-0.7}$ & $ 0.00$ & sync & & $ 6$ & $ 7$ & $ 708$ \\
SPT-S J052850$-$5300.3\tablenotemark{h} & $ 82.212$ & $ -53.006$ & & $ 1.13$ & $ 1.44$ & $1.11^{+1.41}_{-0.80}$ & & $ 5.09$ & $ 17.14$ & $1.59^{+8.06}_{-1.06}$ & & $ 6.7$ & $2.3^{+2.0}_{-3.2}$ & $ 0.58$ & dust & & $ 45$ & $ 856$ & $ 611$ \\
SPT-S J052852$-$5457.6 & $ 82.220$ & $ -54.960$ & & $ 6.87$ & $ 8.75$ & $8.19^{+1.36}_{-1.36}$ & & $ 2.41$ & $ 8.11$ & $6.85^{+3.12}_{-2.67}$ & & $ -0.2$ & $-0.4^{+1.2}_{-1.4}$ & $ 0.01$ & sync & & $ 30$ & $1278$ & $1317$ \\
SPT-S J052903$-$5436.6 & $ 82.263$ & $ -54.611$ & & $ 7.34$ & $ 9.34$ & $9.17^{+1.31}_{-1.31}$ & & $ 10.96$ & $ 36.89$ & $35.38^{+4.77}_{-4.77}$ & & $ 3.7$ & $3.7^{+0.5}_{-0.5}$ & $ 1.00$ & dust & & $ 121$ & $ 543$ & $1757$ \\
SPT-S J052904$-$5538.8 & $ 82.271$ & $ -55.647$ & & $ 6.44$ & $ 8.21$ & $7.28^{+1.37}_{-1.38}$ & & $ 0.00$ & $ -0.02$ & $3.56^{+1.66}_{-0.94}$ & & $ NaN$ & $-2.0^{+1.1}_{-0.7}$ & $ 0.00$ & sync & & $ 4$ & $1165$ & $1122$ \\
SPT-S J053058$-$5951.3 & $ 82.743$ & $ -59.856$ & & $ 12.66$ & $ 15.46$ & $15.20^{+1.36}_{-1.39}$ & & $ 5.67$ & $ 18.11$ & $17.27^{+3.63}_{-3.63}$ & & $ 0.4$ & $0.4^{+0.6}_{-0.7}$ & $ 0.01$ & sync & & $ 6$ & $1326$ & $ 910$ \\
SPT-S J053107$-$5543.5 & $ 82.780$ & $ -55.725$ & & $ 8.76$ & $ 11.16$ & $10.68^{+1.36}_{-1.38}$ & & $ 2.02$ & $ 6.78$ & $6.52^{+2.97}_{-2.05}$ & & $ -1.4$ & $-1.4^{+1.1}_{-1.0}$ & $ 0.00$ & sync & & $ 7$ & $1320$ & $ 272$ \\
SPT-S J053117$-$5504.4 & $ 82.822$ & $ -55.074$ & & $ 28.10$ & $ 35.80$ & $35.54^{+1.95}_{-1.95}$ & & $ 7.84$ & $ 26.40$ & $25.67^{+4.09}_{-4.16}$ & & $ -0.8$ & $-0.9^{+0.4}_{-0.5}$ & $ 0.00$ & sync & & $ 4$ & $ 33$ & $ 752$ \\
SPT-S J053147$-$5414.4 & $ 82.946$ & $ -54.241$ & & $ 6.66$ & $ 8.49$ & $7.73^{+1.37}_{-1.37}$ & & $ 0.79$ & $ 2.67$ & $4.18^{+2.16}_{-1.23}$ & & $ -3.1$ & $-1.7^{+1.2}_{-0.9}$ & $ 0.00$ & sync & & $ 8$ & $ 802$ & $ 412$ \\
SPT-S J053205$-$5434.6 & $ 83.024$ & $ -54.577$ & & $ 1.09$ & $ 1.39$ & $1.12^{+1.43}_{-0.81}$ & & $ 5.14$ & $ 17.30$ & $1.61^{+8.40}_{-1.07}$ & & $ 6.9$ & $2.4^{+2.0}_{-3.2}$ & $ 0.59$ & dust & & $ 493$ & $ 263$ & $ 343$ \\
SPT-S J053208$-$5310.5 & $ 83.035$ & $ -53.176$ & & $ 41.98$ & $ 53.48$ & $53.19^{+2.51}_{-2.51}$ & & $ 14.79$ & $ 49.78$ & $48.98^{+5.62}_{-5.51}$ & & $ -0.2$ & $-0.2^{+0.3}_{-0.3}$ & $ 0.00$ & sync & & $ 2$ & $ 73$ & $ 148$ \\
SPT-S J053244$-$5812.1 & $ 83.186$ & $ -58.202$ & & $ 6.02$ & $ 7.35$ & $6.72^{+1.31}_{-1.34}$ & & $ 2.23$ & $ 7.14$ & $5.81^{+2.77}_{-2.38}$ & & $ -0.1$ & $-0.3^{+1.3}_{-1.5}$ & $ 0.03$ & sync & & $ 9$ & $ 604$ & $1034$ \\
SPT-S J053248$-$5721.6 & $ 83.201$ & $ -57.360$ & & $ 4.91$ & $ 6.00$ & $4.96^{+1.43}_{-1.87}$ & & $ 1.97$ & $ 6.31$ & $4.56^{+2.71}_{-2.22}$ & & $ 0.1$ & $0.1^{+1.6}_{-1.8}$ & $ 0.13$ & sync & & $ 330$ & $ 375$ & $ 611$ \\
SPT-S J053250$-$5047.1 & $ 83.212$ & $ -50.786$ & & $ 10.53$ & $ 13.58$ & $13.35^{+1.43}_{-1.35}$ & & $ 12.20$ & $ 42.26$ & $40.77^{+5.23}_{-5.23}$ & & $ 3.1$ & $3.0^{+0.4}_{-0.4}$ & $ 1.00$ & dust & & $ 424$ & $1002$ & $ 441$ \\
SPT-S J053310$-$5453.3 & $ 83.293$ & $ -54.889$ & & $ 4.94$ & $ 6.29$ & $5.51^{+1.41}_{-1.43}$ & & $ 4.12$ & $ 13.88$ & $12.61^{+3.59}_{-3.75}$ & & $ 2.2$ & $2.3^{+1.0}_{-1.1}$ & $ 0.71$ & dust & & $ 8$ & $ 736$ & $ 300$ \\
SPT-S J053312$-$5238.5\tablenotemark{a} & $ 83.302$ & $ -52.642$ & & $ 1.65$ & $ 2.11$ & $1.73^{+1.37}_{-1.21}$ & & $ 5.04$ & $ 16.97$ & $2.34^{+10.12}_{-1.62}$ & & $ 5.7$ & $2.4^{+2.0}_{-3.4}$ & $ 0.58$ & dust & & $ 12$ & $ 744$ & $ 9$ \\
SPT-S J053319$-$5022.0 & $ 83.332$ & $ -50.367$ & & $ 5.33$ & $ 6.87$ & $5.96^{+1.42}_{-1.57}$ & & $ 1.47$ & $ 5.09$ & $4.39^{+2.74}_{-1.82}$ & & $ -0.8$ & $-0.7^{+1.5}_{-1.5}$ & $ 0.04$ & sync & & $ 8$ & $ 597$ & $ 783$ \\
SPT-S J053324$-$5549.6 & $ 83.352$ & $ -55.827$ & & $ 55.02$ & $ 70.08$ & $69.71^{+3.15}_{-3.15}$ & & $ 17.54$ & $ 59.06$ & $58.28^{+6.18}_{-6.31}$ & & $ -0.5$ & $-0.5^{+0.3}_{-0.3}$ & $ 0.00$ & sync & & $ 5$ & $1224$ & $ 975$ \\
SPT-S J053327$-$5050.8 & $ 83.366$ & $ -50.847$ & & $ 7.82$ & $ 10.09$ & $9.48^{+1.38}_{-1.38}$ & & $ 1.35$ & $ 4.69$ & $5.32^{+2.59}_{-1.59}$ & & $ -2.1$ & $-1.6^{+1.1}_{-1.0}$ & $ 0.00$ & sync & & $ 12$ & $ 793$ & $ 392$ \\
SPT-S J053345$-$5818.1\tablenotemark{a,i} & $ 83.441$ & $ -58.302$ & & $ 7.24$ & $ 8.84$ & $8.38^{+1.32}_{-1.30}$ & & $ 3.66$ & $ 11.71$ & $10.48^{+3.32}_{-3.45}$ & & $ 0.8$ & $0.6^{+0.9}_{-1.1}$ & $ 0.10$ & sync & & $ 36$ & $ 4$ & $ 514$ \\
SPT-S J053412$-$5924.3 & $ 83.553$ & $ -59.406$ & & $ 15.34$ & $ 18.73$ & $18.46^{+1.44}_{-1.44}$ & & $ 3.96$ & $ 12.64$ & $11.85^{+3.37}_{-3.13}$ & & $ -1.1$ & $-1.2^{+0.7}_{-0.8}$ & $ 0.00$ & sync & & $ 5$ & $ 49$ & $1561$ \\
SPT-S J053426$-$5113.2 & $ 83.610$ & $ -51.221$ & & $ 10.88$ & $ 14.02$ & $13.61^{+1.42}_{-1.42}$ & & $ 2.15$ & $ 7.44$ & $7.46^{+3.00}_{-2.02}$ & & $ -1.7$ & $-1.6^{+0.9}_{-0.9}$ & $ 0.00$ & sync & & $ 1$ & $ 473$ & $ 742$ \\
SPT-S J053458$-$5439.0 & $ 83.744$ & $ -54.651$ & & $ 20.37$ & $ 25.95$ & $25.71^{+1.65}_{-1.65}$ & & $ 7.12$ & $ 23.96$ & $23.18^{+4.04}_{-4.04}$ & & $ -0.2$ & $-0.3^{+0.5}_{-0.5}$ & $ 0.00$ & sync & & $ 3$ & $ 570$ & $ 907$ \\
SPT-S J053527$-$5013.6 & $ 83.864$ & $ -50.227$ & & $ 0.62$ & $ 0.79$ & $0.65^{+1.08}_{-0.44}$ & & $ 4.87$ & $ 16.88$ & $1.06^{+2.09}_{-0.66}$ & & $ 8.3$ & $2.0^{+2.2}_{-2.8}$ & $ 0.55$ & dust & & $ 30$ & $ 645$ & $ 419$ \\
SPT-S J053631$-$5258.8 & $ 84.130$ & $ -52.981$ & & $ 1.18$ & $ 1.50$ & $0.94^{+1.28}_{-0.66}$ & & $ 4.72$ & $ 15.89$ & $1.27^{+3.28}_{-0.77}$ & & $ 6.4$ & $1.8^{+2.4}_{-2.9}$ & $ 0.51$ & dust & & $ 167$ & $ 865$ & $ 252$ \\
SPT-S J053644$-$5310.5 & $ 84.185$ & $ -53.176$ & & $ 1.15$ & $ 1.46$ & $0.90^{+1.24}_{-0.62}$ & & $ 4.67$ & $ 15.71$ & $1.22^{+2.77}_{-0.77}$ & & $ 6.5$ & $1.7^{+2.4}_{-2.8}$ & $ 0.51$ & dust & & $ 483$ & $ 951$ & $ 690$ \\
SPT-S J053646$-$5714.1 & $ 84.194$ & $ -57.236$ & & $ 12.01$ & $ 15.30$ & $14.99^{+1.43}_{-1.43}$ & & $ 3.92$ & $ 13.20$ & $12.18^{+3.61}_{-3.53}$ & & $ -0.4$ & $-0.6^{+0.8}_{-1.0}$ & $ 0.00$ & sync & & $ 3$ & $ 353$ & $ 826$ \\
SPT-S J053726$-$5434.4\tablenotemark{j} & $ 84.360$ & $ -54.574$ & & $ 6.36$ & $ 8.11$ & $7.36^{+1.35}_{-1.38}$ & & $ 1.07$ & $ 3.61$ & $4.32^{+2.40}_{-1.40}$ & & $ -2.2$ & $-1.5^{+1.3}_{-1.0}$ & $ 0.00$ & sync & & $ 83$ & $1366$ & $ 192$ \\
SPT-S J053748$-$5718.4 & $ 84.453$ & $ -57.308$ & & $ 12.53$ & $ 15.96$ & $15.66^{+1.44}_{-1.44}$ & & $ 5.32$ & $ 17.91$ & $17.01^{+3.73}_{-3.83}$ & & $ 0.3$ & $0.2^{+0.6}_{-0.7}$ & $ 0.00$ & sync & & $ 0$ & $ 16$ & $ 884$ \\
SPT-S J053816$-$5030.8 & $ 84.569$ & $ -50.514$ & & $ 6.81$ & $ 8.78$ & $8.54^{+1.35}_{-1.35}$ & & $ 9.04$ & $ 31.31$ & $29.68^{+4.52}_{-4.58}$ & & $ 3.5$ & $3.4^{+0.6}_{-0.6}$ & $ 1.00$ & dust & & $ 190$ & $ 559$ & $ 454$ \\
SPT-S J053819$-$5227.7 & $ 84.580$ & $ -52.462$ & & $ 8.32$ & $ 10.60$ & $10.19^{+1.35}_{-1.37}$ & & $ 3.68$ & $ 12.40$ & $11.15^{+3.49}_{-3.58}$ & & $ 0.4$ & $0.3^{+0.8}_{-1.1}$ & $ 0.03$ & sync & & $ 10$ & $ 3$ & $ 409$ \\
SPT-S J053823$-$5625.6 & $ 84.598$ & $ -56.427$ & & $ 5.22$ & $ 6.64$ & $5.78^{+1.42}_{-1.54}$ & & $ 2.10$ & $ 7.08$ & $5.41^{+2.87}_{-2.44}$ & & $ 0.2$ & $0.0^{+1.5}_{-1.7}$ & $ 0.09$ & sync & & $ 14$ & $ 648$ & $ 669$ \\
SPT-S J053834$-$5911.0 & $ 84.643$ & $ -59.184$ & & $ 6.31$ & $ 7.70$ & $7.16^{+1.31}_{-1.33}$ & & $ 3.59$ & $ 11.46$ & $10.17^{+3.33}_{-3.46}$ & & $ 1.1$ & $1.0^{+1.0}_{-1.2}$ & $ 0.23$ & sync & & $ 14$ & $ 414$ & $ 275$ \\
SPT-S J053849$-$5955.5 & $ 84.706$ & $ -59.926$ & & $ 4.07$ & $ 4.97$ & $4.47^{+1.28}_{-1.28}$ & & $ 4.74$ & $ 15.14$ & $8.96^{+5.91}_{-6.70}$ & & $ 3.0$ & $1.8^{+1.6}_{-3.4}$ & $ 0.51$ & dust & & $ 609$ & $ 142$ & $ 232$ \\
SPT-S J053857$-$5712.2 & $ 84.738$ & $ -57.204$ & & $ 4.72$ & $ 6.01$ & $4.57^{+1.66}_{-3.40}$ & & $ 1.67$ & $ 5.61$ & $3.77^{+2.78}_{-2.12}$ & & $ -0.2$ & $-0.0^{+2.0}_{-1.8}$ & $ 0.16$ & sync & & $ 14$ & $ 490$ & $1361$ \\
SPT-S J053909$-$5511.0\tablenotemark{a} & $ 84.789$ & $ -55.183$ & & $ 17.25$ & $ 21.97$ & $21.69^{+1.57}_{-1.57}$ & & $ 3.69$ & $ 12.42$ & $11.88^{+3.38}_{-2.88}$ & & $ -1.6$ & $-1.6^{+0.7}_{-0.8}$ & $ 0.00$ & sync & & $ 8$ & $ 843$ & $ 832$ \\
SPT-S J053942$-$5612.8\tablenotemark{a} & $ 84.927$ & $ -56.214$ & & $ 99.72$ & $ 127.03$ & $126.35^{+5.48}_{-5.24}$ & & $ 29.04$ & $ 97.77$ & $96.58^{+9.29}_{-9.29}$ & & $ -0.7$ & $-0.7^{+0.2}_{-0.3}$ & $ 0.00$ & sync & & $ 5$ & $ 689$ & $1244$ \\
SPT-S J054020$-$5356.0 & $ 85.085$ & $ -53.935$ & & $ 4.59$ & $ 5.84$ & $2.07^{+3.24}_{-1.80}$ & & $ 0.33$ & $ 1.11$ & $1.79^{+2.03}_{-1.47}$ & & $ -4.5$ & $-0.6^{+2.9}_{-1.7}$ & $ 0.21$ & sync & & $ 19$ & $ 759$ & $1145$ \\
SPT-S J054025$-$5303.7 & $ 85.105$ & $ -53.063$ & & $ 23.75$ & $ 30.26$ & $30.04^{+1.76}_{-1.82}$ & & $ 6.57$ & $ 22.13$ & $21.36^{+3.91}_{-3.97}$ & & $ -0.9$ & $-0.9^{+0.5}_{-0.6}$ & $ 0.00$ & sync & & $ 2$ & $ 503$ & $ 643$ \\
SPT-S J054030$-$5356.5 & $ 85.126$ & $ -53.942$ & & $ 15.42$ & $ 19.64$ & $19.35^{+1.51}_{-1.51}$ & & $ 3.86$ & $ 12.98$ & $12.17^{+3.53}_{-3.20}$ & & $ -1.1$ & $-1.3^{+0.7}_{-0.8}$ & $ 0.00$ & sync & & $ 5$ & $ 810$ & $1226$ \\
SPT-S J054045$-$5418.3 & $ 85.191$ & $ -54.306$ & & $ 259.54$ & $ 330.59$ & $328.84^{+13.64}_{-13.02}$ & & $ 78.29$ & $ 263.58$ & $260.65^{+24.17}_{-23.55}$ & & $ -0.6$ & $-0.6^{+0.2}_{-0.2}$ & $ 0.00$ & sync & & $ 5$ & $ 56$ & $ 191$ \\
SPT-S J054120$-$5738.3 & $ 85.337$ & $ -57.640$ & & $ 2.24$ & $ 2.73$ & $2.35^{+1.18}_{-1.39}$ & & $ 4.93$ & $ 15.74$ & $3.73^{+9.35}_{-2.71}$ & & $ 4.8$ & $2.4^{+1.9}_{-3.5}$ & $ 0.57$ & dust & & $ 294$ & $ 657$ & $ 461$ \\
SPT-S J054123$-$5752.8 & $ 85.348$ & $ -57.882$ & & $ 4.27$ & $ 5.21$ & $4.74^{+1.26}_{-1.29}$ & & $ 4.78$ & $ 15.28$ & $9.75^{+5.50}_{-7.25}$ & & $ 2.9$ & $1.8^{+1.5}_{-3.4}$ & $ 0.52$ & dust & & $ 73$ & $ 798$ & $1293$ \\
SPT-S J054135$-$5016.9 & $ 85.400$ & $ -50.282$ & & $ 4.98$ & $ 6.41$ & $5.36^{+1.49}_{-1.88}$ & & $ 1.84$ & $ 6.38$ & $4.62^{+2.78}_{-2.19}$ & & $ -0.0$ & $-0.1^{+1.6}_{-1.7}$ & $ 0.10$ & sync & & $ 91$ & $1277$ & $1082$ \\
SPT-S J054223$-$5142.9 & $ 85.598$ & $ -51.716$ & & $ 70.43$ & $ 90.80$ & $90.32^{+3.92}_{-3.92}$ & & $ 20.28$ & $ 70.27$ & $69.38^{+7.15}_{-7.15}$ & & $ -0.7$ & $-0.7^{+0.3}_{-0.3}$ & $ 0.00$ & sync & & $ 4$ & $ 9$ & $ 914$ \\
SPT-S J054254$-$5122.0 & $ 85.728$ & $ -51.368$ & & $ 2.48$ & $ 3.20$ & $2.36^{+1.38}_{-1.43}$ & & $ 4.50$ & $ 15.61$ & $2.24^{+8.66}_{-1.35}$ & & $ 4.3$ & $0.8^{+3.0}_{-2.7}$ & $ 0.41$ & sync & & $ 408$ & $ 187$ & $ 853$ \\
SPT-S J054343$-$5813.3 & $ 85.931$ & $ -58.223$ & & $ 4.67$ & $ 5.70$ & $4.74^{+1.41}_{-1.66}$ & & $ 3.02$ & $ 9.63$ & $7.83^{+3.32}_{-3.56}$ & & $ 1.4$ & $1.5^{+1.4}_{-1.6}$ & $ 0.43$ & sync & & $ 19$ & $ 469$ & $ 375$ \\
SPT-S J054357$-$5532.1 & $ 85.991$ & $ -55.536$ & & $ 14.32$ & $ 18.24$ & $17.97^{+1.47}_{-1.47}$ & & $ 4.56$ & $ 15.34$ & $14.38^{+3.73}_{-3.66}$ & & $ -0.5$ & $-0.6^{+0.6}_{-0.8}$ & $ 0.00$ & sync & & $ 10$ & $ 6$ & $ 155$ \\
SPT-S J054407$-$5444.2 & $ 86.032$ & $ -54.737$ & & $ 0.00$ & $ -0.41$ & $0.41^{+0.64}_{-0.27}$ & & $ 4.83$ & $ 16.27$ & $0.81^{+1.25}_{-0.46}$ & & $ NaN$ & $2.4^{+1.9}_{-2.7}$ & $ 0.61$ & dust & & $ 405$ & $ 531$ & $ 800$ \\
SPT-S J054431$-$5826.2 & $ 86.131$ & $ -58.438$ & & $ 5.88$ & $ 7.18$ & $6.46^{+1.31}_{-1.36}$ & & $ 1.53$ & $ 4.87$ & $4.49^{+2.60}_{-1.70}$ & & $ -1.1$ & $-0.9^{+1.4}_{-1.3}$ & $ 0.02$ & sync & & $ 10$ & $ 416$ & $ 553$ \\
SPT-S J054434$-$5402.9 & $ 86.144$ & $ -54.049$ & & $ 0.21$ & $ 0.27$ & $0.51^{+0.75}_{-0.35}$ & & $ 4.57$ & $ 15.40$ & $0.88^{+1.33}_{-0.52}$ & & $ 11.0$ & $2.1^{+2.1}_{-2.8}$ & $ 0.57$ & dust & & $ 622$ & $ 694$ & $ 851$ \\
SPT-S J054510$-$5635.1 & $ 86.294$ & $ -56.586$ & & $ 0.51$ & $ 0.64$ & $0.60^{+0.88}_{-0.41}$ & & $ 4.62$ & $ 15.57$ & $0.95^{+1.58}_{-0.55}$ & & $ 8.7$ & $2.0^{+2.2}_{-2.8}$ & $ 0.54$ & dust & & $ 428$ & $ 510$ & $1406$ \\
SPT-S J054545$-$5138.0 & $ 86.439$ & $ -51.635$ & & $ 7.85$ & $ 10.12$ & $9.33^{+1.39}_{-1.39}$ & & $ 0.10$ & $ 0.34$ & $4.30^{+1.69}_{-0.97}$ & & $ -9.3$ & $-2.2^{+1.0}_{-0.6}$ & $ 0.00$ & sync & & $ 10$ & $ 38$ & $ 819$ \\
SPT-S J054554$-$5815.0 & $ 86.478$ & $ -58.251$ & & $ 4.58$ & $ 5.59$ & $2.16^{+2.96}_{-1.88}$ & & $ 0.43$ & $ 1.36$ & $1.80^{+1.97}_{-1.45}$ & & $ -3.8$ & $-0.6^{+2.8}_{-1.7}$ & $ 0.20$ & sync & & $ 272$ & $ 508$ & $ 426$ \\
SPT-S J054644$-$5858.3 & $ 86.684$ & $ -58.973$ & & $ 1.59$ & $ 1.94$ & $2.12^{+1.13}_{-1.43}$ & & $ 5.41$ & $ 17.28$ & $6.21^{+8.26}_{-5.25}$ & & $ 6.0$ & $3.4^{+1.2}_{-3.5}$ & $ 0.71$ & dust & & $ 301$ & $ 509$ & $1209$ \\
SPT-S J054716$-$5104.2\tablenotemark{k} & $ 86.818$ & $ -51.070$ & & $ 7.40$ & $ 9.75$ & $9.52^{+1.38}_{-1.38}$ & & $ 7.86$ & $ 27.22$ & $25.49^{+4.39}_{-4.39}$ & & $ 2.8$ & $2.7^{+0.6}_{-0.6}$ & $ 0.94$ & dust & & $ 350$ & $1255$ & $ 11$ \\
SPT-S J054733$-$5335.6 & $ 86.888$ & $ -53.594$ & & $ 0.01$ & $ 0.02$ & $0.49^{+0.83}_{-0.32}$ & & $ 4.96$ & $ 16.96$ & $0.91^{+1.59}_{-0.54}$ & & $ 19.0$ & $2.3^{+2.0}_{-2.8}$ & $ 0.59$ & dust & & $ 600$ & $ 723$ & $ 322$ \\
SPT-S J054804$-$5955.0 & $ 87.020$ & $ -59.917$ & & $ 6.15$ & $ 7.67$ & $6.89^{+1.34}_{-1.38}$ & & $ 0.93$ & $ 3.06$ & $3.99^{+2.26}_{-1.28}$ & & $ -2.5$ & $-1.5^{+1.3}_{-1.0}$ & $ 0.00$ & sync & & $ 8$ & $ 403$ & $ 433$ \\
SPT-S J054830$-$5218.5 & $ 87.126$ & $ -52.308$ & & $ 19.55$ & $ 25.67$ & $25.41^{+1.68}_{-1.68}$ & & $ 8.19$ & $ 27.97$ & $27.24^{+4.25}_{-4.30}$ & & $ 0.2$ & $0.2^{+0.4}_{-0.5}$ & $ 0.00$ & sync & & $ 3$ & $ 457$ & $ 455$ \\
SPT-S J054846$-$5748.8 & $ 87.194$ & $ -57.814$ & & $ 0.79$ & $ 0.99$ & $0.68^{+0.97}_{-0.47}$ & & $ 4.52$ & $ 14.80$ & $1.04^{+1.72}_{-0.64}$ & & $ 7.4$ & $1.8^{+2.3}_{-2.8}$ & $ 0.52$ & dust & & $ 244$ & $ 916$ & $1032$ \\
SPT-S J054901$-$5752.4 & $ 87.258$ & $ -57.874$ & & $ 5.89$ & $ 7.34$ & $6.61^{+1.34}_{-1.38}$ & & $ 1.61$ & $ 5.28$ & $4.73^{+2.71}_{-1.82}$ & & $ -0.9$ & $-0.9^{+1.4}_{-1.3}$ & $ 0.02$ & sync & & $ 5$ & $ 728$ & $ 929$ \\
SPT-S J054903$-$5741.3 & $ 87.265$ & $ -57.689$ & & $ 7.92$ & $ 9.87$ & $9.44^{+1.33}_{-1.33}$ & & $ 3.67$ & $ 12.01$ & $10.77^{+3.40}_{-3.49}$ & & $ 0.5$ & $0.4^{+0.9}_{-1.1}$ & $ 0.05$ & sync & & $ 2$ & $ 760$ & $ 839$ \\
SPT-S J054912$-$5026.6 & $ 87.304$ & $ -50.444$ & & $ 9.19$ & $ 12.11$ & $11.68^{+1.41}_{-1.43}$ & & $ 2.51$ & $ 8.70$ & $7.91^{+3.34}_{-2.63}$ & & $ -0.9$ & $-1.1^{+1.0}_{-1.1}$ & $ 0.00$ & sync & & $ 10$ & $ 28$ & $ 591$ \\
SPT-S J054941$-$5645.0 & $ 87.422$ & $ -56.751$ & & $ 1.03$ & $ 1.35$ & $0.80^{+1.16}_{-0.55}$ & & $ 4.52$ & $ 15.45$ & $1.12^{+1.97}_{-0.70}$ & & $ 6.6$ & $1.6^{+2.4}_{-2.8}$ & $ 0.49$ & sync & & $ 221$ & $ 235$ & $1029$ \\
SPT-S J054943$-$5246.4 & $ 87.432$ & $ -52.774$ & & $ 163.42$ & $ 214.59$ & $213.45^{+8.85}_{-8.45}$ & & $ 51.06$ & $ 174.45$ & $172.41^{+16.09}_{-15.69}$ & & $ -0.6$ & $-0.6^{+0.2}_{-0.2}$ & $ 0.00$ & sync & & $ 4$ & $ 22$ & $ 369$ \\
SPT-S J054951$-$5047.8 & $ 87.465$ & $ -50.798$ & & $ 5.58$ & $ 7.35$ & $6.50^{+1.43}_{-1.50}$ & & $ 1.47$ & $ 5.08$ & $4.61^{+2.77}_{-1.80}$ & & $ -1.0$ & $-0.9^{+1.5}_{-1.4}$ & $ 0.03$ & sync & & $ 15$ & $ 530$ & $ 670$ \\
SPT-S J054953$-$5358.4 & $ 87.472$ & $ -53.975$ & & $ 4.59$ & $ 6.03$ & $5.17^{+1.44}_{-1.47}$ & & $ 4.24$ & $ 14.48$ & $13.15^{+3.66}_{-3.85}$ & & $ 2.4$ & $2.6^{+1.1}_{-1.1}$ & $ 0.79$ & dust & & $ 189$ & $1149$ & $1087$ \\
SPT-S J055002$-$5356.6 & $ 87.509$ & $ -53.943$ & & $ 2.69$ & $ 3.54$ & $3.88^{+1.02}_{-0.91}$ & & $ 6.15$ & $ 21.02$ & $17.28^{+4.02}_{-4.49}$ & & $ 4.9$ & $4.1^{+0.6}_{-0.9}$ & $ 0.96$ & dust & & $ 101$ & $1069$ & $ 951$ \\
SPT-S J055009$-$5732.3 & $ 87.540$ & $ -57.540$ & & $ 523.43$ & $ 652.66$ & $649.19^{+26.92}_{-25.70}$ & & $ 173.64$ & $ 568.60$ & $562.31^{+51.40}_{-50.17}$ & & $ -0.4$ & $-0.4^{+0.2}_{-0.2}$ & $ 0.00$ & sync & & $ 2$ & $ 6$ & $ 624$ \\
SPT-S J055046$-$5304.9 & $ 87.694$ & $ -53.082$ & & $ 17.12$ & $ 22.48$ & $22.24^{+1.60}_{-1.60}$ & & $ 5.37$ & $ 18.36$ & $17.47^{+3.84}_{-3.84}$ & & $ -0.6$ & $-0.7^{+0.6}_{-0.7}$ & $ 0.00$ & sync & & $ 3$ & $ 211$ & $ 879$ \\
SPT-S J055116$-$5334.4\tablenotemark{l} & $ 87.817$ & $ -53.574$ & & $ 4.83$ & $ 6.34$ & $6.10^{+1.32}_{-1.28}$ & & $ 6.87$ & $ 23.48$ & $21.46^{+4.09}_{-4.23}$ & & $ 3.6$ & $3.4^{+0.8}_{-0.8}$ & $ 0.98$ & dust & & $ 14$ & $ 732$ & $ 4$ \\
SPT-S J055117$-$5007.4 & $ 87.823$ & $ -50.124$ & & $ 2.29$ & $ 3.01$ & $2.25^{+1.39}_{-1.45}$ & & $ 4.65$ & $ 16.11$ & $2.34^{+9.52}_{-1.48}$ & & $ 4.6$ & $1.3^{+2.7}_{-3.0}$ & $ 0.47$ & sync & & $ 267$ & $ 636$ & $ 480$ \\
SPT-S J055119$-$5545.5 & $ 87.830$ & $ -55.760$ & & $ 4.59$ & $ 6.03$ & $3.93^{+2.05}_{-3.48}$ & & $ 1.26$ & $ 4.31$ & $2.88^{+2.76}_{-2.15}$ & & $ -0.9$ & $-0.2^{+2.4}_{-1.8}$ & $ 0.19$ & sync & & $ 8$ & $ 911$ & $1645$ \\
SPT-S J055133$-$5655.3 & $ 87.890$ & $ -56.923$ & & $ 4.60$ & $ 6.03$ & $3.77^{+2.15}_{-3.37}$ & & $ 1.11$ & $ 3.80$ & $2.69^{+2.65}_{-2.06}$ & & $ -1.3$ & $-0.3^{+2.5}_{-1.8}$ & $ 0.20$ & sync & & $ 14$ & $ 457$ & $ 499$ \\
SPT-S J055135$-$5902.7 & $ 87.897$ & $ -59.046$ & & $ 6.35$ & $ 7.92$ & $7.69^{+1.29}_{-1.29}$ & & $ 8.73$ & $ 28.59$ & $26.99^{+4.23}_{-4.29}$ & & $ 3.5$ & $3.4^{+0.6}_{-0.6}$ & $ 1.00$ & dust & & $ 7$ & $ 103$ & $ 4$ \\
SPT-S J055138$-$5058.0 & $ 87.912$ & $ -50.968$ & & $ 2.91$ & $ 3.84$ & $5.03^{+0.94}_{-0.76}$ & & $ 9.05$ & $ 31.36$ & $26.73^{+4.23}_{-4.23}$ & & $ 5.7$ & $4.6^{+0.3}_{-0.5}$ & $ 1.00$ & dust & & $ 260$ & $1281$ & $ 76$ \\
SPT-S J055152$-$5526.5\tablenotemark{a} & $ 87.971$ & $ -55.442$ & & $ 27.63$ & $ 36.28$ & $36.02^{+1.97}_{-1.97}$ & & $ 9.33$ & $ 31.86$ & $31.12^{+4.49}_{-4.49}$ & & $ -0.4$ & $-0.4^{+0.4}_{-0.4}$ & $ 0.00$ & sync & & $ 4$ & $ 556$ & $ 733$ \\
SPT-S J055201$-$5951.4 & $ 88.005$ & $ -59.858$ & & $ 0.49$ & $ 0.61$ & $0.57^{+0.87}_{-0.39}$ & & $ 4.56$ & $ 14.93$ & $0.94^{+1.54}_{-0.56}$ & & $ 8.7$ & $2.0^{+2.1}_{-2.8}$ & $ 0.55$ & dust & & $ 255$ & $ 226$ & $ 424$ \\
SPT-S J055232$-$5349.4\tablenotemark{a} & $ 88.135$ & $ -53.823$ & & $ 9.99$ & $ 13.12$ & $12.73^{+1.43}_{-1.45}$ & & $ 2.78$ & $ 9.48$ & $8.65^{+3.40}_{-2.80}$ & & $ -0.9$ & $-1.0^{+0.9}_{-1.1}$ & $ 0.00$ & sync & & $ 5$ & $ 526$ & $ 971$ \\
SPT-S J055233$-$5242.7 & $ 88.139$ & $ -52.712$ & & $ 1.61$ & $ 2.12$ & $1.27^{+1.41}_{-0.91}$ & & $ 4.57$ & $ 15.62$ & $1.45^{+4.42}_{-0.91}$ & & $ 5.4$ & $1.3^{+2.6}_{-2.8}$ & $ 0.46$ & sync & & $ 464$ & $ 647$ & $1162$ \\
SPT-S J055241$-$5238.5 & $ 88.172$ & $ -52.643$ & & $ 0.81$ & $ 1.06$ & $0.68^{+1.04}_{-0.46}$ & & $ 4.51$ & $ 15.42$ & $1.03^{+1.68}_{-0.61}$ & & $ 7.3$ & $1.7^{+2.3}_{-2.8}$ & $ 0.51$ & dust & & $ 487$ & $ 756$ & $1155$ \\
SPT-S J055302$-$5548.7 & $ 88.259$ & $ -55.812$ & & $ 3.64$ & $ 4.79$ & $4.21^{+1.38}_{-1.32}$ & & $ 4.78$ & $ 16.33$ & $9.75^{+6.21}_{-7.65}$ & & $ 3.3$ & $2.1^{+1.6}_{-3.7}$ & $ 0.56$ & dust & & $ 529$ & $ 667$ & $1820$ \\
SPT-S J055320$-$5007.3 & $ 88.335$ & $ -50.122$ & & $ 2.69$ & $ 3.54$ & $2.95^{+1.33}_{-1.51}$ & & $ 4.76$ & $ 16.48$ & $4.25^{+10.04}_{-3.03}$ & & $ 4.2$ & $1.9^{+2.2}_{-3.5}$ & $ 0.52$ & dust & & $ 543$ & $ 937$ & $1002$ \\
SPT-S J055331$-$5131.7 & $ 88.380$ & $ -51.529$ & & $ 0.96$ & $ 1.26$ & $0.82^{+1.24}_{-0.59}$ & & $ 4.78$ & $ 16.57$ & $1.16^{+2.76}_{-0.71}$ & & $ 7.0$ & $1.9^{+2.3}_{-2.9}$ & $ 0.52$ & dust & & $ 263$ & $ 341$ & $ 351$ \\
SPT-S J055359$-$5051.8 & $ 88.498$ & $ -50.864$ & & $ 4.80$ & $ 6.33$ & $4.91^{+1.67}_{-3.44}$ & & $ 1.44$ & $ 5.00$ & $3.70^{+2.80}_{-1.97}$ & & $ -0.6$ & $-0.3^{+1.9}_{-1.7}$ & $ 0.13$ & sync & & $ 4$ & $ 98$ & $ 761$ \\
SPT-S J055421$-$5018.6 & $ 88.588$ & $ -50.311$ & & $ 11.08$ & $ 14.60$ & $14.26^{+1.46}_{-1.46}$ & & $ 4.67$ & $ 16.19$ & $15.14^{+3.77}_{-3.83}$ & & $ 0.3$ & $0.2^{+0.7}_{-0.8}$ & $ 0.00$ & sync & & $ 4$ & $ 658$ & $ 217$ \\
SPT-S J055422$-$5304.2 & $ 88.596$ & $ -53.071$ & & $ 2.20$ & $ 2.88$ & $2.34^{+1.33}_{-1.52}$ & & $ 4.83$ & $ 16.52$ & $2.93^{+10.28}_{-2.01}$ & & $ 4.7$ & $2.0^{+2.2}_{-3.4}$ & $ 0.53$ & dust & & $ 327$ & $ 274$ & $1578$ \\
SPT-S J055455$-$5541.3 & $ 88.731$ & $ -55.690$ & & $ 5.91$ & $ 7.77$ & $7.04^{+1.41}_{-1.44}$ & & $ 1.85$ & $ 6.30$ & $5.38^{+2.93}_{-2.13}$ & & $ -0.6$ & $-0.7^{+1.3}_{-1.4}$ & $ 0.02$ & sync & & $ 6$ & $ 90$ & $1353$ \\
SPT-S J055512$-$5044.9 & $ 88.802$ & $ -50.750$ & & $ 6.98$ & $ 9.20$ & $8.60^{+1.40}_{-1.41}$ & & $ 1.85$ & $ 6.40$ & $5.86^{+3.04}_{-2.09}$ & & $ -1.0$ & $-1.0^{+1.2}_{-1.2}$ & $ 0.00$ & sync & & $ 11$ & $ 803$ & $ 603$ \\
SPT-S J055528$-$5154.3 & $ 88.868$ & $ -51.906$ & & $ 5.00$ & $ 6.59$ & $4.89^{+1.74}_{-4.14}$ & & $ 0.22$ & $ 0.76$ & $2.70^{+1.94}_{-1.62}$ & & $ -5.9$ & $-1.3^{+2.0}_{-1.2}$ & $ 0.09$ & sync & & $ 9$ & $ 796$ & $ 717$ \\
SPT-S J055545$-$5605.1 & $ 88.939$ & $ -56.085$ & & $ 5.00$ & $ 6.56$ & $5.19^{+1.61}_{-2.88}$ & & $ 0.87$ & $ 2.99$ & $3.25^{+2.39}_{-1.54}$ & & $ -2.1$ & $-1.0^{+1.8}_{-1.4}$ & $ 0.08$ & sync & & $ 52$ & $ 398$ & $ 501$ \\
SPT-S J055555$-$5521.8 & $ 88.979$ & $ -55.364$ & & $ 4.65$ & $ 6.10$ & $1.49^{+3.90}_{-1.26}$ & & $ 0.00$ & $ -0.34$ & $1.57^{+1.86}_{-1.30}$ & & $ NaN$ & $-0.8^{+3.0}_{-1.6}$ & $ 0.20$ & sync & & $ 17$ & $ 546$ & $ 810$ \\
SPT-S J055701$-$5902.6 & $ 89.258$ & $ -59.044$ & & $ 4.97$ & $ 6.19$ & $5.09^{+1.46}_{-1.99}$ & & $ 1.61$ & $ 5.26$ & $4.02^{+2.66}_{-1.89}$ & & $ -0.4$ & $-0.3^{+1.7}_{-1.7}$ & $ 0.10$ & sync & & $ 5$ & $ 295$ & $ 546$ \\
SPT-S J055732$-$5359.3\tablenotemark{a} & $ 89.384$ & $ -53.988$ & & $ 5.44$ & $ 7.15$ & $6.43^{+1.41}_{-1.49}$ & & $ 2.83$ & $ 9.67$ & $8.00^{+3.35}_{-3.37}$ & & $ 0.8$ & $0.7^{+1.2}_{-1.5}$ & $ 0.18$ & sync & & $ 10$ & $ 192$ & $ 202$ \\
SPT-S J055734$-$5801.0 & $ 89.394$ & $ -58.018$ & & $ 6.37$ & $ 7.94$ & $7.32^{+1.34}_{-1.34}$ & & $ 2.04$ & $ 6.69$ & $5.71^{+2.90}_{-2.25}$ & & $ -0.5$ & $-0.6^{+1.3}_{-1.4}$ & $ 0.01$ & sync & & $ 15$ & $1347$ & $ 602$ \\
SPT-S J055811$-$5029.8 & $ 89.549$ & $ -50.497$ & & $ 50.99$ & $ 67.18$ & $66.83^{+3.02}_{-3.02}$ & & $ 16.10$ & $ 55.77$ & $54.98^{+6.05}_{-6.05}$ & & $ -0.5$ & $-0.5^{+0.3}_{-0.3}$ & $ 0.00$ & sync & & $ 1$ & $ 925$ & $ 137$ \\
SPT-S J055830$-$5326.4 & $ 89.627$ & $ -53.440$ & & $ 8.83$ & $ 11.59$ & $11.01^{+1.41}_{-1.41}$ & & $ 1.35$ & $ 4.60$ & $5.66^{+2.46}_{-1.46}$ & & $ -2.5$ & $-1.8^{+1.0}_{-0.8}$ & $ 0.00$ & sync & & $ 6$ & $ 383$ & $ 365$ \\
SPT-S J055833$-$5835.5 & $ 89.638$ & $ -58.593$ & & $ 6.43$ & $ 8.02$ & $7.45^{+1.33}_{-1.34}$ & & $ 2.63$ & $ 8.63$ & $7.16^{+3.06}_{-2.86}$ & & $ 0.2$ & $-0.0^{+1.2}_{-1.4}$ & $ 0.04$ & sync & & $ 9$ & $ 401$ & $1445$ \\
SPT-S J055854$-$5333.8 & $ 89.728$ & $ -53.564$ & & $ 5.01$ & $ 6.58$ & $5.80^{+1.43}_{-1.48}$ & & $ 3.91$ & $ 13.35$ & $12.01^{+3.63}_{-3.81}$ & & $ 1.9$ & $2.0^{+1.1}_{-1.2}$ & $ 0.62$ & dust & & $ 11$ & $ 455$ & $ 31$ \\
SPT-S J055909$-$5128.1 & $ 89.788$ & $ -51.469$ & & $ 1.25$ & $ 1.64$ & $2.27^{+1.18}_{-1.66}$ & & $ 5.68$ & $ 19.68$ & $8.54^{+7.82}_{-7.64}$ & & $ 6.8$ & $3.8^{+0.9}_{-3.5}$ & $ 0.75$ & dust & & $ 750$ & $1094$ & $ 6$ \\
SPT-S J055923$-$5900.6 & $ 89.846$ & $ -59.010$ & & $ 5.70$ & $ 7.10$ & $6.50^{+1.33}_{-1.39}$ & & $ 4.26$ & $ 13.94$ & $12.82^{+3.50}_{-3.66}$ & & $ 1.8$ & $1.9^{+0.9}_{-1.0}$ & $ 0.58$ & dust & & $ 6$ & $ 514$ & $ 201$ \\
SPT-S J055947$-$5026.8 & $ 89.947$ & $ -50.447$ & & $ 37.59$ & $ 49.52$ & $49.26^{+2.41}_{-2.41}$ & & $ 12.09$ & $ 41.90$ & $41.18^{+5.11}_{-5.11}$ & & $ -0.5$ & $-0.5^{+0.3}_{-0.4}$ & $ 0.00$ & sync & & $ 5$ & $ 12$ & $ 7$ \\
\tablenotetext{a}{Extended source; see Sec.~\ref{sec:extend}.}
\tablenotetext{b}{No SUMSS source within $30$~arcsec, but SUMSS source and ATCA detection within $35$~arcsec; see Sec.~\ref{sec:assoc}.}
\tablenotetext{c}{Associated with NGC~1824; see Sec.~\ref{sec:extend}.}
\tablenotetext{d}{Near deep SZ decrement but not spurious; see Sec.~\ref{sec:assoc}.}
\tablenotetext{e}{Offset between $1.4$~mm and $2.0$~mm emission; see Sec.~\ref{sec:extend}.}
\tablenotetext{f}{Associated with NGC~1853; see Sec.~\ref{sec:extend}.}
\tablenotetext{g}{Possibly spurious detection from sidelobe of strong SZ decrement; see Sec.~\ref{sec:assoc}.}
\tablenotetext{h}{Chance superposition ($45$~arcsec separation) of SUMSS source and new dust-dominated detection; see Sec.~\ref{sec:extend}.}
\tablenotetext{i}{Two SUMSS sources within $35$~arcsec; see Sec.~\ref{sec:assoc}.}
\tablenotetext{j}{No SUMSS source within $1$~arcmin, but associated with HE~0536-5435~[VCV2001]; see Sec.~\ref{sec:assoc}.}
\tablenotetext{k}{Associated with $\beta$ Pictoris; see Sec.~\ref{sec:assoc}.}
\tablenotetext{l}{Associated with ESO~160-~G~002; see Sec.~\ref{sec:extend}.}

\end{deluxetable}
\end{center}
\clearpage
\end{landscape}
\pagestyle{plain}

\end{document}